\pdfoutput=1
\documentclass[
aps,
prd,
twocolumn,
eqsecnum,
floatfix,
superscriptaddress,
showpacs,
longbibliography
]{revtex4-1} 

\usepackage{graphicx}
\usepackage{dcolumn}
\usepackage{amsmath}
\usepackage{amssymb}
\usepackage{xspace}
\usepackage{color}
\usepackage{subfigure}
\usepackage{braket}
\usepackage{multirow}

\newcommand{\val}[2]{\ensuremath{#1 \; \mathrm{#2}\xspace}}
\newcommand{\sci}[2]{\ensuremath{#1 \times 10^{#2}}\xspace}
\renewcommand{\Re}{\operatorname{Re}}
\renewcommand{\Im}{\operatorname{Im}}

\newcommand{\tab}[1]{Tab.~\ref{tab:#1}}
\newcommand{\fig}[1]{Fig.~\ref{fig:#1}}
\newcommand{\figs}[2]{Figs.~\ref{fig:#1}-\ref{fig:#2}}
\newcommand{\eqn}[1]{Eqn.~\ref{eq:#1}}

\newcommand{\Figure}[1]{Figure~\ref{fig:#1}}

\newcommand{\Equation}[1]{Equation~\ref{eq:#1}}
\newcommand{\sect}[1]{Section~\ref{sec:#1}}
\newcommand{\app}[1]{Appendix~\ref{app:#1}}

\newcommand{\sk}{\mbox{Super-K}\xspace}
\newcommand{\cz}{\ensuremath{\cos \theta_z}\xspace}
\newcommand{\UP}{UP-$\mu$\xspace}

\newcommand{\numu}{\ensuremath{\nu_{\mu}}\xspace}

\newcommand{\nue}{\ensuremath{\nu_{e}}\xspace}

\newcommand{\pizero}{\ensuremath{\pi^0}\xspace}

\newcommand{\dm}{\ensuremath{\Delta m^{2}}\xspace}
\newcommand{\dmsq}[1]{\ensuremath{\dm_{#1}}\xspace}

\newcommand{\sn}[1]{\ensuremath{ \sin^{2}(\theta_{#1}) }\xspace }
\newcommand{\snt}[1]{\ensuremath{ \sin^{2}(2\theta_{#1}) }\xspace }
\newcommand{\Hlv}{\ensuremath{H_{LV}}\xspace}

\newcommand{\at}{\ensuremath{a^{T}}\xspace}
\newcommand{\ctt}{\ensuremath{c^{TT}}\xspace}
\newcommand{\ats}[1]{\ensuremath{\at_{#1}}\xspace}
\newcommand{\ctts}[1]{\ensuremath{\ctt_{#1}}\xspace}
\newcommand{\as}[1]{\ensuremath{\left(a^{T}_{#1}\right)^*}\xspace}
\newcommand{\cs}[1]{\ensuremath{\left(c^{TT}_{#1}\right)^*}\xspace}

\newcommand{\sonetwo}{0.305}

\newcommand{\stonethree}{0.095}
\newcommand{\stwothree}{0.514}

\newcommand{\chisq}{\ensuremath{\chi^{2}}\xspace}

\DeclareMathOperator{\Tr}{Tr}

\newcommand{\fwid}{8cm}

\newcommand{\zwid}{5cm}

\renewcommand{\tabcolsep}{4pt}

\graphicspath{ {fig/} {fig/lv_ratio_a_all/} {fig/lv_ratio_c_all/} }

\begin{document}

\title{Test of Lorentz Invariance with Atmospheric Neutrinos}
\newcommand{\AFFicrr}{\affiliation{Kamioka Observatory, Institute for Cosmic Ray Research, University of Tokyo, Kamioka, Gifu 506-1205, Japan}}
\newcommand{\AFFkashiwa}{\affiliation{Research Center for Cosmic Neutrinos, Institute for Cosmic Ray Research, University of Tokyo, Kashiwa, Chiba 277-8582, Japan}}
\newcommand{\AFFipmu}{\affiliation{Kavli Institute for the Physics and
Mathematics of the Universe (WPI), Todai Institutes for Advanced Study,
University of Tokyo, Kashiwa, Chiba 277-8582, Japan }}
\newcommand{\AFFmad}{\affiliation{Department of Theoretical Physics, University Autonoma Madrid, 28049 Madrid, Spain}}
\newcommand{\AFFubc}{\affiliation{Department of Physics and Astronomy, University of British Columbia, Vancouver, BC, V6T1Z4, Canada}}
\newcommand{\AFFbu}{\affiliation{Department of Physics, Boston University, Boston, MA 02215, USA}}
\newcommand{\AFFbnl}{\affiliation{Physics Department, Brookhaven National Laboratory, Upton, NY 11973, USA}}
\newcommand{\AFFuci}{\affiliation{Department of Physics and Astronomy, University of California, Irvine, Irvine, CA 92697-4575, USA }}
\newcommand{\AFFcsu}{\affiliation{Department of Physics, California State University, Dominguez Hills, Carson, CA 90747, USA}}
\newcommand{\AFFcnm}{\affiliation{Department of Physics, Chonnam National University, Kwangju 500-757, Korea}}
\newcommand{\AFFduke}{\affiliation{Department of Physics, Duke University, Durham NC 27708, USA}}
\newcommand{\AFFfukuoka}{\affiliation{Junior College, Fukuoka Institute of Technology, Fukuoka, Fukuoka 811-0295, Japan}}
\newcommand{\AFFgmu}{\affiliation{Department of Physics, George Mason University, Fairfax, VA 22030, USA }}
\newcommand{\AFFgifu}{\affiliation{Department of Physics, Gifu University, Gifu, Gifu 501-1193, Japan}}
\newcommand{\AFFgist}{\affiliation{GIST College, Gwangju Institute of Science and Technology, Gwangju 500-712, Korea}}
\newcommand{\AFFuh}{\affiliation{Department of Physics and Astronomy, University of Hawaii, Honolulu, HI 96822, USA}}
\newcommand{\AFFkanagawa}{\affiliation{Physics Division, Department of Engineering, Kanagawa University, Kanagawa, Yokohama 221-8686, Japan}}
\newcommand{\AFFkek}{\affiliation{High Energy Accelerator Research Organization (KEK), Tsukuba, Ibaraki 305-0801, Japan }}
\newcommand{\AFFkobe}{\affiliation{Department of Physics, Kobe University, Kobe, Hyogo 657-8501, Japan}}
\newcommand{\AFFkyoto}{\affiliation{Department of Physics, Kyoto University, Kyoto, Kyoto 606-8502, Japan}}
\newcommand{\AFFumd}{\affiliation{Department of Physics, University of Maryland, College Park, MD 20742, USA }}
\newcommand{\AFFmit}{\affiliation{Department of Physics, Massachusetts Institute of Technology, Cambridge, MA 02139, USA}}
\newcommand{\AFFmiyagi}{\affiliation{Department of Physics, Miyagi University of Education, Sendai, Miyagi 980-0845, Japan}}
\newcommand{\AFFnagoya}{\affiliation{Solar Terrestrial Environment Laboratory, Nagoya University, Nagoya, Aichi 464-8602, Japan}}
\newcommand{\AFFpol}{\affiliation{National Centre For Nuclear Research, 00-681 Warsaw, Poland}}
\newcommand{\AFFsuny}{\affiliation{Department of Physics and Astronomy, State University of New York at Stony Brook, NY 11794-3800, USA}}
\newcommand{\AFFniigata}{\affiliation{Department of Physics, Niigata University, Niigata, Niigata 950-2181, Japan }}
\newcommand{\AFFokayama}{\affiliation{Department of Physics, Okayama University, Okayama, Okayama 700-8530, Japan }}
\newcommand{\AFFosaka}{\affiliation{Department of Physics, Osaka University, Toyonaka, Osaka 560-0043, Japan}}
\newcommand{\AFFregina}{\affiliation{Department of Physics, University of Regina, 3737 Wascana Parkway, Regina, SK, S4SOA2, Canada}}
\newcommand{\AFFseoul}{\affiliation{Department of Physics, Seoul National University, Seoul 151-742, Korea}}
\newcommand{\AFFshizuokasc}{\affiliation{Department of Informatics in
Social Welfare, Shizuoka University of Welfare, Yaizu, Shizuoka, 425-8611, Japan}}
\newcommand{\AFFskk}{\affiliation{Department of Physics, Sungkyunkwan University, Suwon 440-746, Korea}}
\newcommand{\AFFtohoku}{\affiliation{Research Center for Neutrino Science, Tohoku University, Sendai, Miyagi 980-8578, Japan}}
\newcommand{\AFFtokyo}{\affiliation{The University of Tokyo, Bunkyo, Tokyo 113-0033, Japan }}
\newcommand{\AFFtorront}{\affiliation{Department of Physics, University of Toronto, 60 St. George Street, Toronto, Ontario, M5S1A7, Canada }}
\newcommand{\AFFtriumf}{\affiliation{TRIUMF, 4004 Wesbrook Mall, Vancouver, BC, V6T2A3, Canada }}
\newcommand{\AFFtokai}{\affiliation{Department of Physics, Tokai University, Hiratsuka, Kanagawa 259-1292, Japan}}
\newcommand{\AFFtit}{\affiliation{Department of Physics, Tokyo Institute
for Technology, Meguro, Tokyo 152-8551, Japan }}
\newcommand{\AFFtsinghua}{\affiliation{Department of Engineering Physics, Tsinghua University, Beijing, 100084, China}}
\newcommand{\AFFwarsaw}{\affiliation{Institute of Experimental Physics, Warsaw University, 00-681 Warsaw, Poland }}
\newcommand{\AFFuw}{\affiliation{Department of Physics, University of Washington, Seattle, WA 98195-1560, USA}}

\AFFicrr
\AFFkashiwa
\AFFmad
\AFFbu
\AFFubc
\AFFbnl
\AFFuci
\AFFcsu
\AFFcnm
\AFFduke
\AFFfukuoka
\AFFgifu
\AFFgist
\AFFuh
\AFFkek
\AFFkobe
\AFFkyoto
\AFFmiyagi
\AFFnagoya
\AFFsuny
\AFFokayama
\AFFosaka
\AFFregina
\AFFseoul
\AFFshizuokasc
\AFFskk
\AFFtokai
\AFFtokyo
\AFFipmu
\AFFtorront
\AFFtriumf
\AFFtsinghua
\AFFuw

\author{K.~Abe}
\AFFicrr
\AFFipmu
\author{Y.~Haga}
\AFFicrr
\author{Y.~Hayato}
\AFFicrr
\AFFipmu
\author{M.~Ikeda}
\AFFicrr
\author{K.~Iyogi}
\AFFicrr 
\author{J.~Kameda}
\author{Y.~Kishimoto}
\author{M.~Miura} 
\author{S.~Moriyama} 
\author{M.~Nakahata}
\AFFicrr
\AFFipmu 
% ...

\collaboration{The Super-Kamiokande Collaboration}
\noaffiliation

\date{\today}

\begin{abstract}
A search for neutrino oscillations induced by Lorentz violation has been performed using 4,438 live-days of Super-Kamiokande atmospheric neutrino data. The Lorentz violation is included in addition to standard three-flavor oscillations using the non-perturbative Standard Model Extension (SME), allowing the use of the full range of neutrino path lengths, ranging from 15 to 12,800 km, and energies ranging from 100 MeV to more than 100 TeV in the search. 
No evidence of Lorentz violation was observed, so limits are set on the renormalizable isotropic SME coefficients in the $e\mu$, $\mu\tau$, and $e\tau$ sectors, improving the existing limits by up to seven orders of magnitude and setting limits for the first time in the neutrino $\mu\tau$ sector of the SME.

\end{abstract}

\maketitle

\section{Introduction}

Symmetry under Lorentz transformations is a fundamental feature of both the standard model of particle physics and the general theory of relativity, but violations of this symmetry at or below the Planck scale, $m_P \approx \val{10^{19}}{GeV}$, have been predicted in a variety of models, including discrete spacetime structure and spacetime foam interactions~\cite{Kostelecky:1988zi,Hawking:1976ra}.  The direct observation of Lorentz violation (LV) would provide access to this Planck-scale physics~\cite{Brustein:2001ik,Colladay:1996iz,Colladay:1998fq,Kostelecky:2003fs}.  The Standard Model Extension~\cite{AmelinoCamelia:2005qa,Bluhm:2005uj,Colladay:1996iz,Colladay:1998fq,Kostelecky:2003fs} (SME) is an observer-independent effective field theory with all the features of the standard model and general relativity plus all possible LV terms.  Lorentz violation can also include violation of charge-parity-time reversal ($CPT$) symmetry~\cite{Greenberg:2002uu}\footnote{While LV can exist without $CPT$ violation, $CPT$ violation requires LV.}.  At experimentally accessible energies, these LV signatures are strongly suppressed by a factor of the order of $m_W/m_P \approx 10^{-17}$, 
the relative magnitudes of the electroweak and Planck scales~\cite{Kostelecky:2004hg}. Despite this suppression, numerous experimental techniques have been employed to search for LV phenomena~\cite{Kostelecky:2008ts,Mattingly:2005re,Kostelecky:1999dx,*Kostelecky:2002zz,*Kostelecky:2005mj,*Kostelecky:2008zz}.

Neutrino oscillations, as an interferometric effect, are a sensitive probe of LV with two possible signatures: sidereal variations, which would be evidence of a preferred spatial direction, and spectral anomalies~\cite{Kostelecky:2011gq,Kostelecky:2003xn,Kostelecky:2003cr}. Previous searches within the SME framework have generally focused on sidereal variations, though they sometimes include time-independent components that only modify the spectrum. They have been performed in short-baseline muon (anti)neutrino beams~\cite{Auerbach:2005tq, AguilarArevalo:2011yi, Adamson:2008aa, Adamson:2012hp}, the long-baseline NuMI neutrino beam~\cite{Adamson:2010rn,Rebel:2013vc}, the Double-Chooz reactor experiment~\cite{Abe:2012gw,Diaz:2013iba}, and in atmospheric neutrinos at Ice Cube~\cite{Abbasi:2010kx}.  These experiments have generally used either the short-baseline~\cite{Kostelecky:2004hg} or perturbative~\cite{Diaz:2009qk} approximations of the SME to set limits on LV parameters and report limits for each parameter independently.  

The Super-Kamiokande experiment (SK)~\cite{Fukuda:2002uc} is a cylindrical, underground, water-Cherenkov detector, with a fiducial mass of \val{22.5}{kton} in the Inner Detector (ID) and an active veto outer detector (OD) for tagging cosmic ray muons entering the detector. 
The atmospheric neutrinos are incident from all directions with path lengths and energies spanning three and six orders of magnitude, respectively.  This wide range in $L$ and $E$ makes atmospheric neutrinos a sensitive tool to probe the coefficients which produce spectral anomalies. 
The SK data cover such a wide range of lengths and energies that the perturbative SME can no longer be used and the exact Hamiltonian must be diagonalized~\cite{Akiri:2013hca}. Since the oscillation of massive neutrinos has been well-established in numerous experiments~\cite{Fukuda:1998mi,Ashie:2004mr,Abe:2010hy,Cleveland:1998nv, Abdurashitov:2009tn, Altmann:2005ix, Hampel:1998xg, Aharmim:2011vm,Abe:2008aa, An:2012eh, Abe:2013sxa, Ahn:2012nd, Ahn:2006zza, Adamson:2013whj, Abe:2012gx, Abe:2012jj, Abe:2013xua, Agafonova:2013dtp}, the exact Hamiltonian includes three-flavor oscillations and MSW matter effects~\cite{Wolfenstein:1977ue,*Mikheyev:1989dy} in addition to LV. We investigate the real and imaginary parts of the lowest-order effective $CPT$-even and $CPT$-odd LV coefficients in the $e\mu$, $\mu\tau$, and $e\tau$ sectors. We are the first neutrino experiment to study isotropic LV in the $\mu\tau$ sector of the SME; all previous experiments sensitive to this sector searched only for sidereal variations.

The neutrino oscillation probability, whether from standard oscillations or Lorentz violation, depends on the initial neutrino flavor, the distance the neutrino travels, $L$, and the neutrino energy, $E$. We separate our data into samples correlated with energy and with enhanced \numu, \nue or neutral current (NC) flavor content. The events fully contained (FC) within the inner detector have the lowest energies.  Events that start in the inner detector but then exit before depositing all their energy are classified as partially contained (PC) and have generally have higher energies.  The up-going muon events (\UP) that enter the detector having deposited some of their energy in the surrounding rock beforehand are the highest energy sample.  This sample only contains up-going events to avoid the overwhelming background of down-going cosmic ray muons.  While the energy of the muons cannot be determined event-by-event, the highest-energy muons will shower inside the detector and can be identified using the method described in~\cite{Desai:2007ra}.  

We then further bin the data using observables correlated with $L$ and $E$.
Instead of path length, we bin the data in zenith angle, \cz, defined as the angle between the event direction and the downward vertical direction.  The neutrinos with the shortest path lengths are downward-going (\cz near 1) and the neutrinos with the longest path lengths are upward-going (\cz near -1).  The simulation which predicts the number of neutrino events in each bin includes a distribution of neutrino production heights based on a model of the atmosphere described in more detail in~\cite{Ashie:2005ik}.  This range of production heights introduces a smearing of the oscillation probability for a given zenith angle for downward-going and horizontal events but is negligible for upward-going events which cross most of the Earth.
For events with one visible Cherenkov ring, we bin in momentum, which is reconstructed using the total amount of light with a $70^\circ$ cone, and then refined using templates from simulation.
For multi-ring events, partially-contained events, and stopping \UP events, we bin in visible energy,
defined as the energy of an electron that would produce the same total amount of light seen in the detector. The data are divided into a total of 480 bins for each run period, which are then combined across run periods before fitting.  The binning is chosen so that enough events are expected in each bin for the fit to be stable.   The binning scheme is largely the same as that used for the standard three-flavor oscillation analysis~\cite{Wendell:2010md}, with some upgrades described in~\cite{sterilepaper}.

The various data samples and the SK-I through SK-IV data used in this analysis are described in detail in~\cite{sterilepaper}, the event generator, Monte Carlo simulation (MC), and reconstruction are described in~\cite{Ashie:2005ik}, and recent improvements are described in~\cite{Abe:2013gga}.

\section{Lorentz violation in neutrino oscillations}\label{sec:theory}

\begin{figure}
  \subfigure[~$\numu \to \numu$, No Lorentz violation]{
        \includegraphics[width=7.4cm,clip]{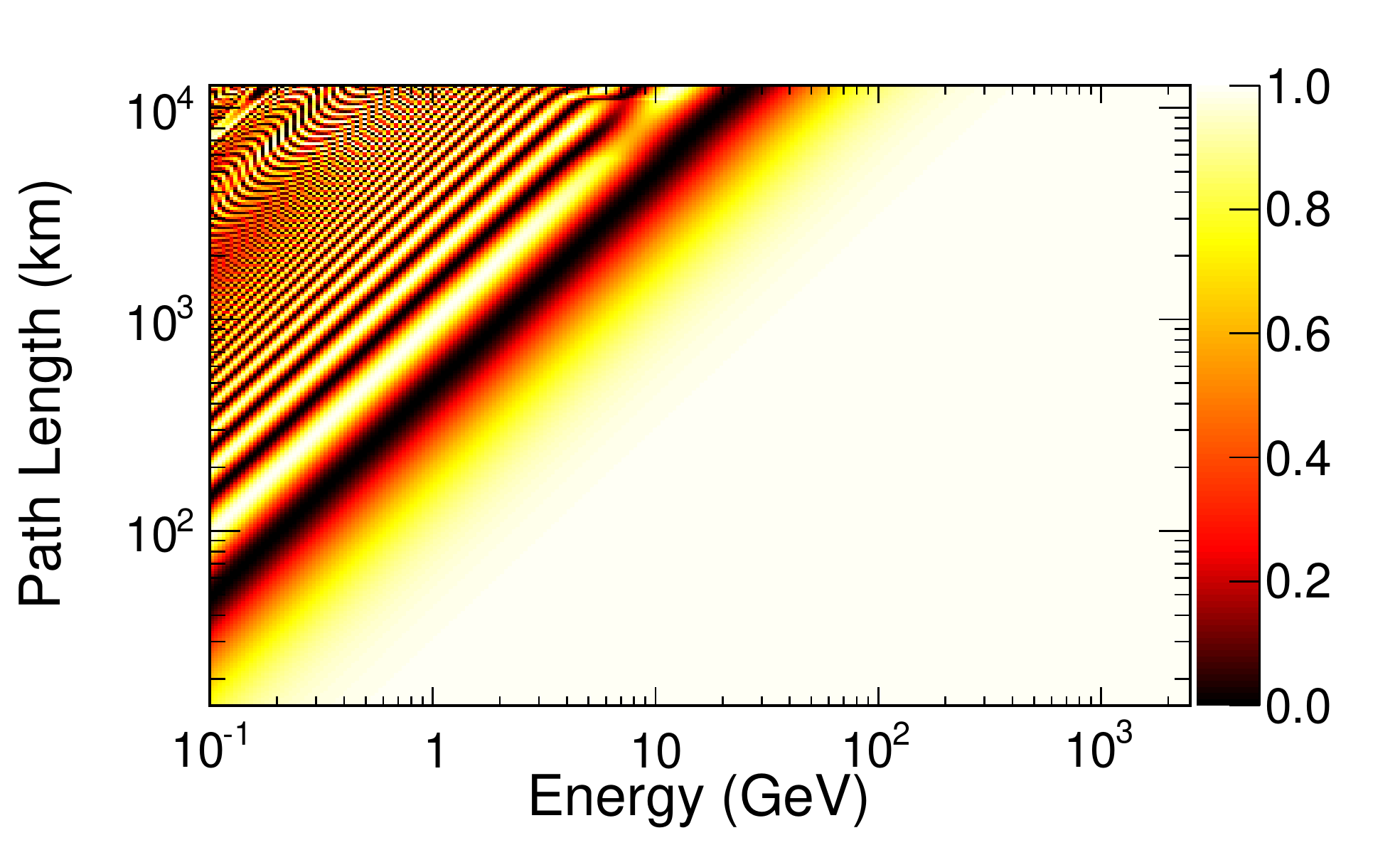} \label{fig:ogrm_nolv}
  }
  \subfigure[~$\numu \to \numu$, $\ats{\mu\tau} = \val{10^{-22}}{GeV}$]{
        \includegraphics[width=7.4cm,clip]{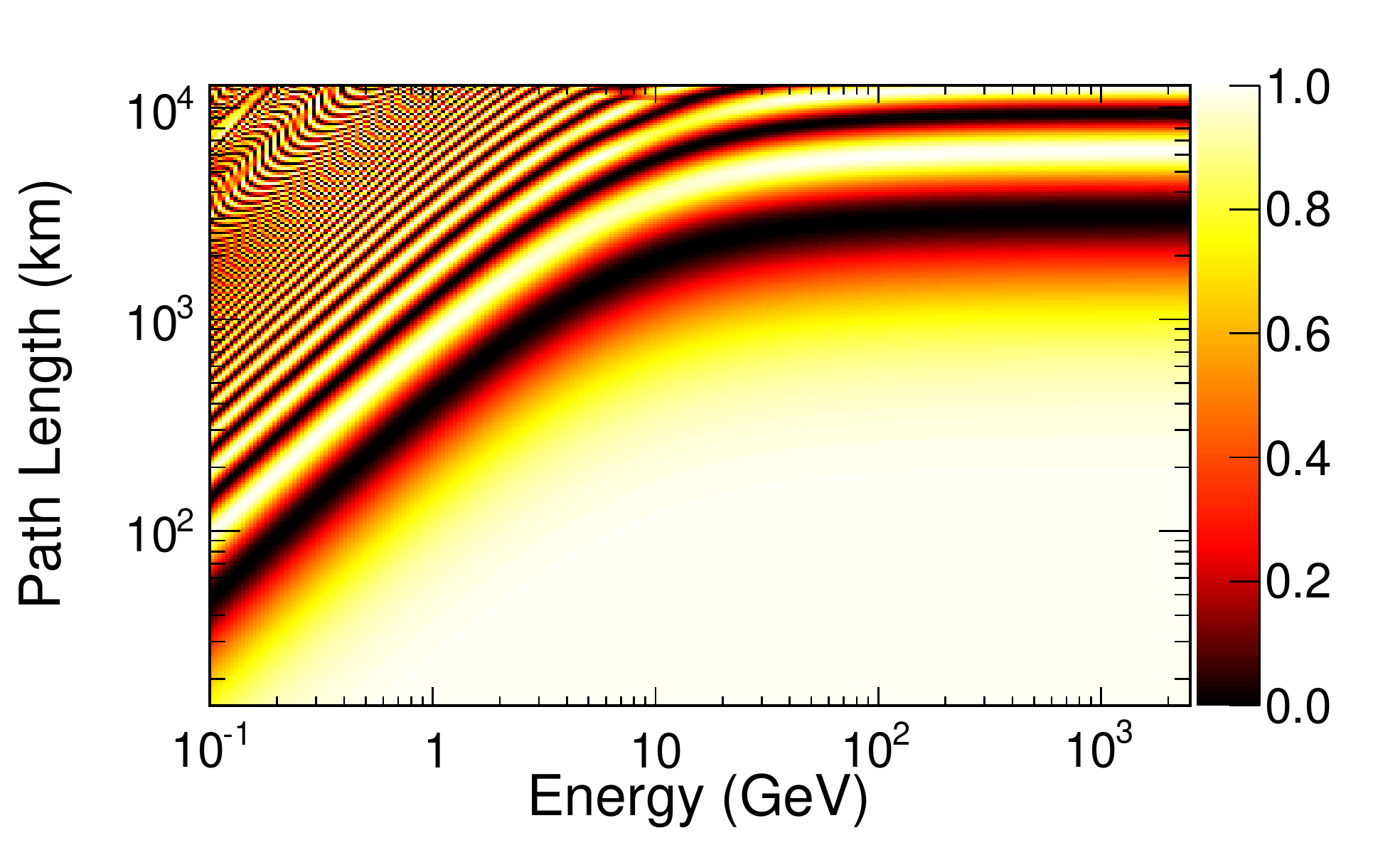} \label{fig:ogrm_a}
  }
  \subfigure[~$\numu \to \numu$, $\ctts{\mu\tau} = \sci{7.5}{-23}$]{
        \includegraphics[width=7.4cm,clip]{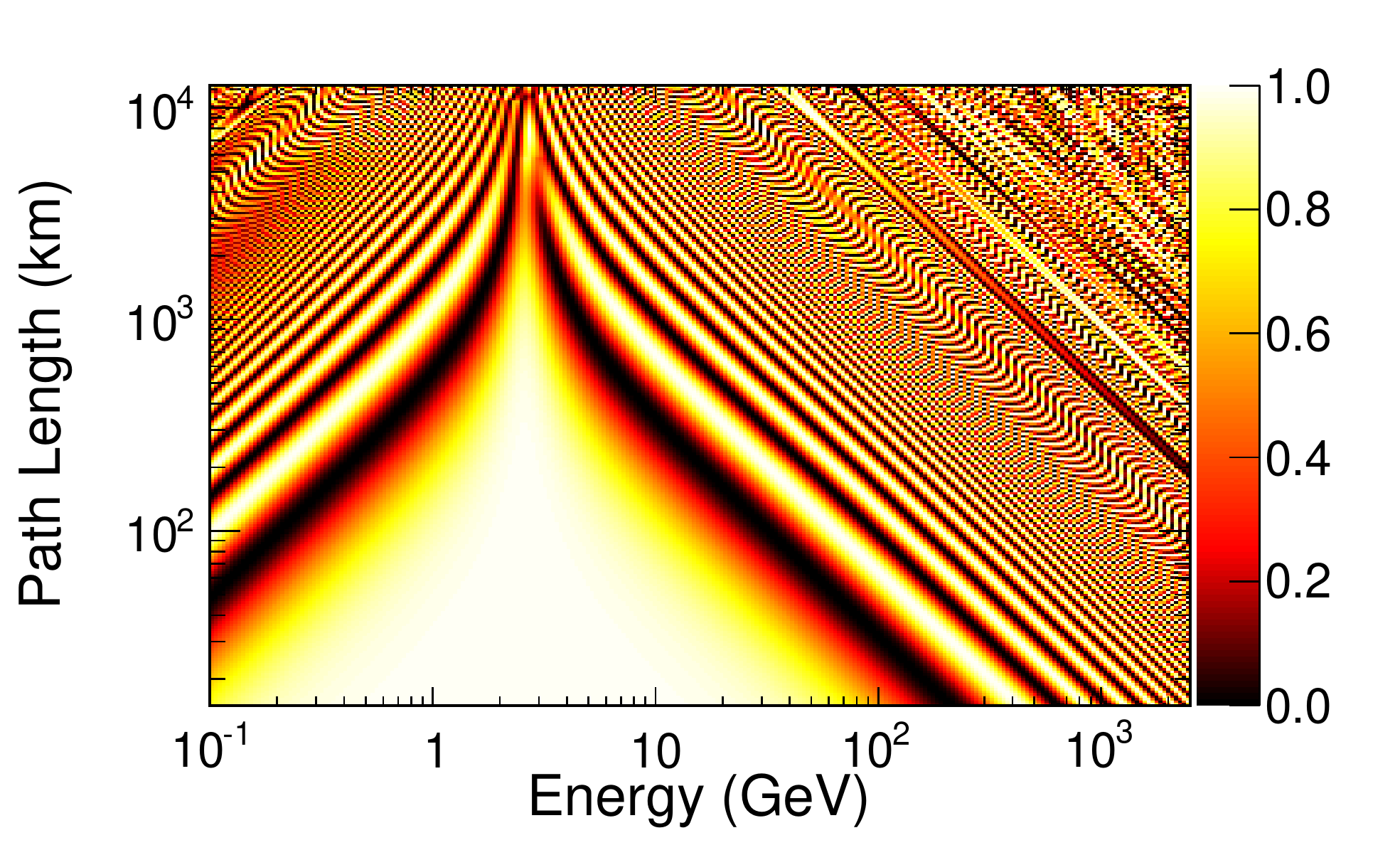} \label{fig:ogrm_c}
  }
  \caption{
    (color online) The $\numu \to \numu$ oscillation probabilities, plotted in path length vs. neutrino energy.  \subref{fig:ogrm_nolv} Standard oscillations appear as lines of constant $L/E$ which have slope 1 on this log-log scale.  Standard three-flavor oscillations are concentrated in the upper-left portion in all three oscillograms, corresponding to low energy and long distance. \subref{fig:ogrm_a} The \ats{\mu\tau} coefficient introduces oscillations proportional to $L$, which appear as horizontal lines (constant $L$) at high energies.  \subref{fig:ogrm_c} The \ctts{\mu\tau} coefficient introduces $LE$ oscillations which appear as lines with slope minus one.  Oscillograms for all sectors, as well as the $\mu \to e$ probabilities are shown in \app{oscillogram}.
  }
  \label{fig:oscillogram_example}
\end{figure}

\begin{table}
\centering
\begin{tabular}{lcccc}
\hline\hline
Coefficient                      & Unit & $d$ & $CPT$ & Oscillation Effect \\
\hline
\multicolumn{5}{l}{\bf Isotropic} \\
\hspace{1em}$a_{\alpha\beta}^{T}$            & GeV  & 3                      & odd            & $\propto L$  \\
\hspace{1em}$c_{\alpha\beta}^{TT}$           &  -   & 4                      & even           & $\propto LE$ \\
\multicolumn{5}{l}{\bf Directional} \\
\hspace{1em}$a_{\alpha\beta}^{X}, a_{\alpha\beta}^{Y}, a_{\alpha\beta}^{Z}$  & GeV  & 3 & odd            & sidereal variation \\
\hspace{1em}$c_{\alpha\beta}^{XX}, c_{\alpha\beta}^{YZ}, \ldots$             &  -   & 4 & even           & sidereal variation \\
\hline\hline
\end{tabular}
\caption{Lorentz-violating coefficients and their properties. The last row includes all possible combinations of $X,Y,Z$, and $T$ except $TT$. $d$ refers to the dimension of the operator. $\alpha$ and $\beta$ range over the neutrino flavors, $e$, $\mu$, and $\tau$. The $X$, $Y$, and $Z$ indicate coefficients which introduce effects in a particular direction in a Lorentz-violating preferred reference frame.  The $T$ and $TT$ terms are not associated with any direction and thus introduce isotropic distortions in the oscillation pattern.}
\label{tab:coefficients}
\end{table}

In the SME, Lorentz violation is included with neutrino oscillations by adding an LV term, \Hlv, to the standard neutrino Hamiltonian,
\begin{equation}
H = U M U^\dagger + V_e + \Hlv,
\end{equation}
where $U$ is the PMNS mixing matrix~\cite{Maki:1962mu}, $M$ is the neutrino mass matrix, 
\begin{equation}
M = \frac{1}{2E}
\left(\begin{array}{ccc}
0 & 0 & 0 \\
0 & \dmsq{21} & 0 \\
0 & 0 & \dmsq{31}
\end{array}\right),
\end{equation}
and $V_e$ is the electron potential which introduces matter effects~\cite{Wolfenstein:1977ue,*Mikheyev:1989dy},
\begin{equation}
V_e = \pm \sqrt{2} G_F 
\left(\begin{array}{ccc}
N_e & 0 & 0 \\
0 & 0 & 0 \\
0 & 0 & 0
\end{array}\right),
\end{equation}
where $G_F$ is the Fermi constant and $N_e$ is the average electron density along the neutrino's path, calculated using the four-layer PREM model of the density profile of the Earth~\cite{Dziewonski:1981xy}.

The Lorentz-violating Hamiltonian, \Hlv, has many possible terms with complex coefficients summarized in \tab{coefficients}, broadly categorized as isotropic or directional. While in principle atmospheric neutrinos are sensitive to the sidereal variations induced by the directional terms~\cite{Abbasi:2010kx}, in this analysis we focus only on the isotropic terms which introduce spectral variations which oscillate depending on $L$ and $LE$ (as opposed to the $L/E$ dependence of standard oscillations~\cite{Wang:2007zzl}). The diagonal elements of \Hlv have also been neglected since they cannot be observed in oscillations, giving,
\begin{align}
\Hlv =& 
\left(\begin{array}{ccc}
0          & \ats{e\mu}     & \ats{e\tau} \\
\as{e\mu}  & 0            & \ats{\mu\tau} \\
\as{e\tau} & \as{\mu\tau} & 0  \\
\end{array}\right) \nonumber\\
& -\frac{4E}{3} 
\left(\begin{array}{ccc}
0          & \ctts{e\mu}     & \ctts{e\tau} \\
\cs{e\mu}  & 0            & \ctts{\mu\tau} \\
\cs{e\tau} & \cs{\mu\tau} & 0  \\
\end{array}\right),
\end{align}
for neutrinos.  For antineutrinos, the \at parameters go to $-(\at)^*$ and the \ctt parameters go to $(\ctt)^*$, which is equivalent to $\Re\left(\at \right) \to -\Re\left(\at \right)$ and $\Im\left(\ctt \right) \to -\Im\left(\ctt \right)$.

\begin{figure*}
 \begin{center} 
 \includegraphics[width=\zwid,clip]{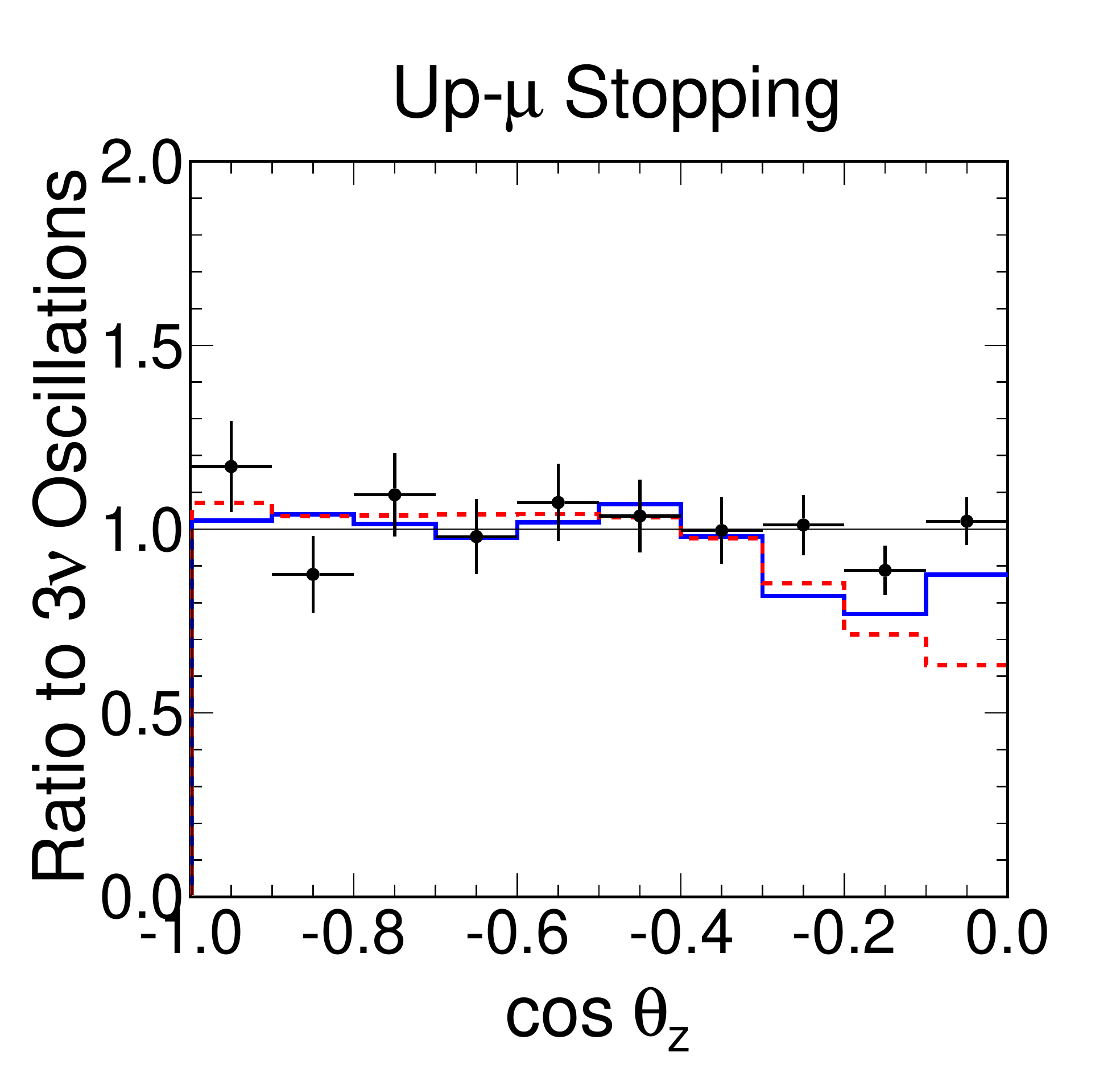}
 \includegraphics[width=\zwid,clip]{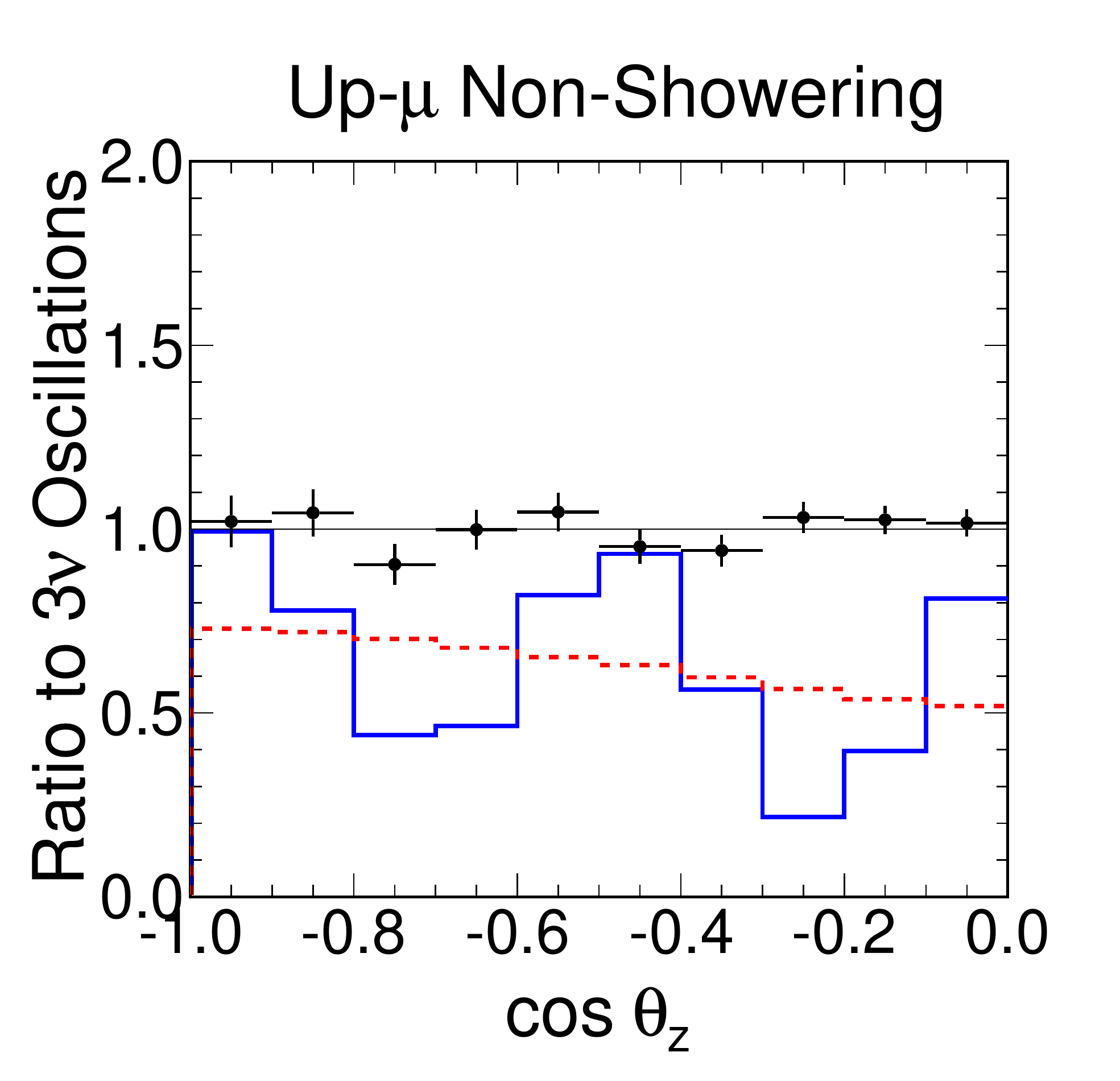}
 \includegraphics[width=\zwid,clip]{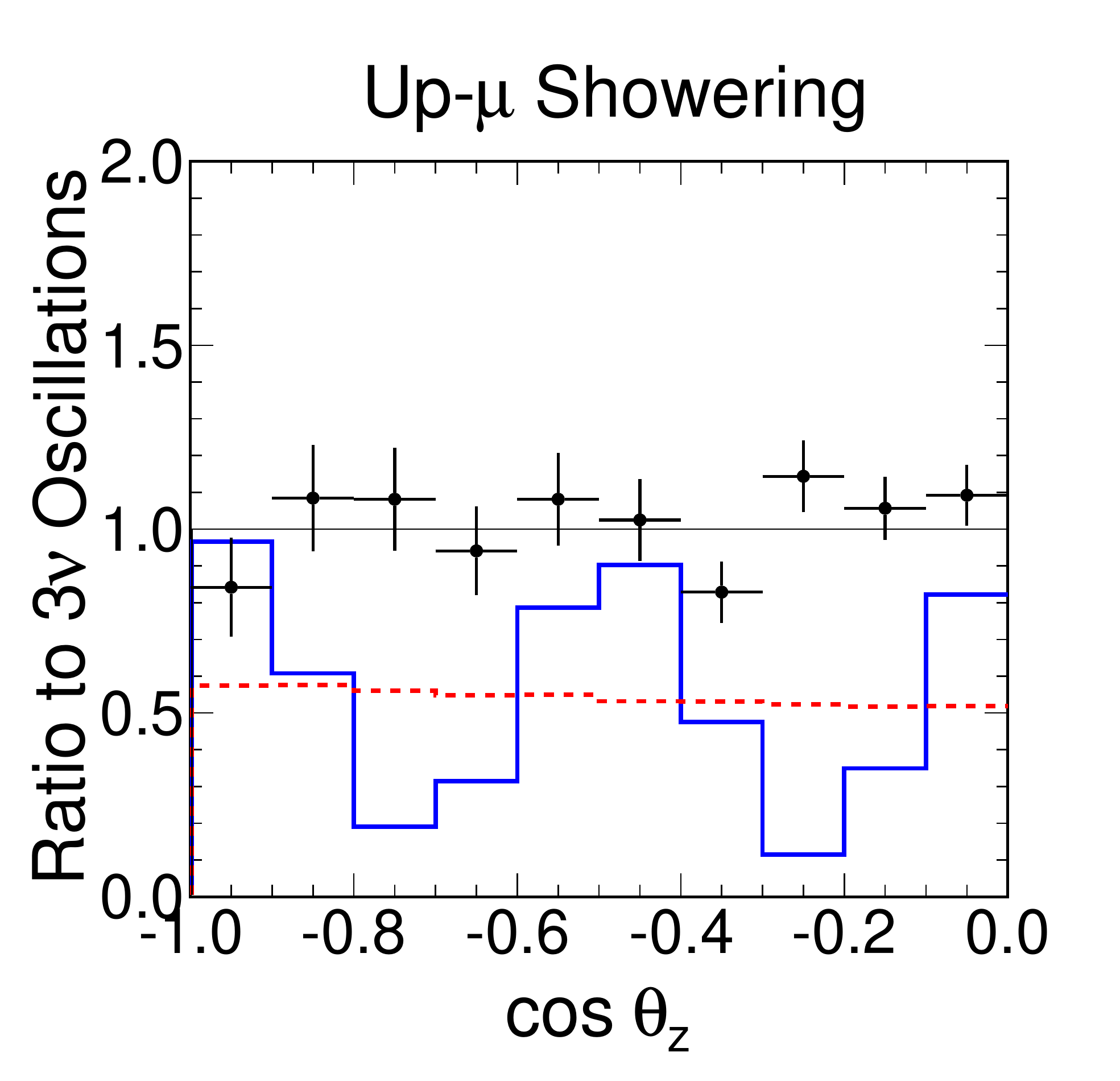}
 \caption{ (color online) Ratios of the summed SK-I through SK-IV \cz distributions relative to standard three-flavor oscillations for the \UP sub-samples, which are the most sensitive to the effects of LV.  The stopping sub-sample (left) contains neutrinos with energies peaking around \val{10}{GeV}, the non-showering sub-sample (center) peaks around \val{100}{GeV}, and the showering sub-sample (right) peaks around \val{1}{TeV}.  The black points represent the data with statistical errors.  The lines corresponds to the MC prediction including Lorentz-violating effects, with $\ats{\mu\tau} = \val{10^{-22}}{GeV}$ in solid blue and $\ctts{\mu\tau} = \sci{7.5}{-23}$ in dashed red.
 }
 \label{fig:zenith_example}
 \end{center}
\end{figure*}

A common approach in neutrino oscillation experiments is to treat \Hlv as a small perturbation $\delta h$ on the standard model Hamiltonian~\cite{Diaz:2009qk} and calculate the lowest order non-zero variations in the oscillation probability (first order when standard oscillations are present, second order when they are not, such as at short baselines). The sensitivity of \sk to this model was evaluated~\cite{Akiri:2013hca}.  However, for this approach to be valid the perturbation must be small, defined as $|\delta h| \ll 1/L$. If we take the condition as $|\delta h| < 10\% \times (1/L)$, then more than 30\% of the events in SK fail this perturbative condition for $\at = \val{\sci{5}{-24}}{GeV}$, resulting in unphysical oscillation probabilities greater than one and less than zero. Since the events failing the condition belong to the samples most sensitive to Lorentz-violation effects, this model was deemed inappropriate for the \sk atmospheric neutrino analysis.  Instead, we use an exact diagonalization of $H$ which produces bounded oscillation probabilities in \sk samples for all values of the Lorentz-violating coefficients, shown in detail in \app{diagonalization}. 
The accuracy of this calculation was ensured by confirming that the oscillation probabilities from the full diagonalization matched the standard three-flavor oscillation calculation used in SK (based on~\cite{Barger:1980tf}) and the perturbative calculations for parameter values that were valid in the perturbative scheme.  

\Figure{oscillogram_example} shows examples of the \numu survival probability vs. energy and path length for the \at and \ctt parameters in the $\mu\tau$ sector. Standard oscillations appear as lines of constant $L/E$, which have slope one on these log-log plots.
The \ats{\mu\tau} and \ats{e\mu} coefficients create oscillation patterns in \numu disappearance that depend only on length. These oscillations will appear as horizontal lines, which can be seen in \fig{ogrm_a} at high energies where there are no $L/E$ oscillations.  The distance (or equivalently \cz) at which the LV oscillation begin is set by the value of \at.
The \ctts{\mu\tau} and \ctts{e\mu} coefficients introduce $LE$ oscillations which will appear as lines with slope minus one, which can be seen at high energies in \fig{ogrm_c}.  The value of \ctt controls the energy the new oscillations begin at.

The samples most sensitive to the high-energy $\mu\tau$ signatures are the \UP samples.  \Figure{zenith_example} shows the zenith-angle distributions of the three \UP data samples, as ratios relative to standard oscillations, compared with the MC predictions corresponding $\ats{\mu\tau} = \val{10^{-22}}{GeV}$ and $\ctts{\mu\tau} = \sci{7.5}{-23}$ (the same as the examples in  \fig{oscillogram_example}). The length-only oscillations from \ats{\mu\tau} appear as large, zenith-dependent oscillations in the non-showering and showering \UP samples since \cz is monotonically (though not linearly) related to distance.  The fast oscillations at high energy from \ctts{\mu\tau} create significant extra \numu disappearance at all \cz's in the same through-going samples.

Plots of the \numu survival probabilities and the $\numu \to \nue$ oscillation probabilities for all the \at and \ctt parameters can be seen in \app{oscillogram} and the zenith distributions of all the samples compared with the data can be seen in \app{zeniths}.  
Both \ats{e\mu} and \ctts{e\mu} behave much like their $\mu\tau$ counterpart in the highest energy samples, but also introduce some smaller but significant changes in the lower energy $e$-like and $\mu$-like samples that would allow the effects of the two sectors to be distinguished from one another.  The \ats{e\tau} and \ctts{e\tau} terms, on the other hand, behave quite differently from the other sectors: they reduce or eliminate $L/E$ oscillations that should otherwise occur at medium and higher energies.  So, instead of extra \numu disappearance there is less.  They also enhance the \nue appearance signal at lower energies. 
Oscillograms are only shown for non-zero real parts of the parameters, but the real and imaginary parts produce similar oscillation effects in the high-energy regions where LV-induced oscillations are dominant.  The influence of the imaginary parts is only in this high-energy region while the real parts also introduce small modifications in the low-energy oscillation probability.

\section{Lorentz-violating oscillation analysis}

\begin{table*}
\renewcommand{\tabcolsep}{2pt}
\newcommand{\tsp}{\hspace{1.2em}}
\centering
\begin{tabular}{l@{\tsp}c@{\tsp}cc@{\tsp}cc@{\tsp}c@{\tsp}ccc}
\hline\hline
\multicolumn{2}{l}{LV Parameter}           & 
\multicolumn{2}{c}{Limit at $95\%$ C.L.}     & 
\multicolumn{2}{c}{Best Fit}               & 
No LV $\Delta \chisq$                      & 
\multicolumn{3}{c}{Previous Limit}\\
\hline
\multirow{4}{*}{$e\mu    $} & $\Re\left(\at \right)$ & $1.8 \times 10^{-23} $ & GeV  & $1.0 \times 10^{-23} $ & GeV  & \multirow{2}{*}{1.4} & \multirow{2}{*}{$4.2 \times 10^{-20} $}  & \multirow{2}{*}{GeV }  & \multirow{2}{*}{\cite{Katori:2012pe}} \\
                            & $\Im\left(\at \right)$ & $1.8 \times 10^{-23} $ & GeV  & $4.6 \times 10^{-24} $ & GeV  &                             &    &  &  \\
                            & $\Re\left(\ctt\right)$ & $8.0 \times 10^{-27} $ &      & $1.0 \times 10^{-28} $ &      & \multirow{2}{*}{0.0} & \multirow{2}{*}{$9.6 \times 10^{-20} $}  & \multirow{2}{*}{    }  & \multirow{2}{*}{\cite{Katori:2012pe}} \\
                            & $\Im\left(\ctt\right)$ & $8.0 \times 10^{-27} $ &      & $1.0 \times 10^{-28} $ &      &                             &    &  &  \\
\hline
\multirow{4}{*}{$e\tau   $} & $\Re\left(\at \right)$ & $4.1 \times 10^{-23} $ & GeV  & $2.2 \times 10^{-24} $ & GeV  & \multirow{2}{*}{0.0} & \multirow{2}{*}{$7.8 \times 10^{-20} $}  & \multirow{2}{*}{GeV }  & \multirow{2}{*}{\cite{Katori:2013jca}} \\
                            & $\Im\left(\at \right)$ & $2.8 \times 10^{-23} $ & GeV  & $1.0 \times 10^{-28} $ & GeV  &                             &    &  &  \\
                            & $\Re\left(\ctt\right)$ & $9.3 \times 10^{-25} $ &      & $1.0 \times 10^{-28} $ &      & \multirow{2}{*}{0.3} & \multirow{2}{*}{$1.3 \times 10^{-17} $}  & \multirow{2}{*}{    }  & \multirow{2}{*}{\cite{Katori:2013jca}} \\
                            & $\Im\left(\ctt\right)$ & $1.0 \times 10^{-24} $ &      & $3.5 \times 10^{-25} $ &      &                             &    &  &  \\
\hline
\multirow{4}{*}{$\mu\tau $} & $\Re\left(\at \right)$ & $6.5 \times 10^{-24} $ & GeV  & $3.2 \times 10^{-24} $ & GeV  & \multirow{2}{*}{0.9} & \multirow{2}{*}{$-                   $}  & \multirow{2}{*}{    }  & \multirow{2}{*}{} \\
                            & $\Im\left(\at \right)$ & $5.1 \times 10^{-24} $ & GeV  & $1.0 \times 10^{-28} $ & GeV  &                             &    &  &  \\
                            & $\Re\left(\ctt\right)$ & $4.4 \times 10^{-27} $ &      & $1.0 \times 10^{-28} $ &      & \multirow{2}{*}{0.1} & \multirow{2}{*}{$-                   $}  & \multirow{2}{*}{    }  & \multirow{2}{*}{} \\
                            & $\Im\left(\ctt\right)$ & $4.2 \times 10^{-27} $ &      & $7.5 \times 10^{-28} $ &      &                             &    &  &  \\
\hline\hline
\end{tabular}
\caption{Summary of the results of the six fits for Lorentz-violating parameters (the real and imaginary parts of each parameter are fit simultaneously). The upper limits at the 95\% confidence level (C.L.) and best fits are shown, as well as the $\Delta \chisq$ between the best fit and the hypothesis of no Lorentz violation.  The most significant exclusion of Lorentz invariance is in the \ats{e\mu} fit, which is still consistent with no LV at the 68\% confidence level.  Since the parameters are scanned on a logarithmic scale and only positive parameter values are used, $10^{-28}$ is the minimum value considered and is equivalent to no LV.}
\label{tab:results}
\end{table*}

\begin{figure}
  \subfigure[]{
    \includegraphics[width=\fwid,clip]{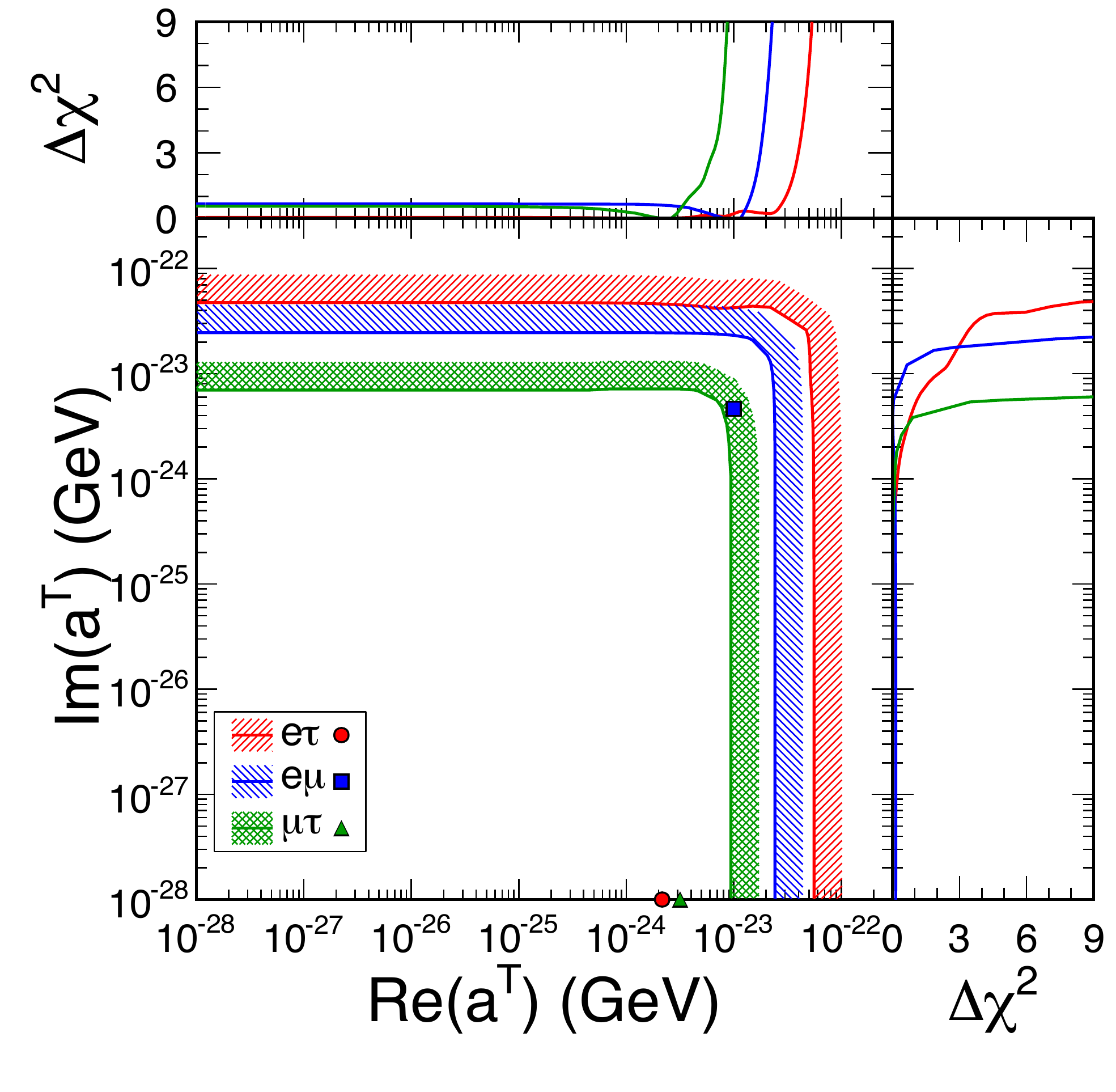} \label{fig:contourA}
  }
  \subfigure[]{
    \includegraphics[width=\fwid,clip]{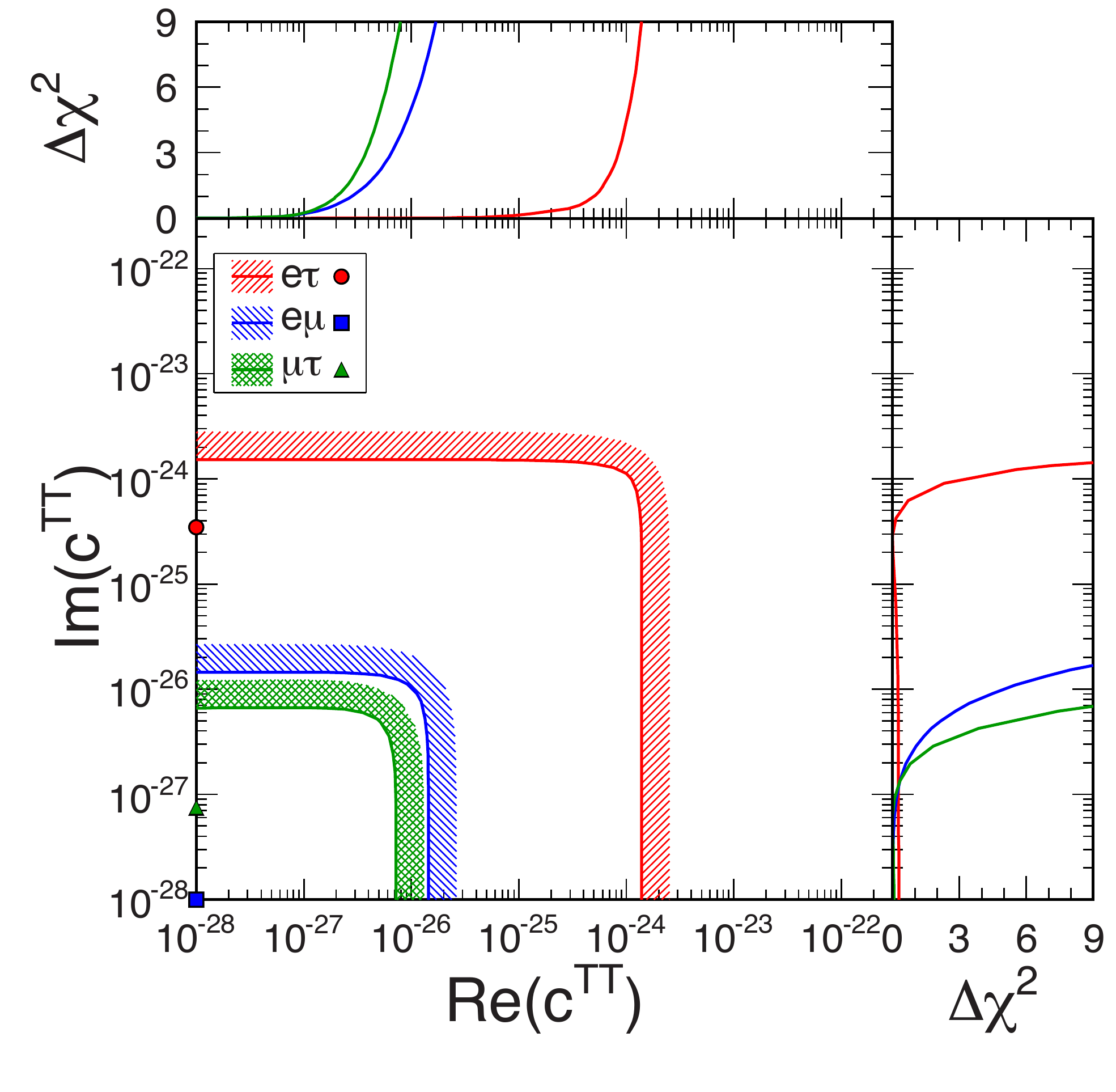} \label{fig:contourC}
  }
  \caption{(color online) Two-dimensional contours at the 95\% confidence level for the real and imaginary parts of \ats{e\tau}, \ats{e\mu}, and \ats{\mu\tau} in \subref{fig:contourA} and \ctts{e\tau}, \ctts{e\mu}, and \ctts{\mu\tau} in \subref{fig:contourC}.  The hashed areas indicate the side of the contour that is excluded.  The best-fit points from the three fits are also shown as markers.  The one-dimensional $\Delta\chisq$ curves are shown in the top and right side plots with the alternate variable profiled out. }
\end{figure}

The three-flavor plus SME oscillation model described in \sect{theory} is fit to the data samples described above using the techniques from~\cite{sterilepaper}. 
The fitter minimizes a ``pulled'' \chisq~\cite{Fogli:2002pt} assuming Poisson-statistics between the MC expectation, calculated for a particular value of the complex coefficient, and the data:
\newcommand{\skn}{\textrm{SK} n}
\newcommand{\ob}{ \ensuremath{\mathcal{O}^{\skn}_{i} }\xspace } 
\newcommand{\ex}{ \ensuremath{E^{\skn}_{i}(\vec \theta) }\xspace } 
\newcommand{\sy}{ \ensuremath{\tilde{E}^{\skn}_{i}(\vec \theta, \vec \epsilon) }\xspace } 
\begin{widetext}
\begin{equation}
\chisq = 2 \sum_i \left( \sum_n \sy - \sum_n \ob 
       + \sum_n \ob \ln \frac{\sum_n \ob}{ \sum_n \sy} \right)
       + \chisq_{\textrm{penalty}}(\vec \epsilon)
\label{eq:chisq}
\end{equation}
\end{widetext}
where $n$ indexes the four SK run periods, $i$ indexes the analysis bins, \ob is the number of observed events in bin $i$ during SK$n$, and \sy is the MC expectation in bin $i$ in SK$n$ with the coefficients being tested, $\vec \theta$, and systematic parameters, $\vec \epsilon$. The expectation in each bin is calculated separately for each run period and then the run periods are summed for the comparison between data and MC.  

The systematic uncertainties are approximated as linear effects on the analysis bins,
\begin{equation}
\sy    =  \ex \left(1 + \sum_j f^{\skn}_{i,j} \frac{\epsilon_j}{\sigma_j}\right) 
\label{eq:linsys}
\end{equation}
where $j$ indexes the systematic errors, \ex is the MC expectation in bin $i$ in SK$n$ without systematic shifts, and $f^{\skn}_{i,j}$ is the fractional change in bin $i$ in SK$n$ due to $\sigma_j$, the 1-sigma change in systematic $j$.  The constraints on these parameters are included as a penalty term in \eqn{chisq}:
\begin{equation}
 \chisq_{\textrm{penalty}}(\vec \epsilon) =  \sum_{j} \left(\frac{\epsilon_j}{\sigma_j}\right)^2.
\end{equation}
The analysis includes 155 systematic error parameters. The uncertainties in the atmospheric neutrino flux, neutrino interaction cross-sections, particle production within nuclei, and the standard PMNS oscillation parameters are shared across all run periods so $f^{\skn}_{i,j}$ is the same in SK-I through SK-IV. The uncertainties related to detector performance: reconstruction, particle identification, energy scale, and fiducial volume uncertainties, differ between run periods since they depend on the specific detector geometry and hardware. For these uncertainties, $f^{\skn}_{i,j}$ will be non-zero in only one run period. A table of all systematic uncertainties included in the analysis can be found in the appendix to~\cite{sterilepaper}.

In order to focus the analysis on the LV coefficients, the standard oscillation parameters are constrained to external measurements and their uncertainties are taken as systematic errors. 
The T2K measurement of \numu disappearance, 
$|\dmsq{32}| = \val{\sci{(2.51 \pm 0.10)}{-3}}{eV^2}$ and 
$\sn{23} = \stwothree \pm 0.055$~\cite{Abe:2014ugx}, 
is used because its narrow-band beam and shorter fixed baseline makes it less sensitive to the Lorentz-violating spectral distortions considered in this analysis.
The mixing angle 
$\snt{13} = \stonethree \pm 0.01$ 
is taken from the 2013 PDG world-average~\cite{PDG}, 
the solar terms are taken from the global fit performed by the SK solar+KamLAND analysis, $\dmsq{21} = \val{\sci{(7.46 \pm 0.19)}{-5}}{eV^2}$, 
$\sn{12} = \sonetwo \pm 0.021$~\cite{Abe:2010hy}.
The $CP$-violating phase $\delta$ and the mass hierarchy (the sign of \dm) are not yet known and so are allowed to float unconstrained.

\Equation{chisq} is minimized with respect to the $\vec \epsilon$ for each choice of $\vec \theta$ in a fit's parameter space. A set of linear equations in $\epsilon_j$'s are derived from \eqn{chisq} using the fact that the derivative $\partial \chisq/\partial \epsilon_j$ is zero at the minimum~\cite{Fogli:2002pt}. These equations can then be solved iteratively to find the minimum profile likelihood for that set of oscillation parameters, building up a map of \chisq vs. $\vec \theta$. The best fit point is defined as the global minimum of this map.

Six fits are performed for the real and imaginary parts of \at and \ctt in the three sectors, $e\mu$, $e\tau$, and $\mu\tau$. The real and imaginary parts of each coefficient are fit simultaneously, but otherwise the coefficients are fit independently following the procedure typical for SME analyses~\cite{Kostelecky:2008ts}.  Tests with fits to high-statistics fake data sets reliably find no LV when none is present and correctly extract the best fit point if a fake data set with an LV signal is used.  However, there is generally some ambiguity between the real and imaginary parts since they produce similar oscillation effects at the energies where LV-oscillations dominate.  The low-energy differences allow the correct parameter to be chosen in fits to simulated data with high statistics, but small fluctuations can easily move the best fit point to just the real part, just the imaginary part, or a combination of the two.

No significant evidence of Lorentz violation is seen in any of the fits.  The most significant exclusion of no LV is for \ats{e\mu}, and it has a $\Delta\chisq = 1.4$, less than $1\sigma$ with two degrees of freedom. The absolute \chisq for the fits range from 538.6 to 540.0 with 480 bins (477 degrees of freedom), corresponding to goodness-of-fit $p$-values around 2.5\%.  The best-fit momentum and zenith distributions for the \at and \ctt fits are shown, compared with the data, in \app{zeniths}.  A summary of the fit results, including upper limits at the 95\% confidence level, best-fit values, and levels of agreement with no Lorentz violation can be seen in \tab{results}. The two-dimensional contours at the 95\% confidence level on \at and \ctt are shown in \figs{contourA}{contourC} respectively.  The limits on the real and imaginary parts of the parameter are slightly different in the $e\tau$ and $\mu\tau$ sectors because these fits found best fit points with different values for the real and imaginary components.

\section{Conclusions}

The large range of energies and path lengths in the atmospheric neutrino sample make it sensitive to a variety of spectral distortions introduced by violations of Lorentz invariance as parameterized by the Standard Model Extension. However, the long distances and high energies make the perturbative approach used in other experiments inappropriate, so we present the first analysis of Lorentz violation in neutrino oscillations where the full, non-perturbative Hamiltonian is used, combined with three-flavor neutrino oscillations.  No evidence of LV is seen, so we set limits on the isotropic parameters \at and \ctt in the $e\mu$, $\mu\tau$, and $e\tau$ sectors. These are the first limits on the isotropic parameters in the $\mu\tau$ sector, and we improve the existing limits~\cite{Kostelecky:2008ts} on \at by 3 orders of magnitude and on \ctt by seven orders of magnitude thanks to the wide range of energies and path lengths of the neutrinos in the atmospheric neutrino samples. Future studies of SK atmospheric neutrino data could also set limits on the directional parameters by searching for sidereal variations in the atmospheric neutrino data.

\section{Acknowledgments}
We would like to thank A. Kostelecky for his advice and support
and we are grateful to J. S. Diaz for working closely with us
to calculate and implement the Lorentz-violating oscillation probabilities.
The authors gratefully acknowledge the cooperation of the Kamioka 
Mining and Smelting Company. Super-K has been built and operated from 
funds provided by the Japanese Ministry of Education, Culture, Sports, 
Science and Technology, the U.S.  Department of Energy, and the 
U.S. National Science Foundation. This work was partially supported by 
the Research Foundation of Korea (BK21 and KNRC), the Korean Ministry 
of Science and Technology, the National Science Foundation of China,
the European Union FP7 (DS laguna-lbno PN-284518 and ITN
invisibles GA-2011-289442)
the National Science and Engineering Research Council (NSERC) of Canada, 
and the Scinet and Westgrid consortia of Compute Canada.

\appendix
\vspace*{1cm}
\section{Neutrino oscillations with Lorentz violation} \label{app:diagonalization}

This appendix shows the full calculation of the neutrino oscillation probabilities with both three-neutrino mixing and Lorentz Violation, without assuming that the baseline is short or that the LV Hamiltonian is small.  Neither of these approximations is valid for SK because of its wide range of path lengths and energies.  

The oscillation probabilities for Lorentz violation plus three-flavor oscillations including matter effects are calculated by diagonalizing the Hamiltonian which includes all these pieces, following the method from~\cite{JSDiaz}. Combining the parts described individually in \sect{theory},
\begin{widetext}
\begin{equation}
H = U 
\left(\begin{array}{ccc}
0 & 0 & 0 \\
0 & \frac{\dmsq{21}}{2E} & 0 \\
0 & 0 & \frac{\dmsq{31}}{2E}
\end{array}\right) U^\dagger
\pm 
\sqrt{2} G_F 
\left(\begin{array}{ccc}
N_e & 0 & 0 \\
0 & 0 & 0 \\
0 & 0 & 0
\end{array}\right)
\pm
\left(\begin{array}{ccc}
0          & \ats{e\mu}     & \ats{e\tau} \\
\as{e\mu}  & 0            & \ats{\mu\tau} \\
\as{e\tau} & \as{\mu\tau} & 0  \\
\end{array}\right) 
 -\frac{4E}{3} 
\left(\begin{array}{ccc}
0          & \ctts{e\mu}     & \ctts{e\tau} \\
\cs{e\mu}  & 0            & \ctts{\mu\tau} \\
\cs{e\tau} & \cs{\mu\tau} & 0  \\
\end{array}\right),
\end{equation}
\end{widetext}
where $U$ is the PMNS mixing matrix, $E$ is the neutrino energy, $G_F$ is Fermi's constant, and $N_e$ is the average electron density along the neutrino's path.  For antineutrinos, the complex conjugates of all terms are taken (though in practice this only affects $\delta_{cp}$ in $U$ and the $a$ and $c$ parameters) and the signs of the \nue matter effect and $a$ matrices are negative.

The next step is to diagonalize this $3 \times 3$ matrix to calculate the new eigenvalues and mixing matrix. Since the Hamiltonian is Hermitian the eigenvalues are guaranteed to be real. They can be calculated below as the roots of a cubic equation,
\begin{equation}
    E_i = -2 \sqrt{Q} \cos \left(\frac{\theta_i}{3}\right) - \frac{a}{3},
\end{equation}
with $i = 0, 1, 2$ and where the components $Q$ and $\theta_i$, 
\begin{align}
    Q =& \frac{a^2 -3b}{9} \\
    \theta_0 =& \cos^{-1} \left(R Q ^{-\frac{3}{2}}\right) \\ 
    \theta_1 =& \theta_0 + 2\pi \\ 
    \theta_2 =& \theta_0 - 2\pi,
\end{align}
can be calculated from the trace and determinant of H:
\begin{align}
    a =& -\Tr(H) \\
    b =& \frac{\Tr(H)^2 - \Tr(H^2)}{2} \\
    c =& -\det(H)\\
    R =& \frac{2a^3 - 9 a b + 27 c}{54}.
\end{align}

The diagonalization also produces a mixing matrix $U$,
\begin{align}
    U_{e i}    =& \frac{B_i^* C_i}{N_i} &
    U_{\mu i}  =& \frac{A_i C_i}{N_i} &
    U_{\tau i} =& \frac{A_i B_i}{N_i},
\end{align}
where
\begin{align}
    A_i =& H_{\mu \tau}(H_{e   e   } - E_i) - H_{\mu e   }H_{e \tau} \\
    B_i =& H_{\tau e  }(H_{\mu \mu } - E_i) - H_{\tau \mu}H_{\mu  e} \\
    C_i =& H_{\mu  e  }(H_{\tau\tau} - E_i) - H_{\mu \tau}H_{\tau e} \\
    N_i^2 =& |A_i B_i|^2 + |A_i C_i|^2 + |B_i C_i|^2.
\end{align}

The oscillation probabilities can then be calculated from,
\begin{equation}
    P_{\alpha\beta} = \left| \Bra{\nu_\beta} e^{-i H L} \Ket{\nu_\alpha} \right|^2
\end{equation}
which expands to
\begin{align}
    P_{\alpha\beta} &= \delta_{\alpha\beta} \nonumber\\
                    &- 4 \sum_{j>i} \Re\left( U_{\beta j} U_{\beta i}^* U_{\alpha j}^* U_{\alpha i}\right) \sin^2 \left(L\Delta E_{ji}/2\right) \nonumber\\
                    &+ 2 \sum_{j>i} \Im\left( U_{\beta j} U_{\beta i}^* U_{\alpha j}^* U_{\alpha i}\right) \sin^2 \left(L\Delta E_{ji}\right),
\end{align}
where $\Delta E_{ji} = E_j - E_i$ are the differences between the eigenvalues.

We can expand further, taking two examples particularly relevant for atmospheric neutrinos. The \numu survival probability,
\begin{equation}
    P_{\mu\mu} = 1 - \sum_{j>i} \frac{|A_j A_i C_j C_i|^2}{N_j^2 N_i^2} \sin^2\left( L\Delta E_{ji}/2 \right),
\end{equation}
and the \nue appearance probability,
\begin{align}
    P_{\mu e} = &-4 \sum_{j>i} \frac{\Re(A_j^* A_i   B_j^* B_i  ) |C_j C_i|^2}{N_j^2 N_i^2} \sin^2\left( L\Delta E_{ji}/2 \right) \nonumber\\
                &+2 \sum_{j>i} \frac{\Im(A_j^* A_i   B_j^* B_i  ) |C_j C_i|^2}{N_j^2 N_i^2} \sin  \left( L\Delta E_{ji} \right). 
\end{align}

\begin{widetext}
\section{Oscillograms}
\label{app:oscillogram}

This appendix includes plots of the $\numu \to \numu$ and $\numu \to \nue$ oscillation probabilities vs. 
neutrino path length 
and neutrino energy for large (\val{10^{-22}}{GeV} and \sci{7.5}{-23}) values of the LV parameters to show what the effects of the six different coefficients.  Oscillograms for standard three-flavor oscillations are included for comparison at the end.  For both \at and \ctt, the $e\mu$ and $\mu\tau$ sectors have similar $\numu \to \numu$ probabilities but different $\numu \to \nue$ probabilities.  For both parameters the $e\tau$ sector has the opposite effect of the other sectors: eliminating standard oscillations instead of introducing non-standard oscillations.

\begin{figure*}[h!]
  \subfigure[$\numu \to \numu$, No Lorentz violation]{
       \includegraphics[width=\fwid,clip]{mutomu_noLV.pdf}
  }
  \subfigure[$\numu \to \nue$, No Lorentz violation]{
       \includegraphics[width=\fwid,clip]{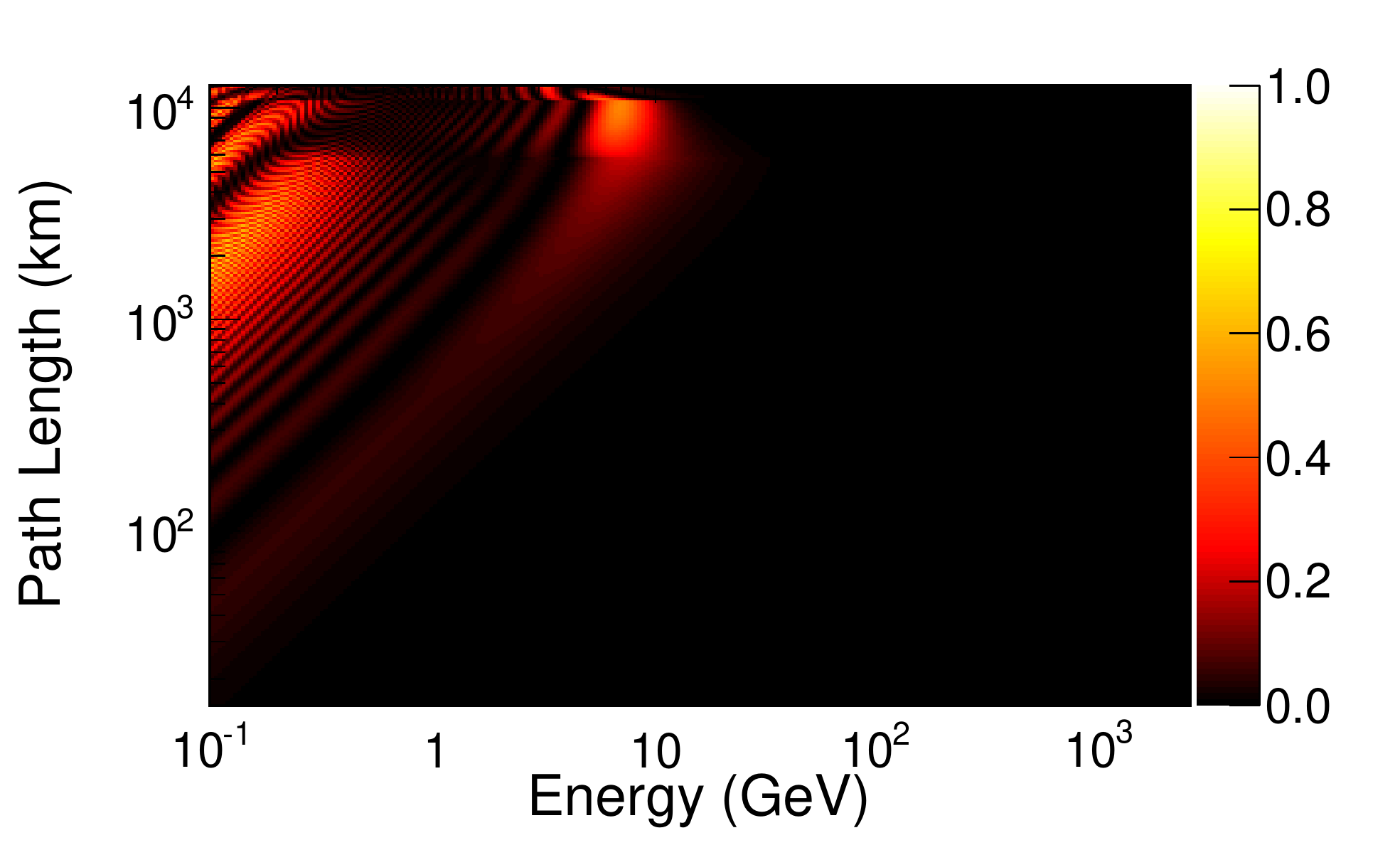}
  }
  \caption{(color online) For comparison, the $\numu \to \numu$ (left) and $\numu \to \nue$ (right) oscillation probabilities, plotted in 
  path length vs. neutrino energy
  for standard three-flavor oscillations.}
  \label{fig:oscillogramthreenu}
\end{figure*}

\begin{figure*}[h!]
  \subfigure[$\numu \to \numu$, $\ats{e\mu} = \val{10^{-22}}{GeV}$]{
        \includegraphics[width=\fwid,clip]{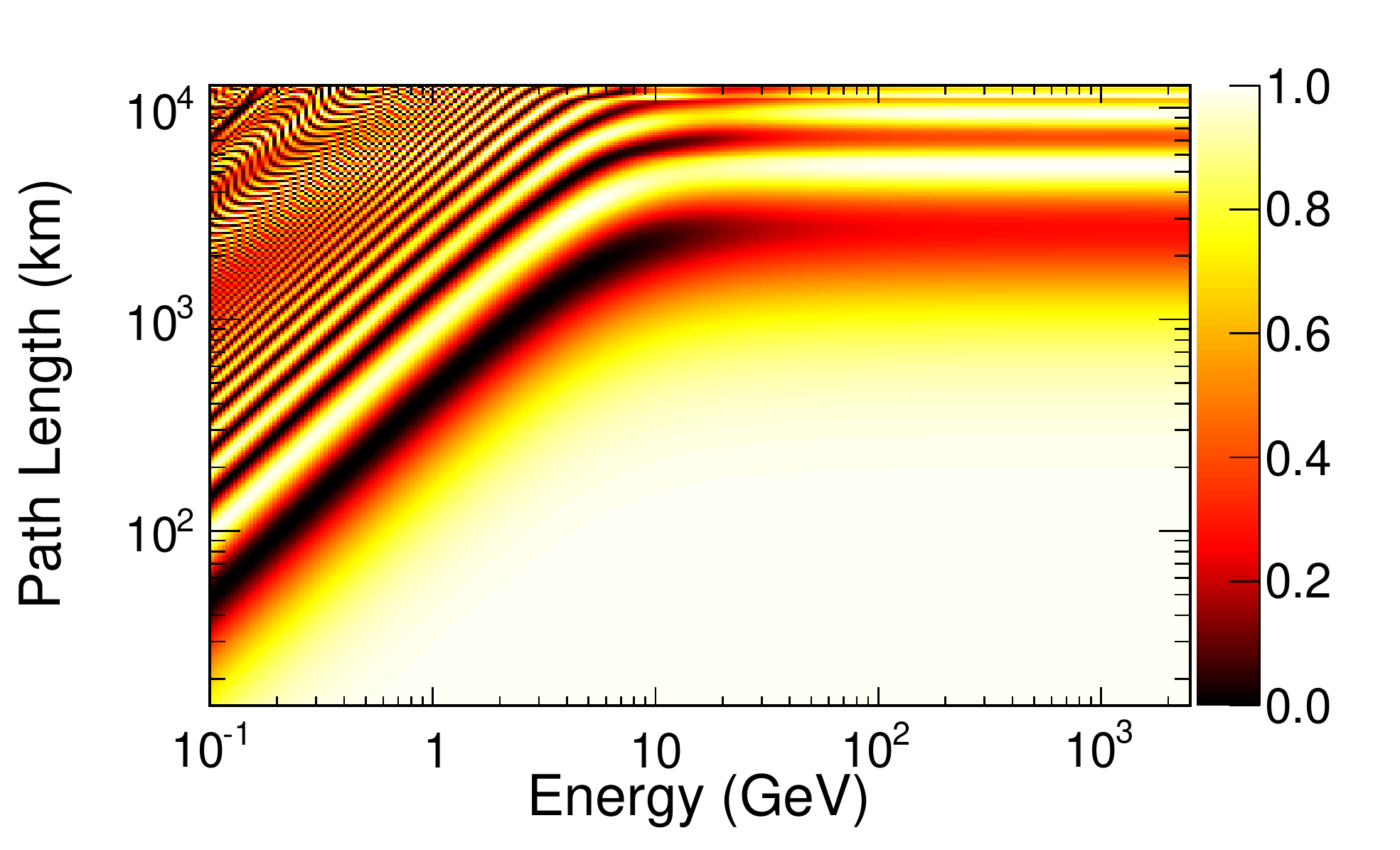}
  }
  \subfigure[$\numu \to \nue$, $\ats{e\mu} = \val{10^{-22}}{GeV}$]{
        \includegraphics[width=\fwid,clip]{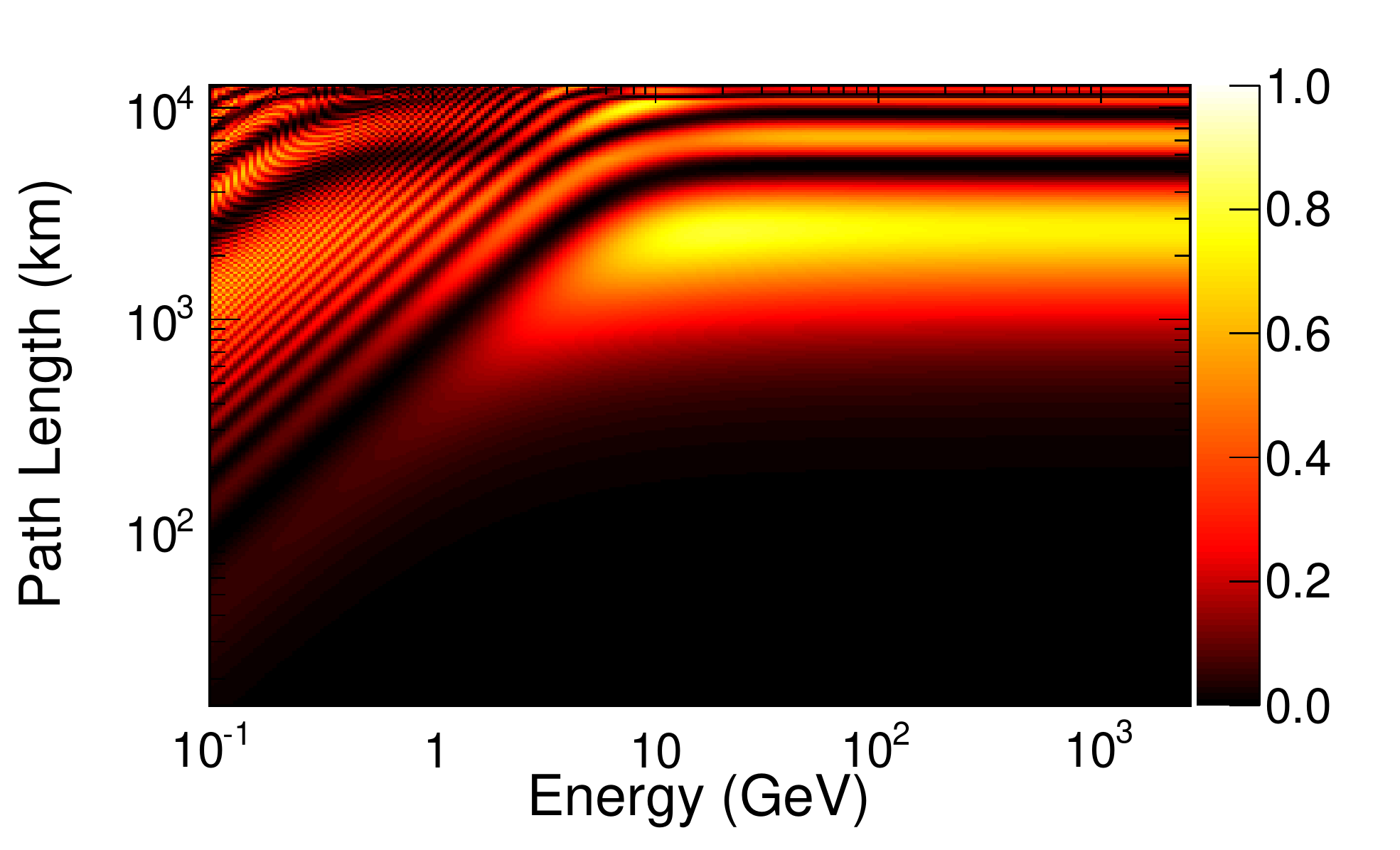}
  }
  \subfigure[$\numu \to \numu$, $\ats{\mu\tau} = \val{10^{-22}}{GeV}$]{
        \includegraphics[width=\fwid,clip]{mutomu_mutau_a_Re_22.pdf}
  }
  \subfigure[$\numu \to \nue$, $\ats{\mu\tau} = \val{10^{-22}}{GeV}$]{
        \includegraphics[width=\fwid,clip]{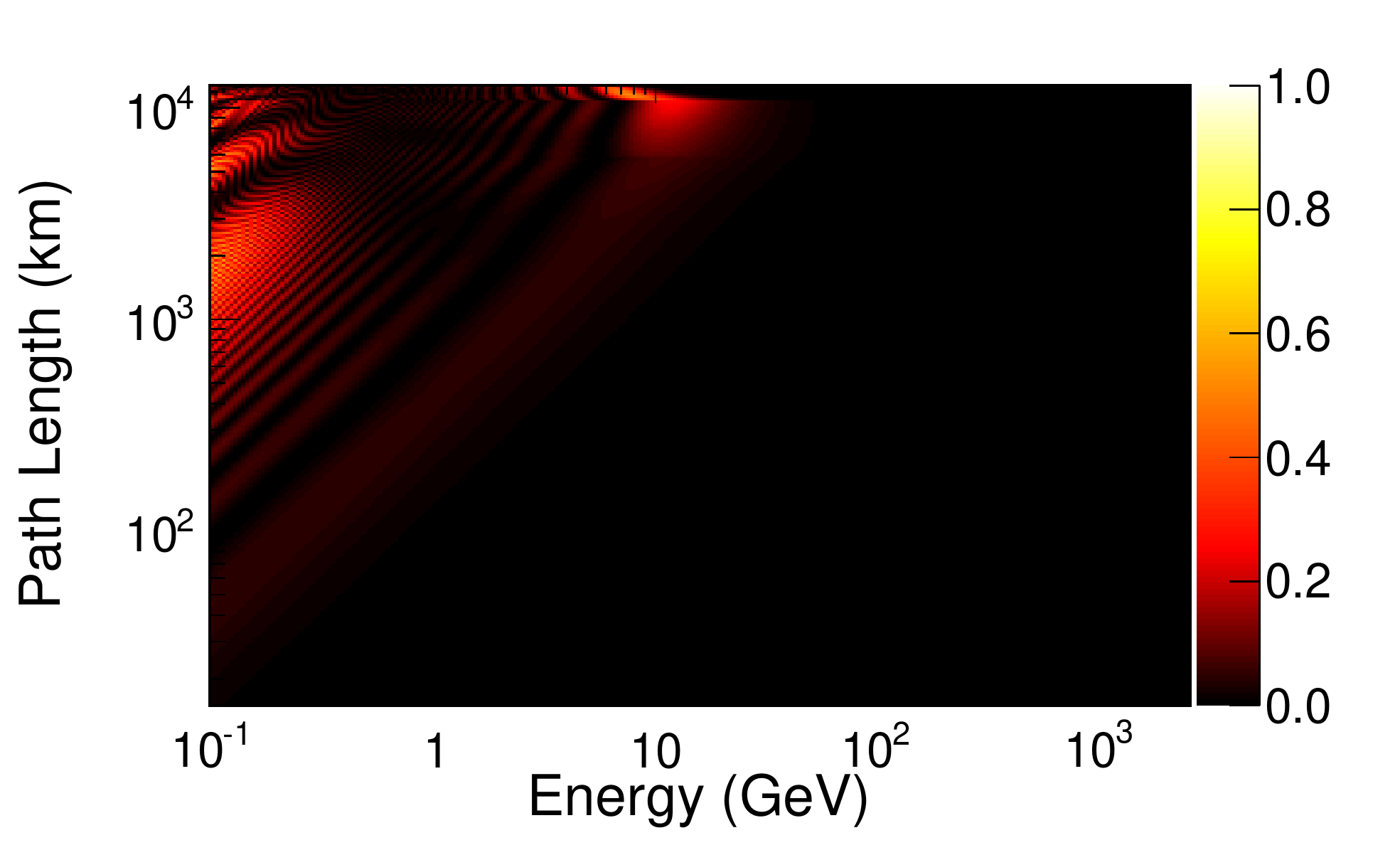}
  }
  \subfigure[$\numu \to \numu$, $\ats{e\tau} = \val{10^{-22}}{GeV}$]{
        \includegraphics[width=\fwid,clip]{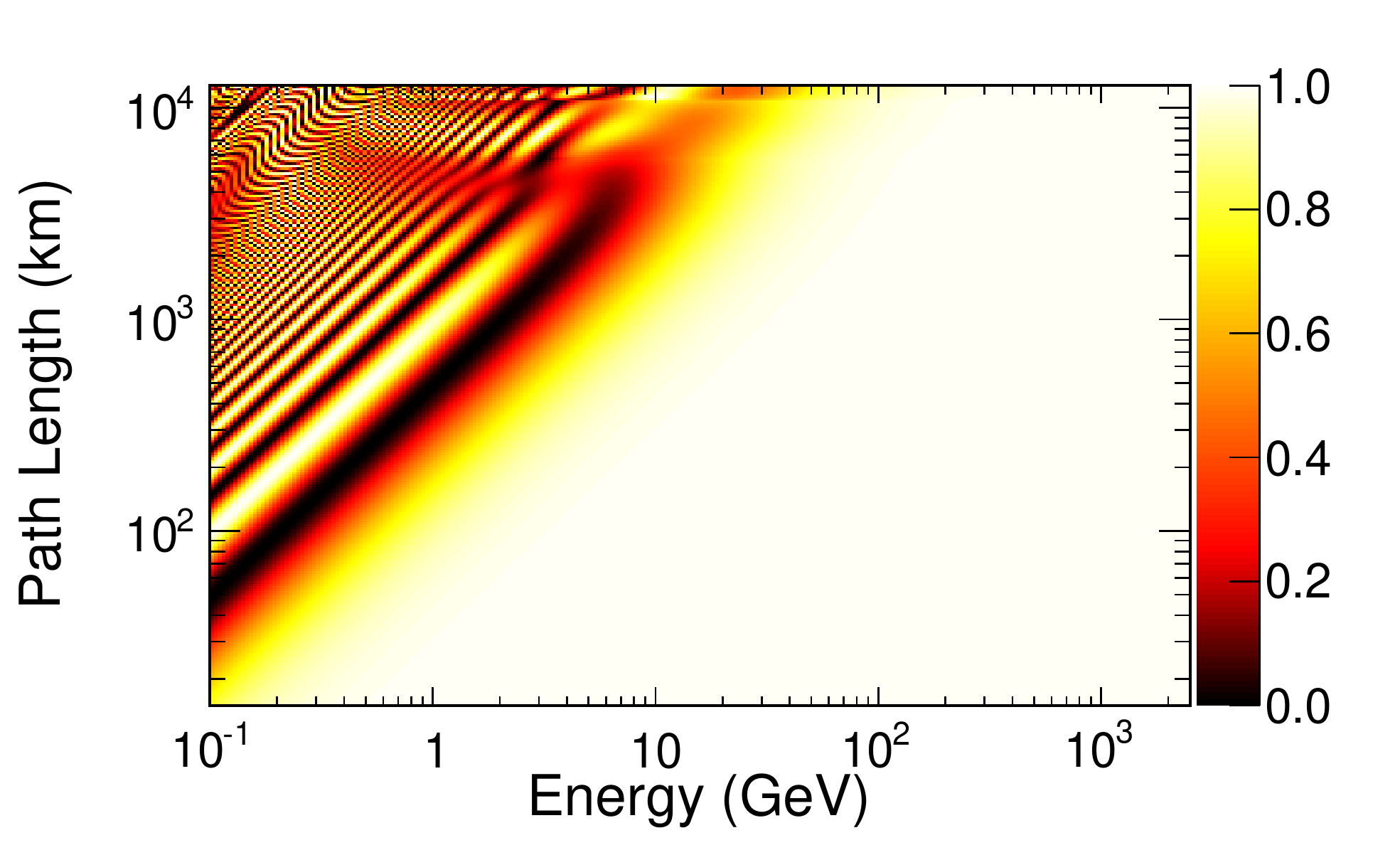}
  }
  \subfigure[$\numu \to \nue$, $\ats{e\tau} = \val{10^{-22}}{GeV}$]{
        \includegraphics[width=\fwid,clip]{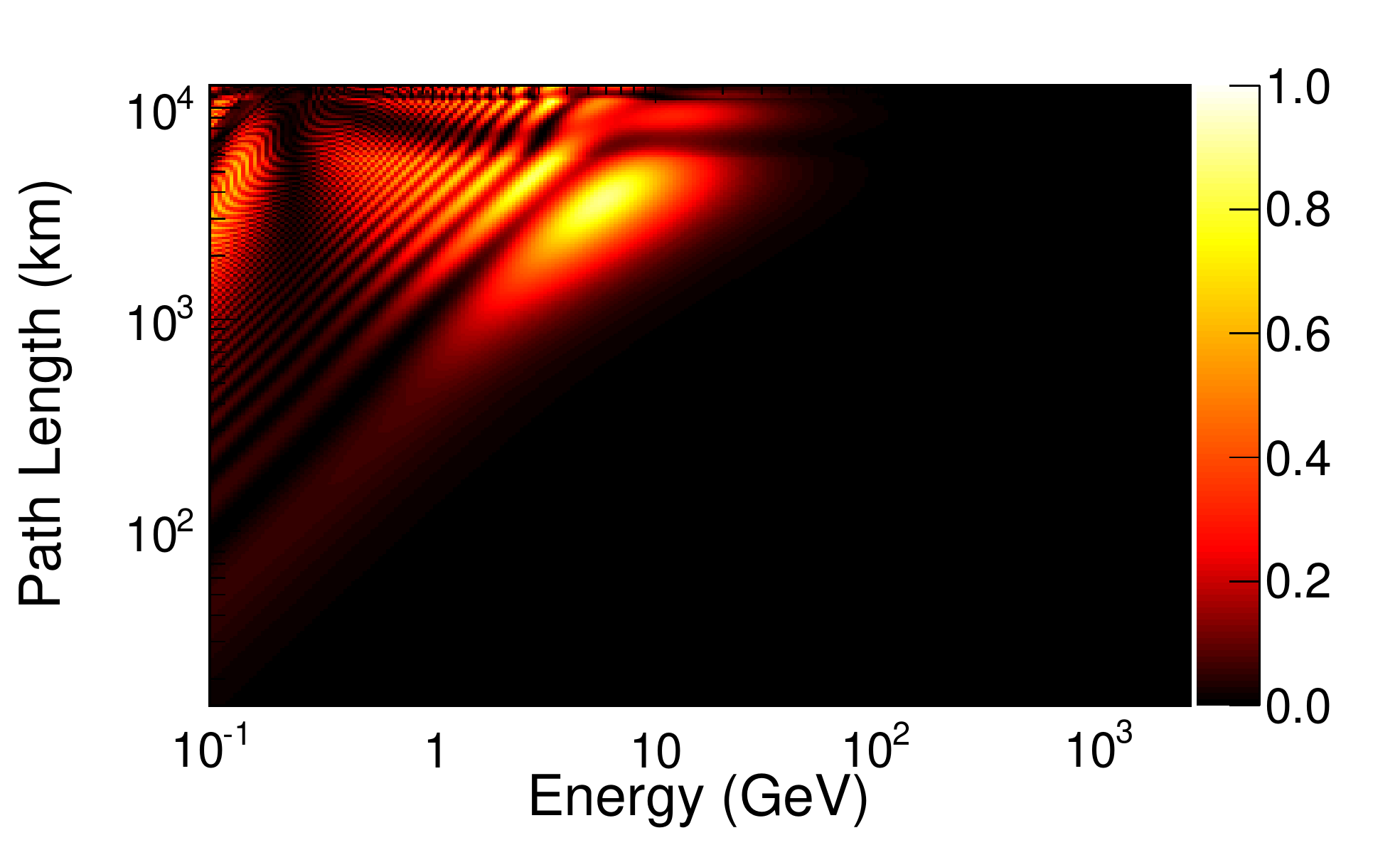}
  }
  \caption{(color online) The $\numu \to \numu$ (left) and $\numu \to \nue$ (right) oscillation probabilities, plotted in 
  path length vs. neutrino energy
  for the \at parameter in the (top to bottom) $e\mu$, $\mu\tau$, and $e\tau$ sectors. The \at coefficients scale terms proportional to $L$, so the distortions get stronger as \cz approaches -1.}
  \label{fig:oscillograma}
\end{figure*}

\begin{figure*}
  \subfigure[$\numu \to \numu$, $\ctts{e\mu} = \sci{7.5}{-23}$]{
        \includegraphics[width=\fwid,clip]{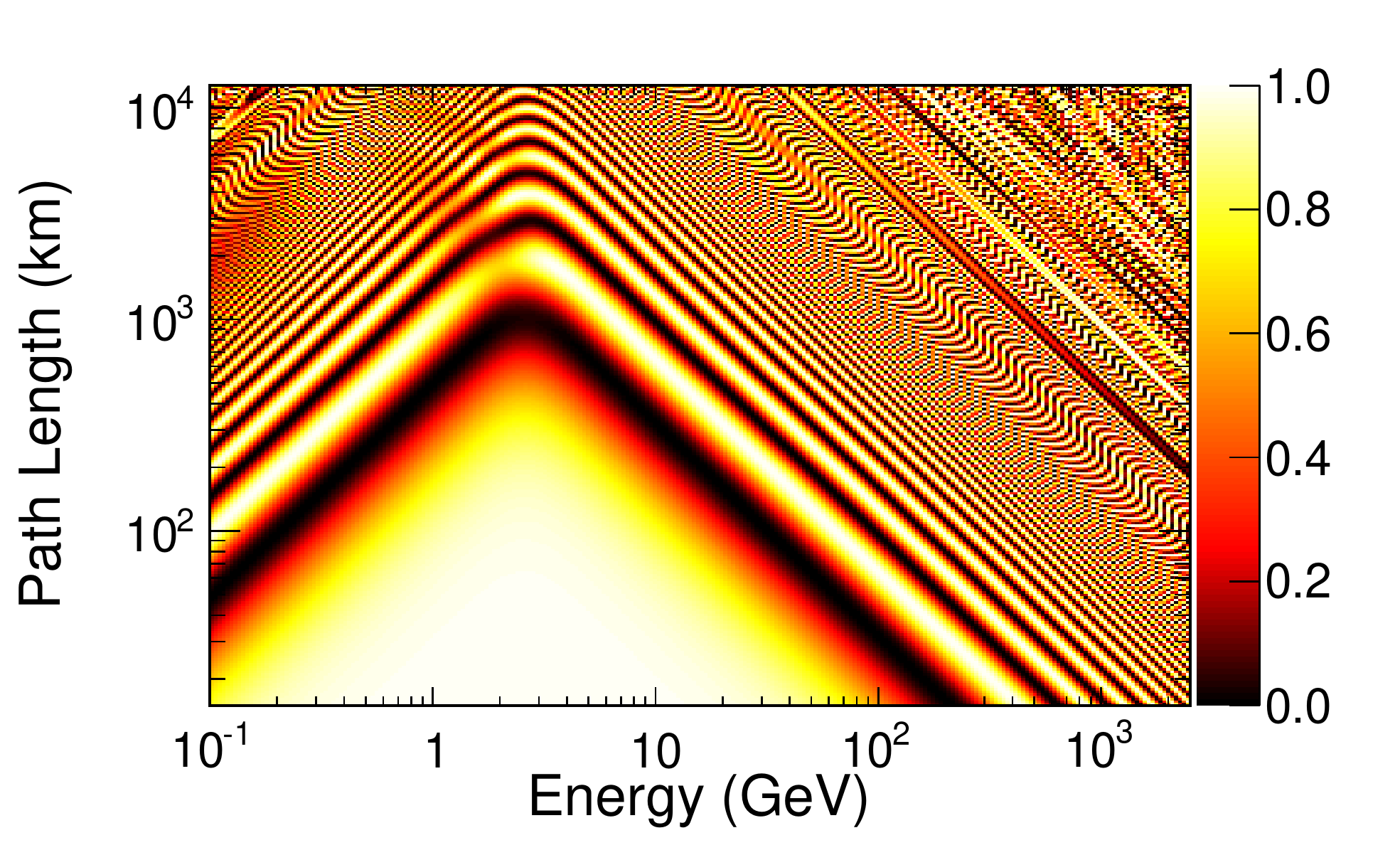}
  }
  \subfigure[$\numu \to \nue$, $\ctts{e\mu} = \sci{7.5}{-23}$]{
        \includegraphics[width=\fwid,clip]{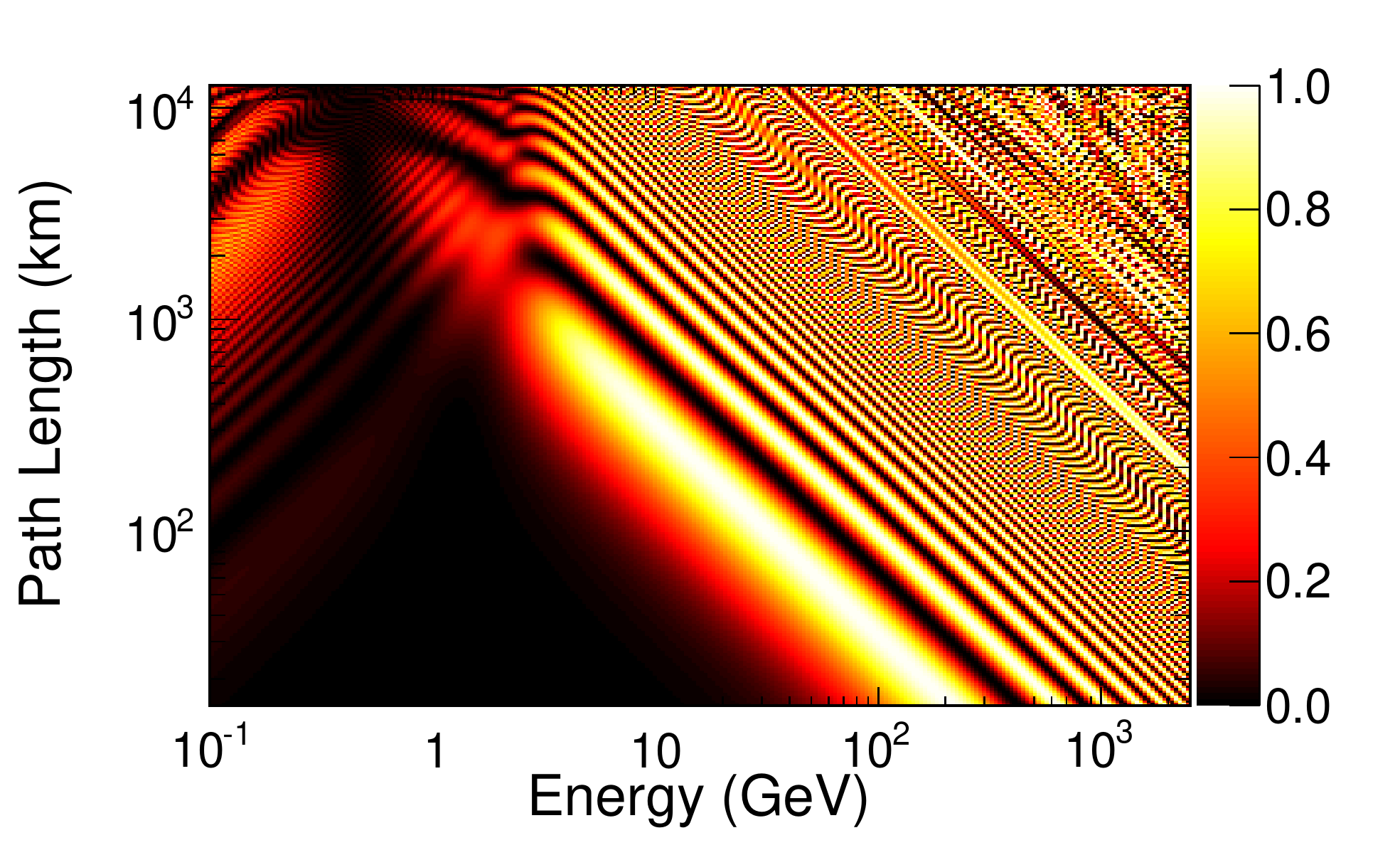}
  }
  \subfigure[$\numu \to \numu$, $\ctts{\mu\tau} = \sci{7.5}{-23}$]{
        \includegraphics[width=\fwid,clip]{mutomu_mutau_c_Re_22.pdf}
  }
  \subfigure[$\numu \to \nue$, $\ctts{\mu\tau} = \sci{7.5}{-23}$]{
        \includegraphics[width=\fwid,clip]{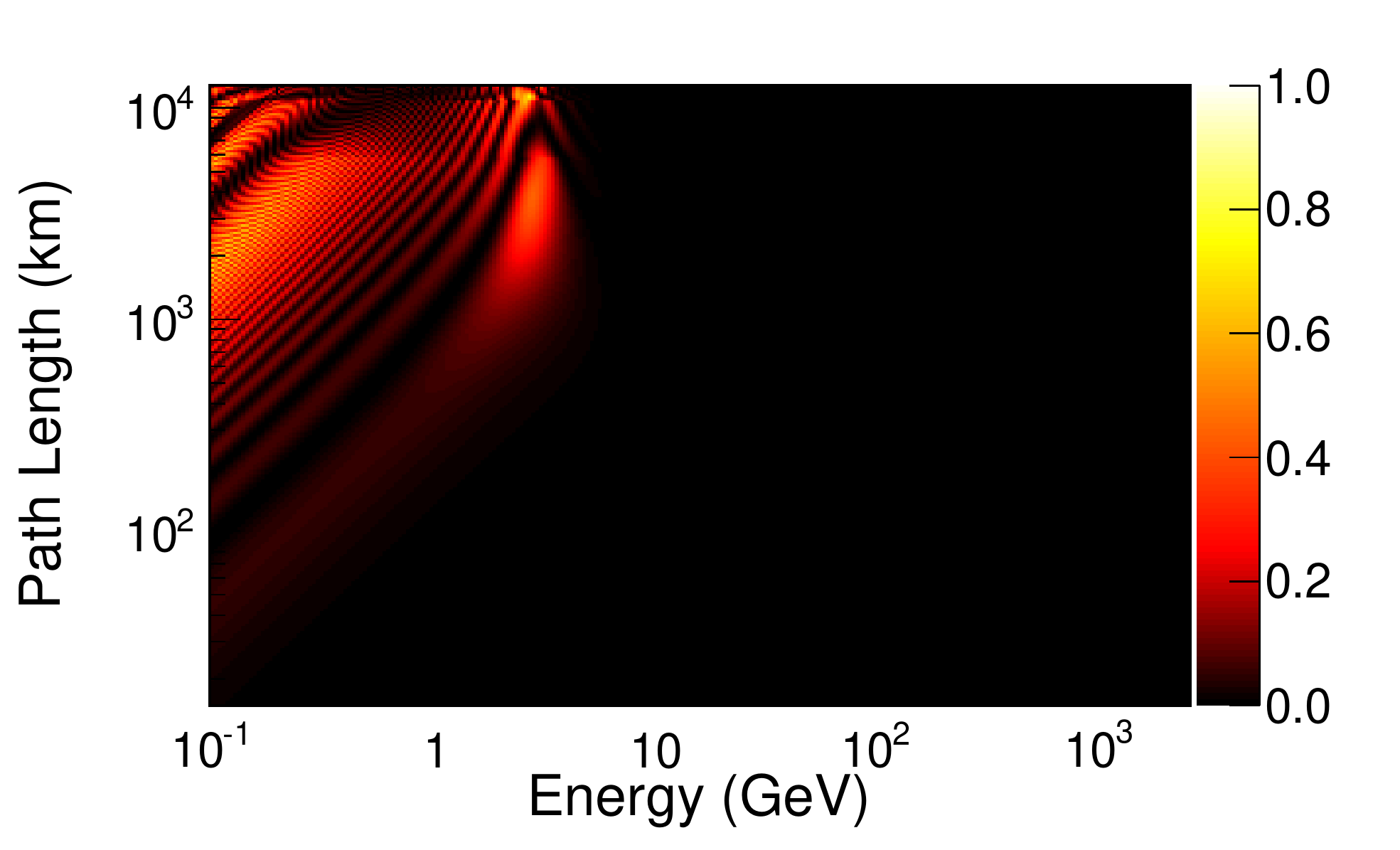}
  }
  \subfigure[$\numu \to \numu$, $\ctts{e\tau} = \sci{7.5}{-23}$]{
        \includegraphics[width=\fwid,clip]{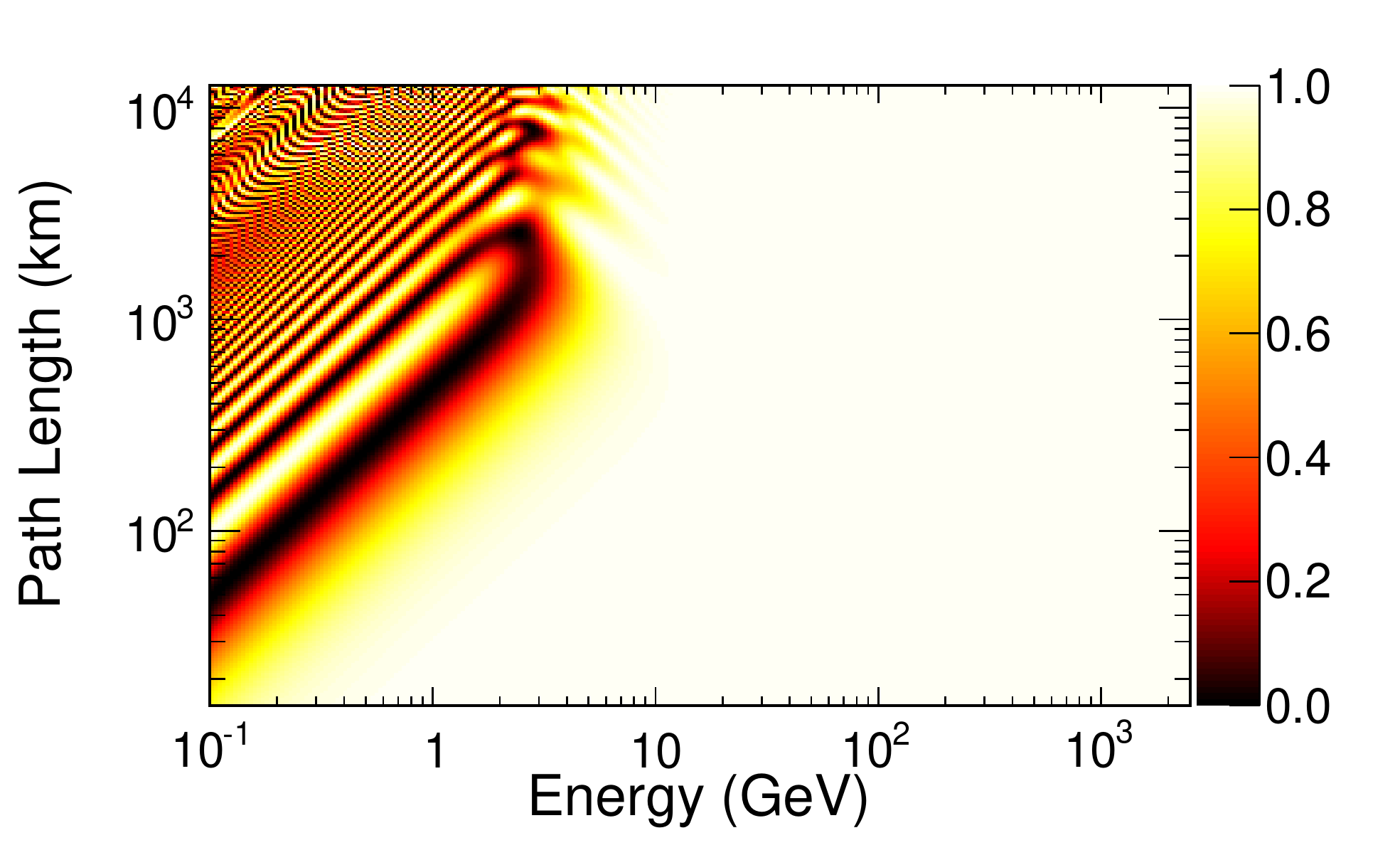}
  }
  \subfigure[$\numu \to \nue$,$\ctts{e\tau} = \sci{7.5}{-23}$]{
        \includegraphics[width=\fwid,clip]{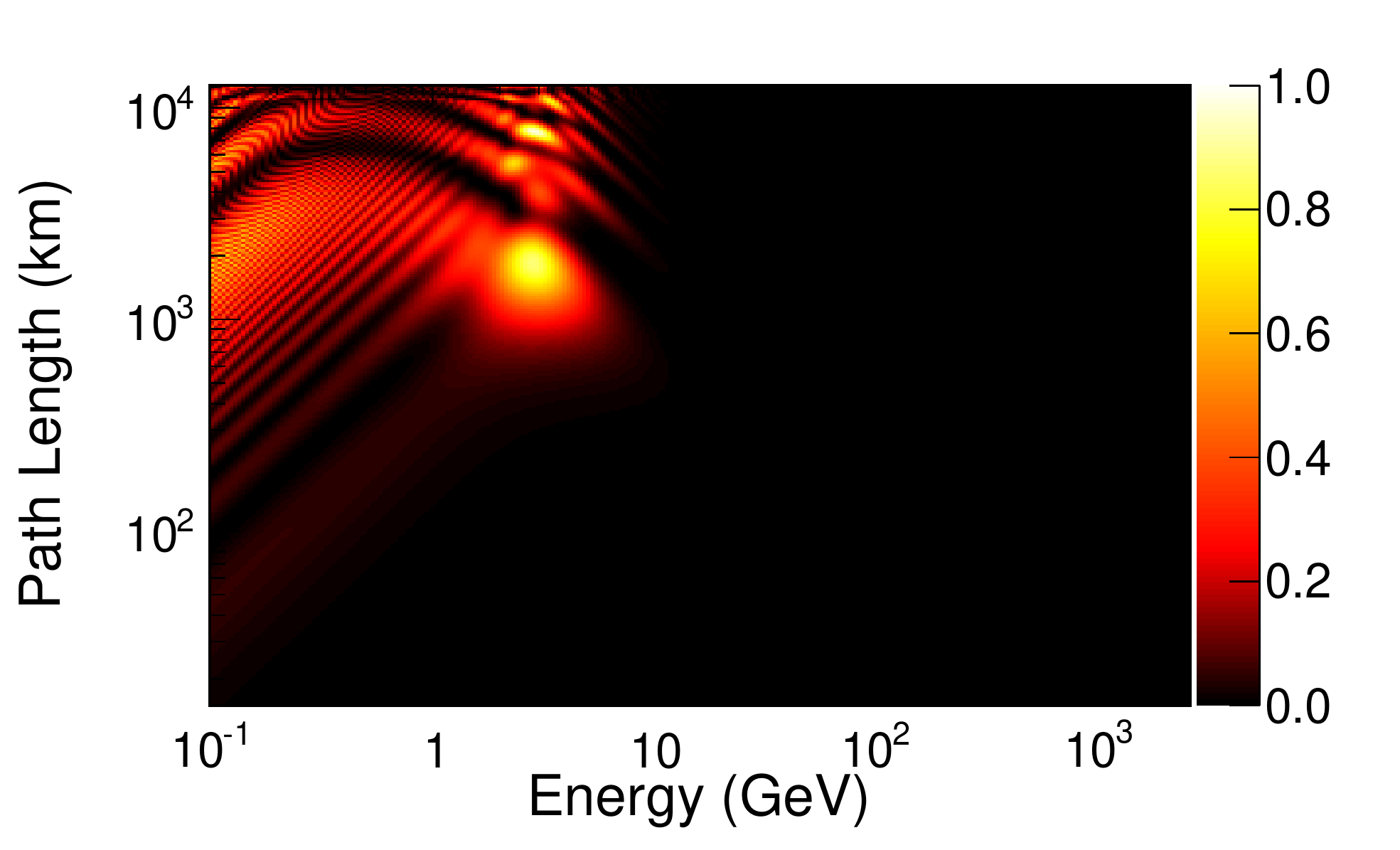}
  }
  \caption{(color online) The $\numu \to \numu$ (left) and $\numu \to \nue$ (right) oscillation probabilities, plotted in 
  path length vs. neutrino energy
  for the \ctt parameter in the (top to bottom) $e\mu$, $\mu\tau$, and $e\tau$ sectors. The \ctt coefficients scale terms proportional to $LE$, so the distortions get stronger at higher energies.}
  \label{fig:oscillogramc}
\end{figure*}

\clearpage

\section{Zenith angle distributions}
\label{app:zeniths}

This appendix includes ratios relative to standard three-flavor oscillations for all of the sub-samples included in the analysis.  The ratios are generally plotted vs. zenith angle (\cz), except for samples which are binned only in energy.  The data is shown as points with statistical error bars.  The dashed lines represent the best fits for each of the 6 fits while the solid lines represent examples of large LV parameters (\val{10^{-22}}{GeV} and \sci{7.5}{-23}), the same as the oscillograms in \app{oscillogram}).  The plots are divided into $e$- and NC-like samples and $\mu$-like samples, and the \at and \ctt fits are shown separately.  

\begin{figure*}[h!]
 \begin{center}
 \includegraphics[width=\zwid,clip]{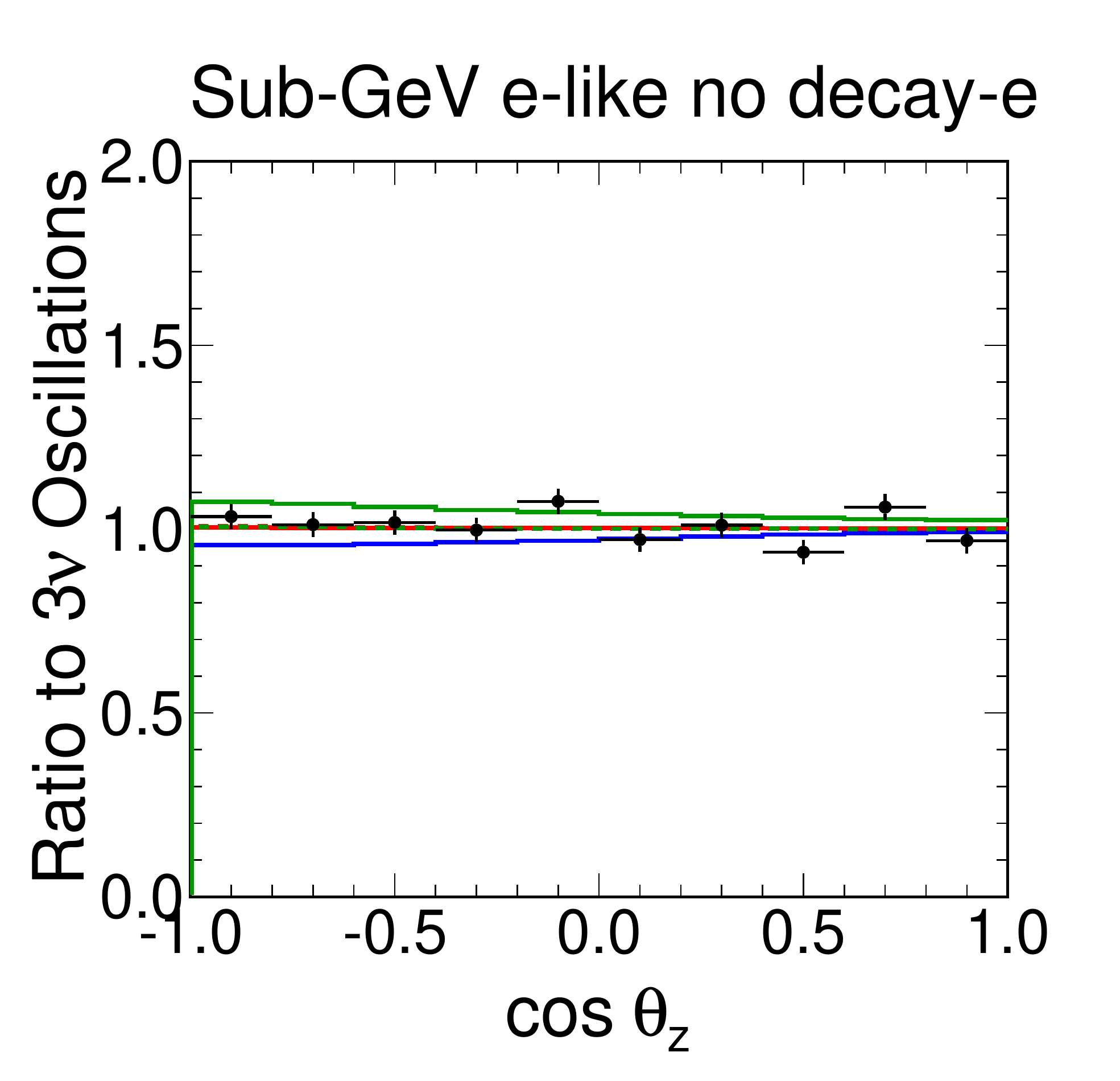}
 \includegraphics[width=\zwid,clip]{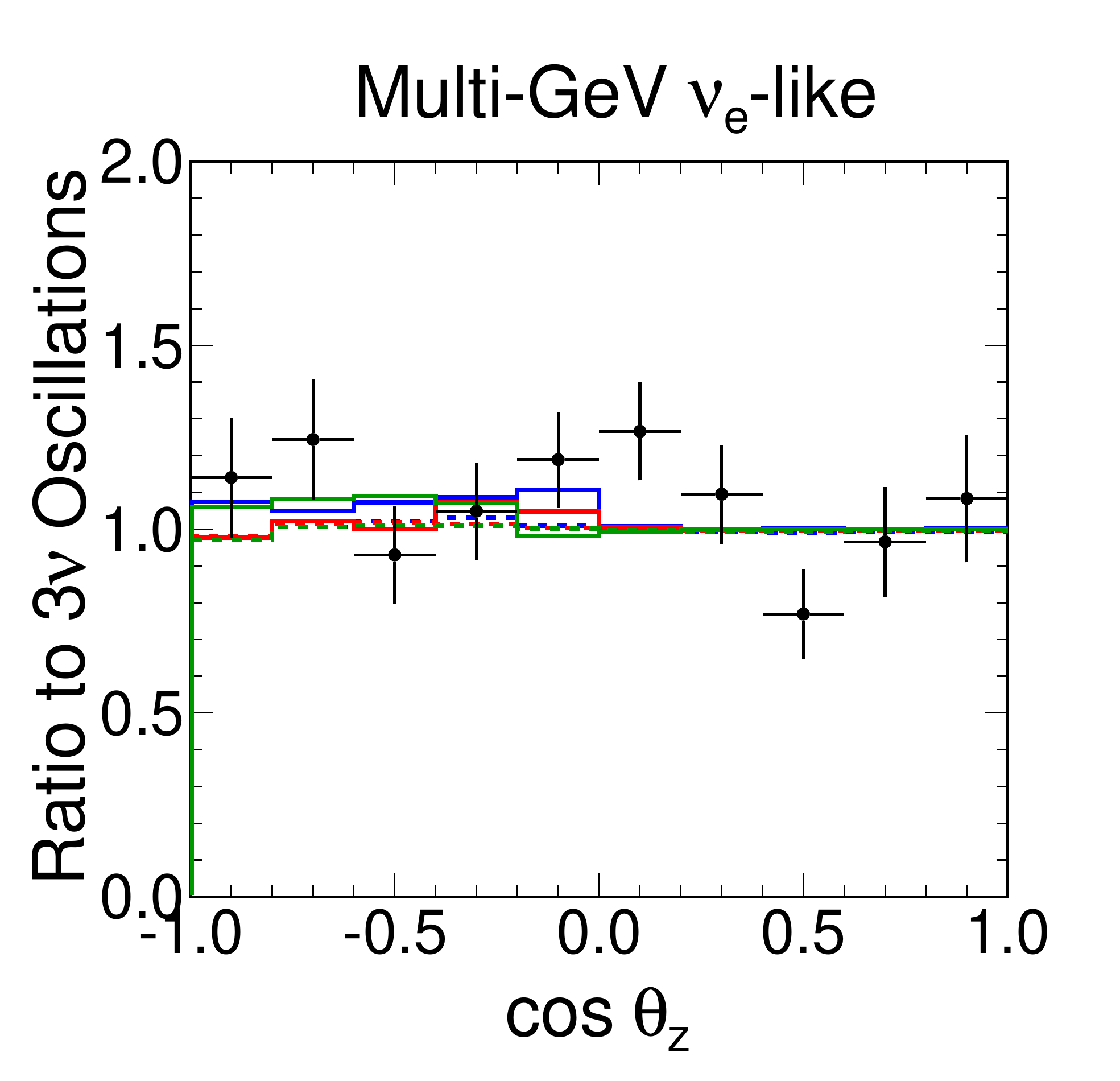}
 \includegraphics[width=\zwid,clip]{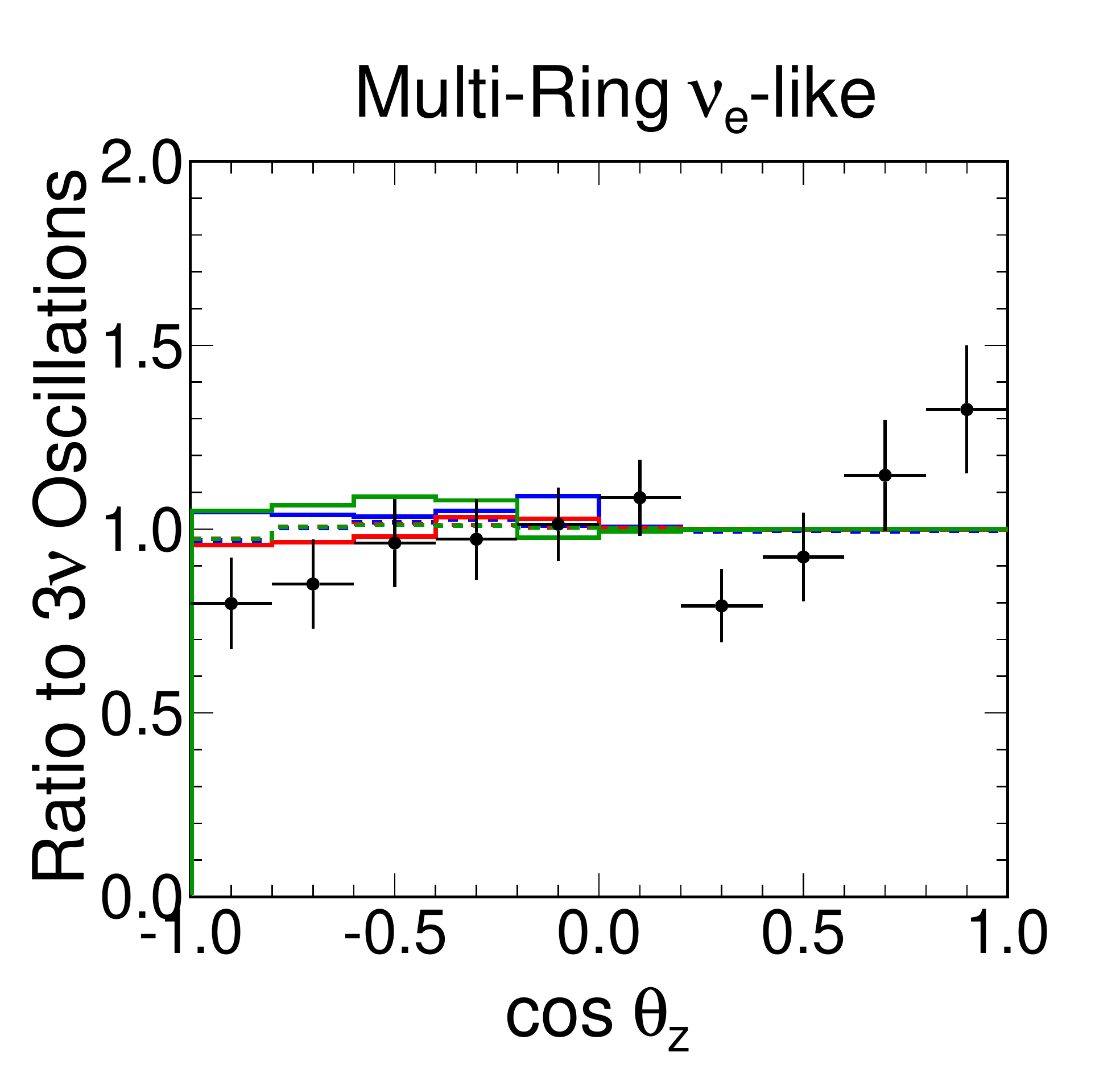} \\
 \includegraphics[width=\zwid,clip]{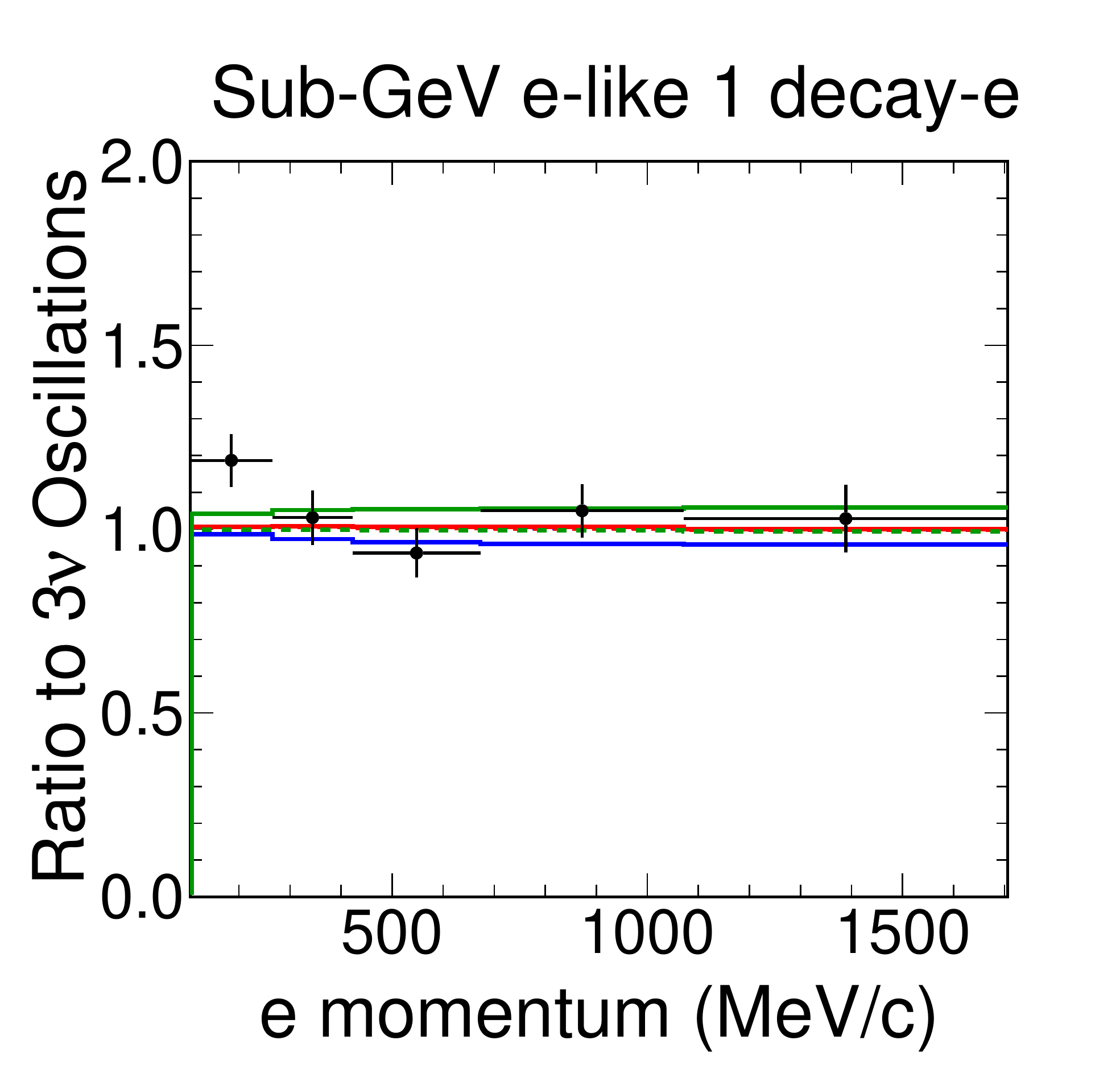} 
 \includegraphics[width=\zwid,clip]{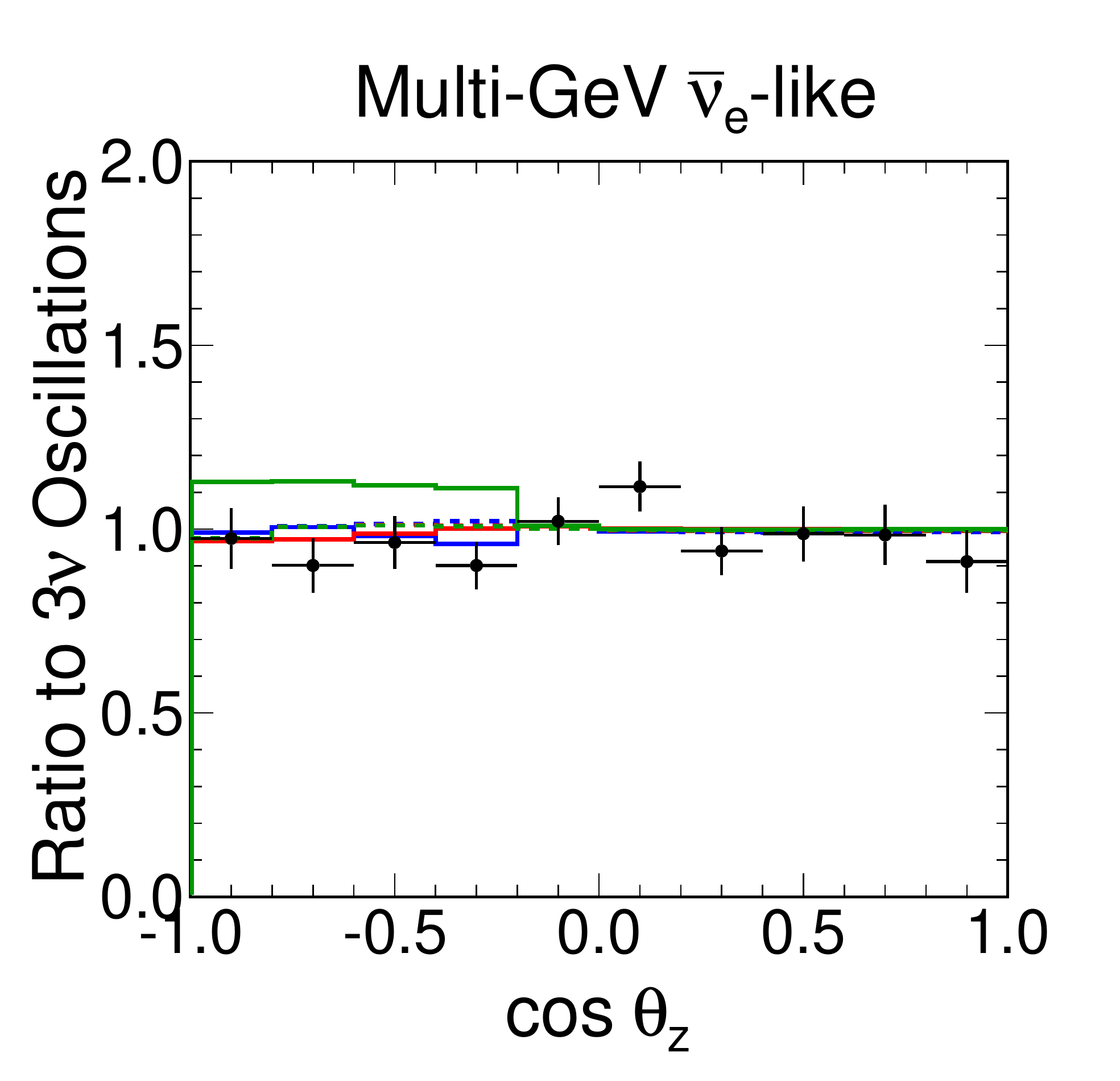}
 \includegraphics[width=\zwid,clip]{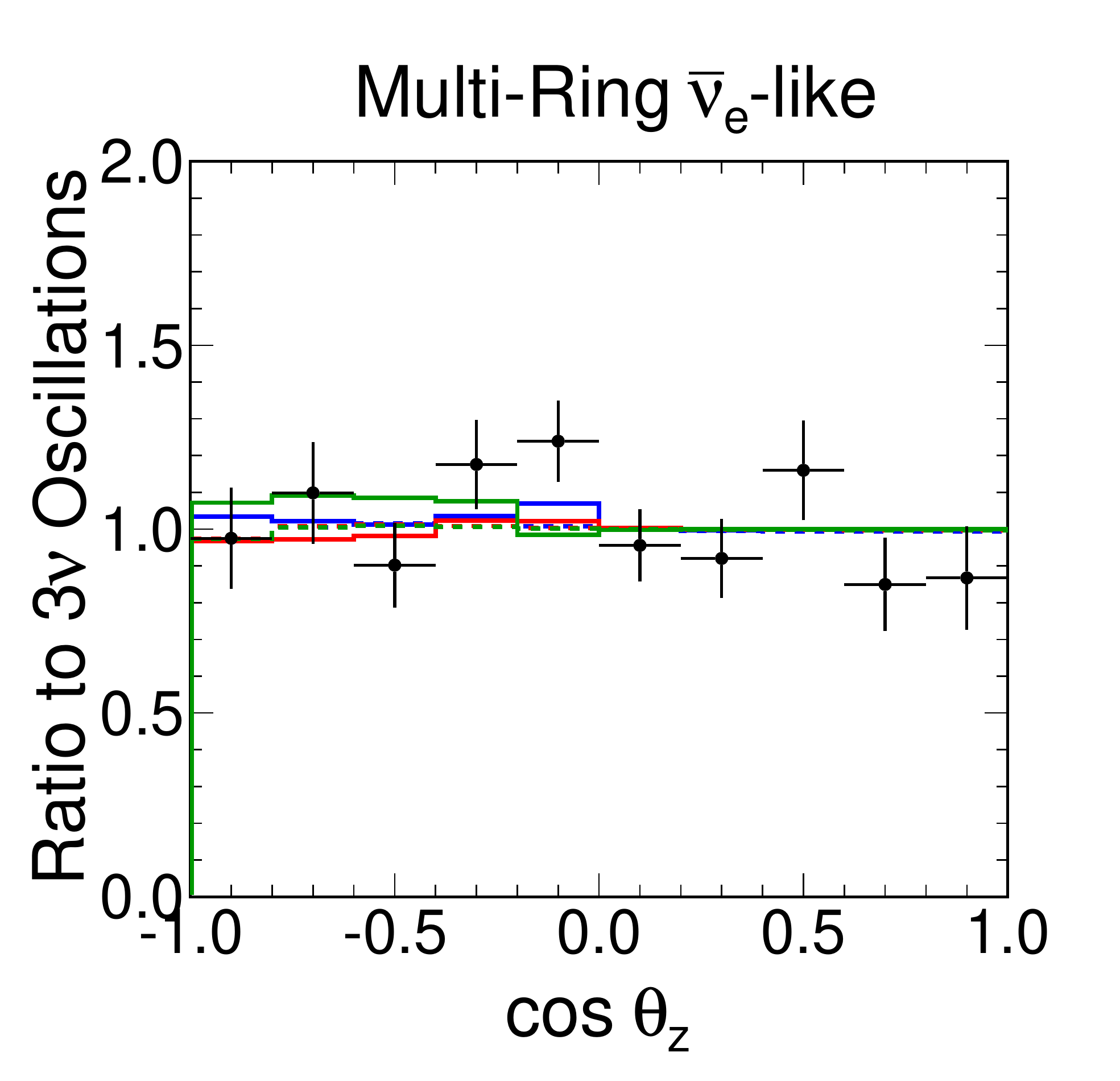}\\
 \includegraphics[width=\zwid,clip]{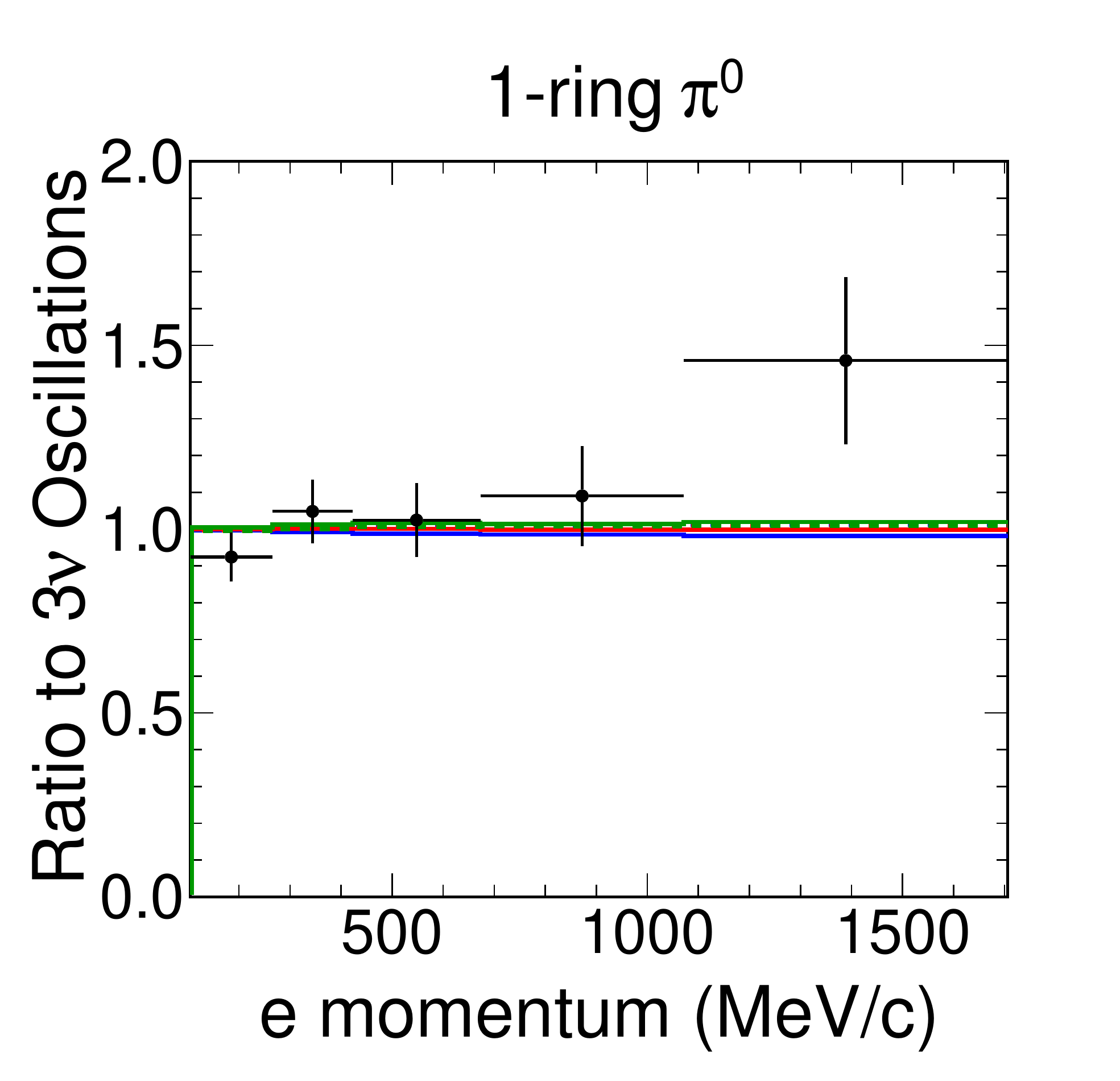}
 \includegraphics[width=\zwid,clip]{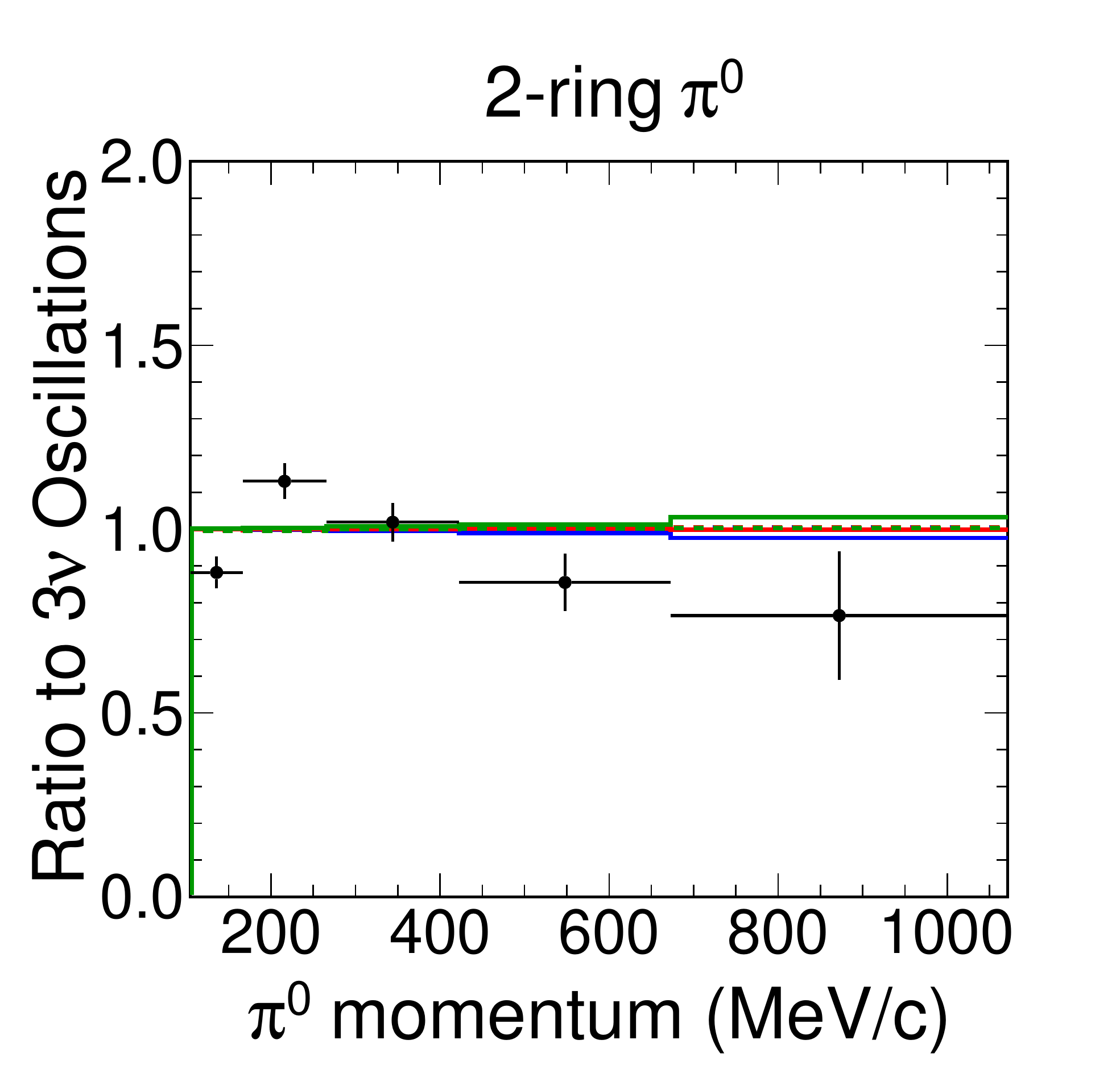}
 \includegraphics[width=\zwid,clip]{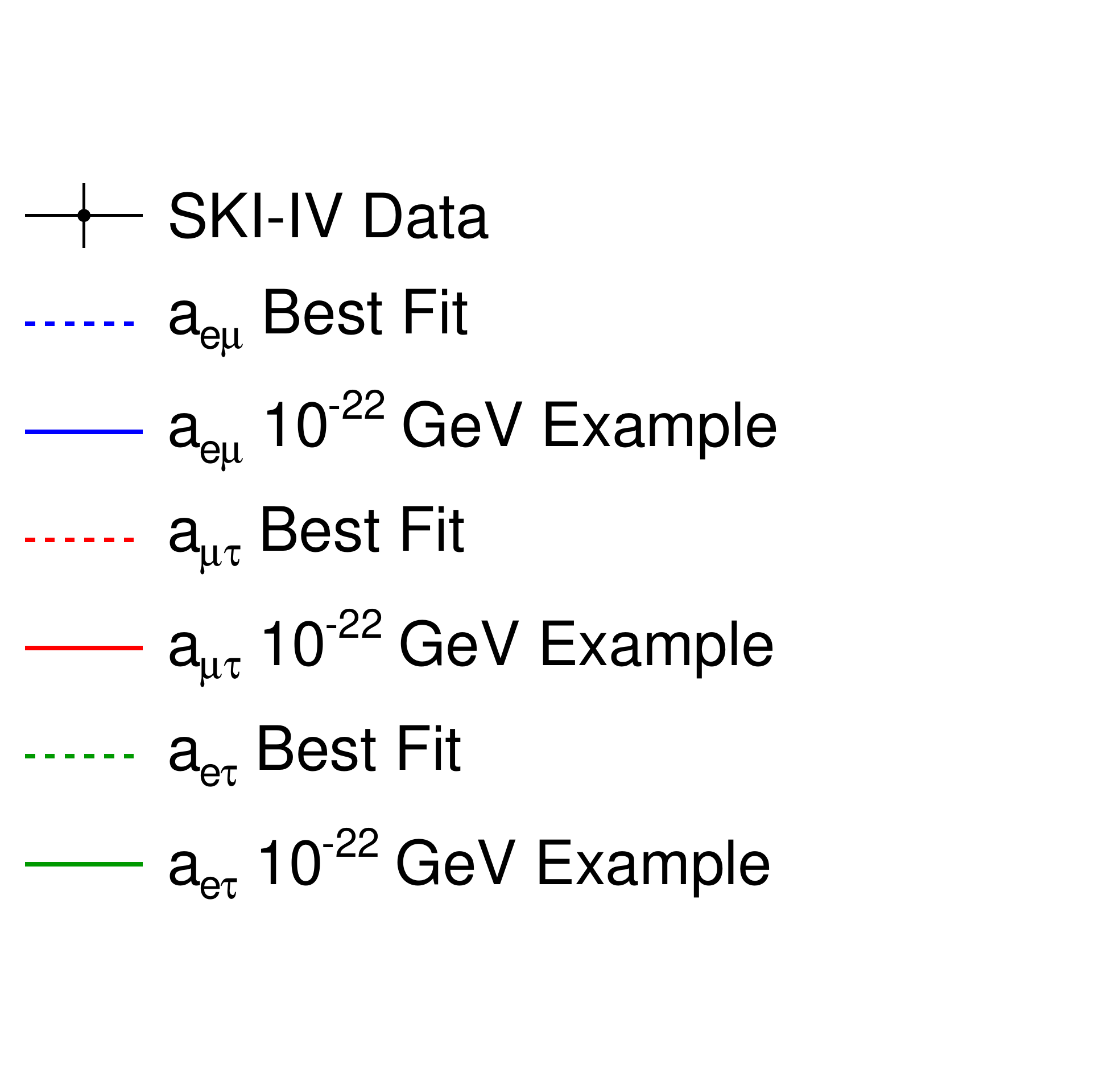}
 \caption{ (color online) Ratios of the summed SK-I through SK-IV \cz or momentum distributions  relative to standard three-flavor oscillations of the $e$- and NC\pizero-like FC sub-samples. They are projected into \cz when binned in both momentum and angle. The black points represent the data with statistical errors. The dashed lines represent the best fits from the three sectors for the \at parameters and the solid lines represent examples of large Lorentz violation ($\at = \val{10^{-22}}{GeV}$, equivalent to \fig{oscillograma}). Significant deviations from unity would indicate Lorentz violation.
 }
 \label{fig:zenitha_e}
 \end{center}
\end{figure*}

\begin{figure*}
 \begin{center}
 \includegraphics[width=\zwid,clip]{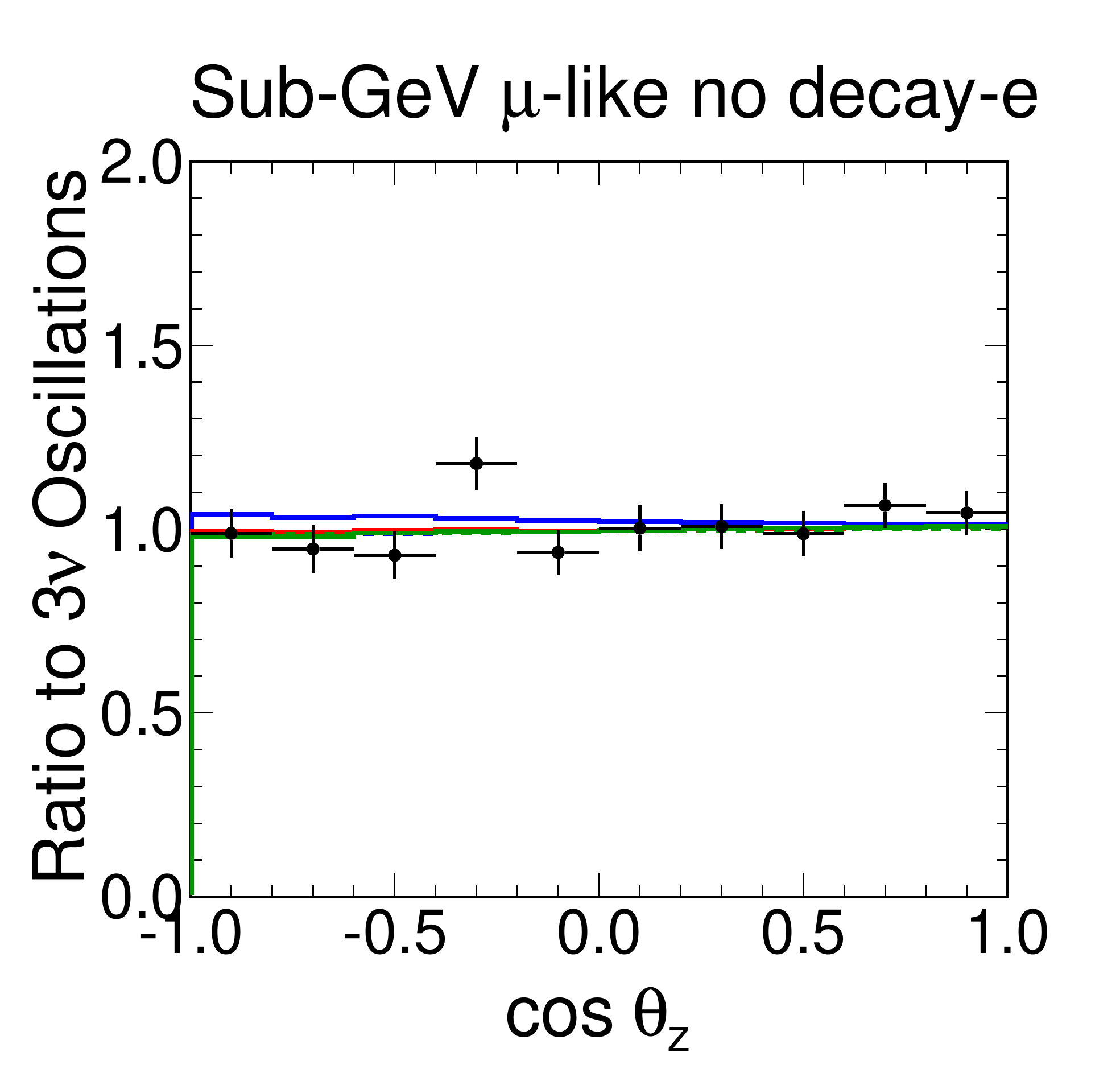}
 \includegraphics[width=\zwid,clip]{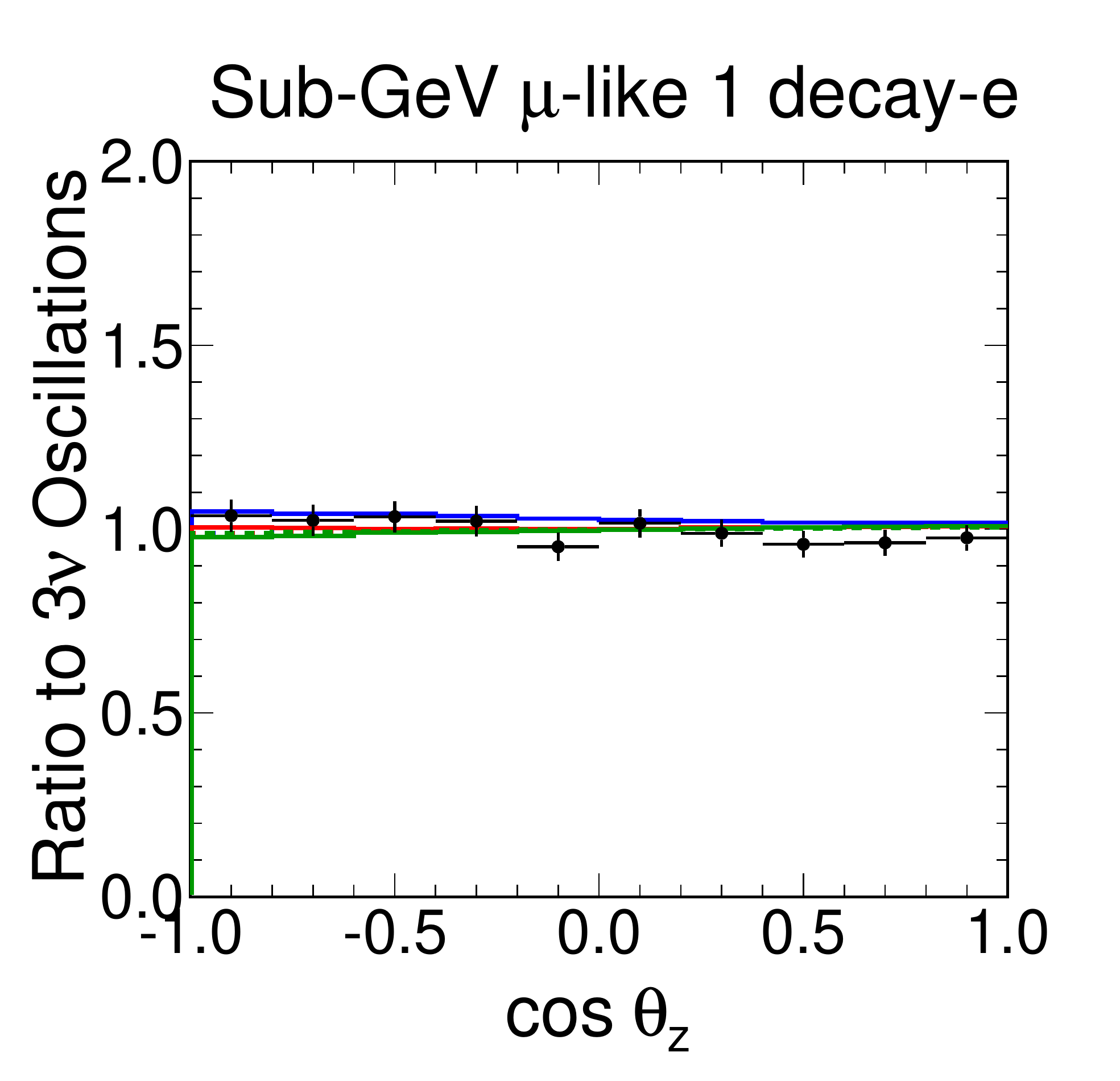}
 \includegraphics[width=\zwid,clip]{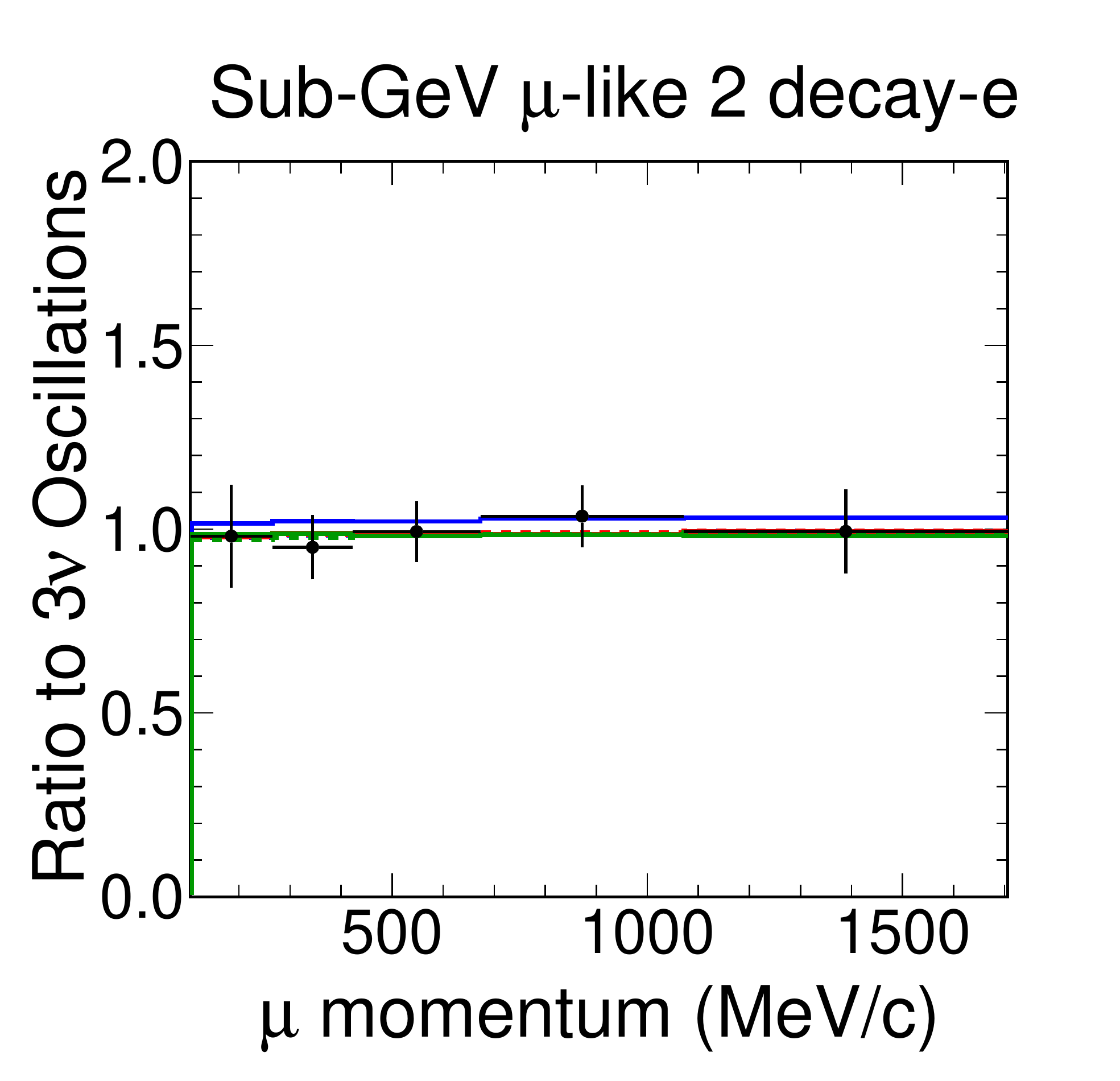} \\
 \includegraphics[width=\zwid,clip]{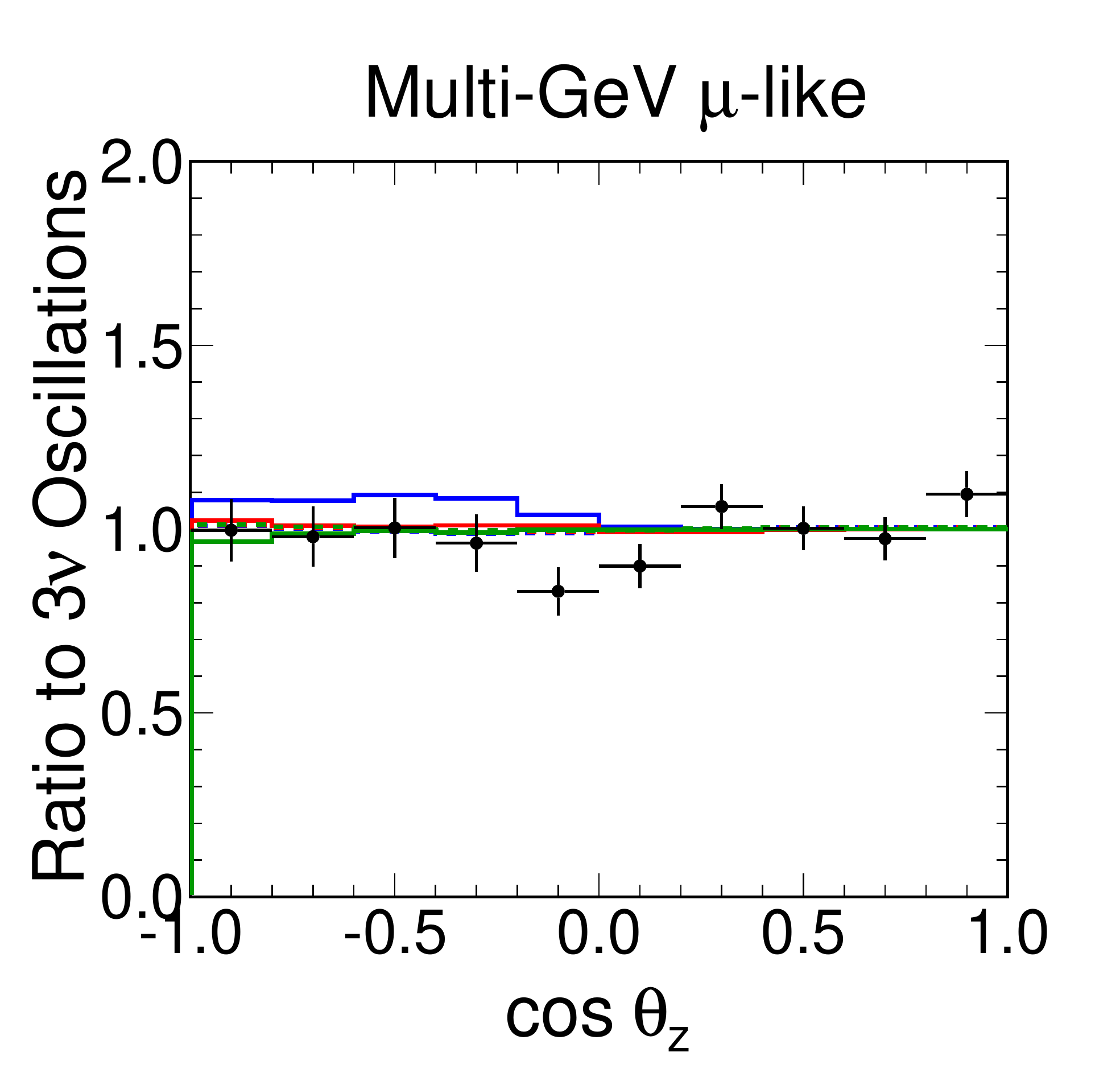}
 \includegraphics[width=\zwid,clip]{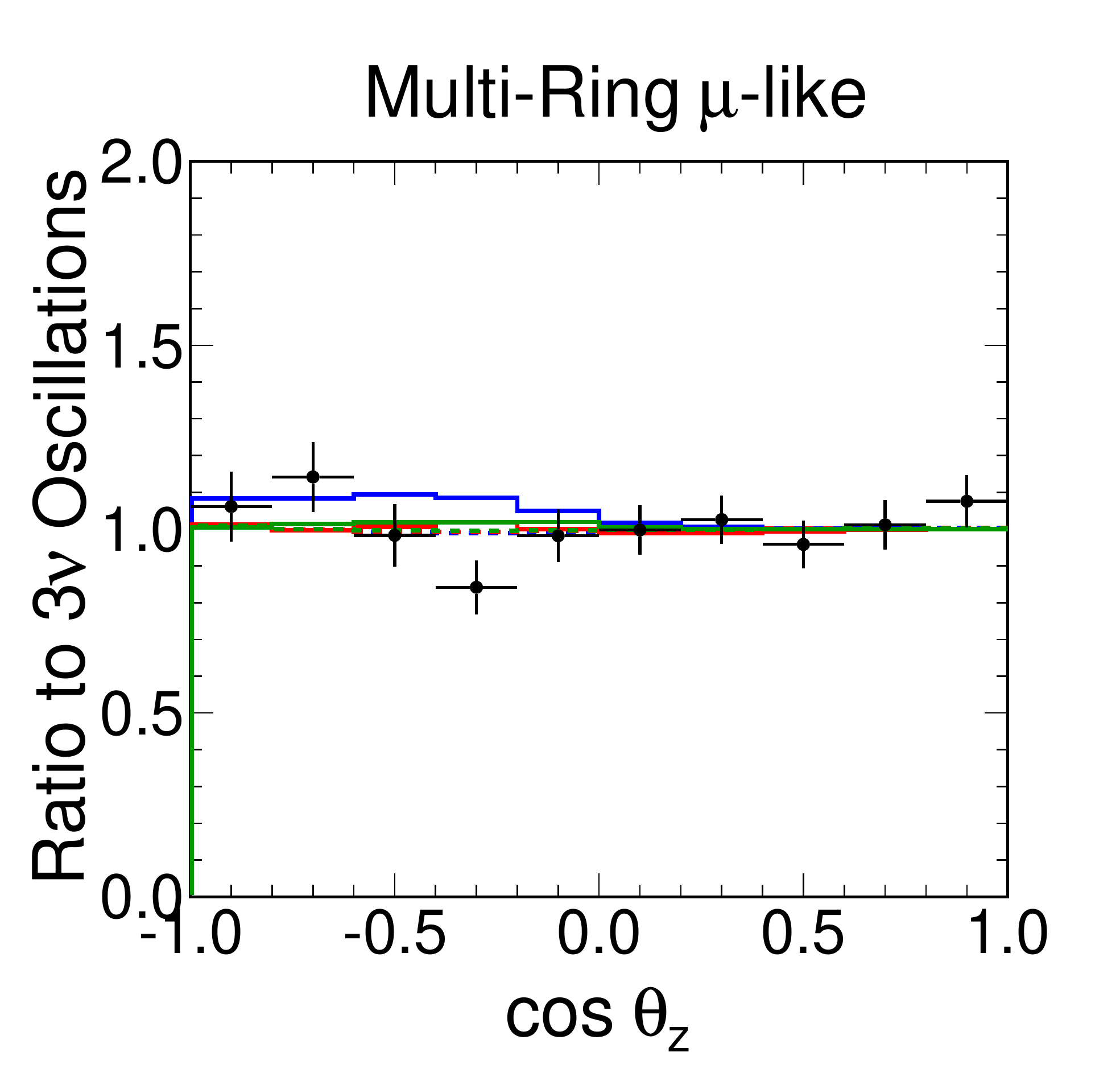}
 \includegraphics[width=\zwid,clip]{lv_ratio_a_legend.pdf} \\
 \includegraphics[width=\zwid,clip]{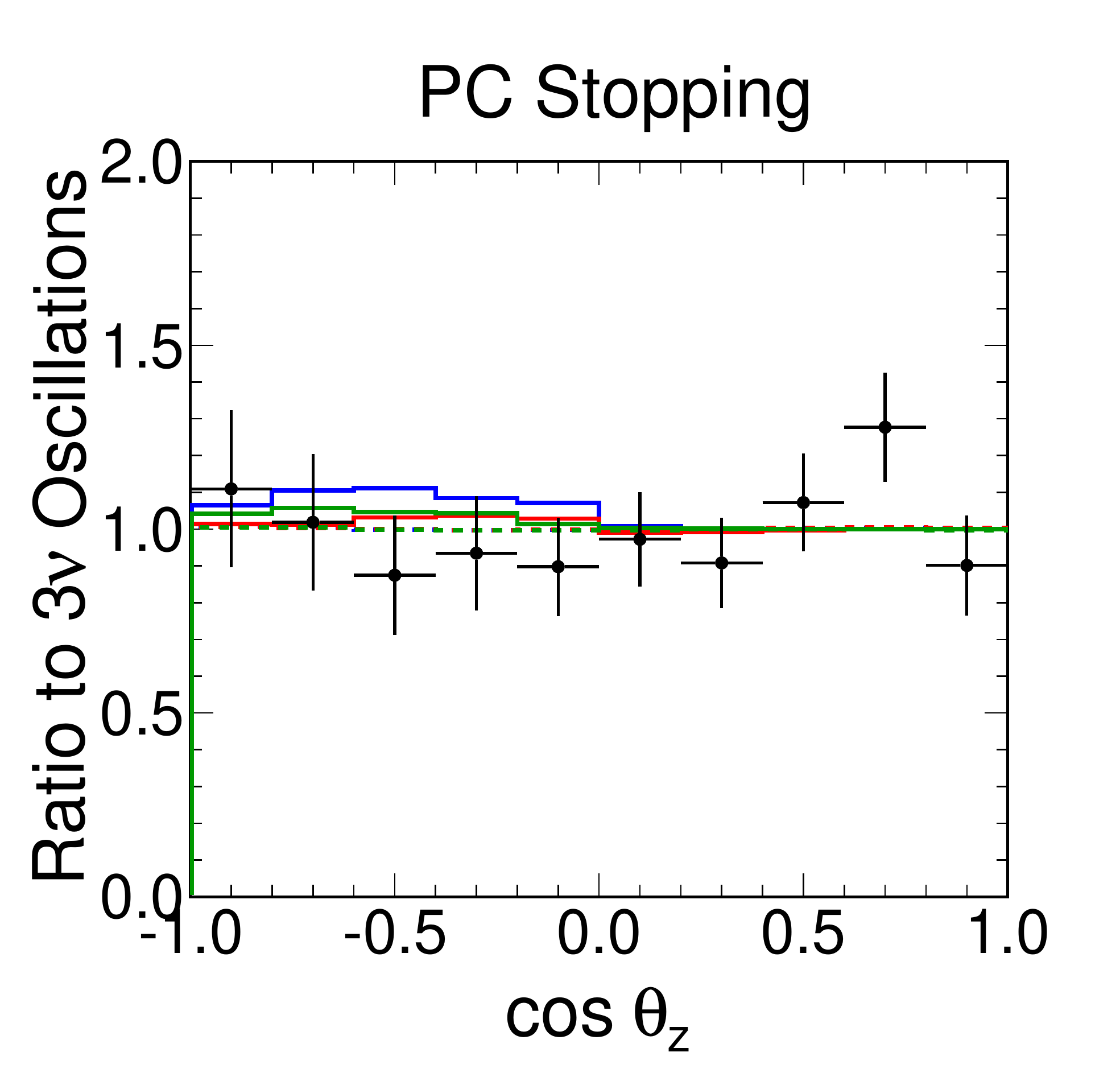}
 \includegraphics[width=\zwid,clip]{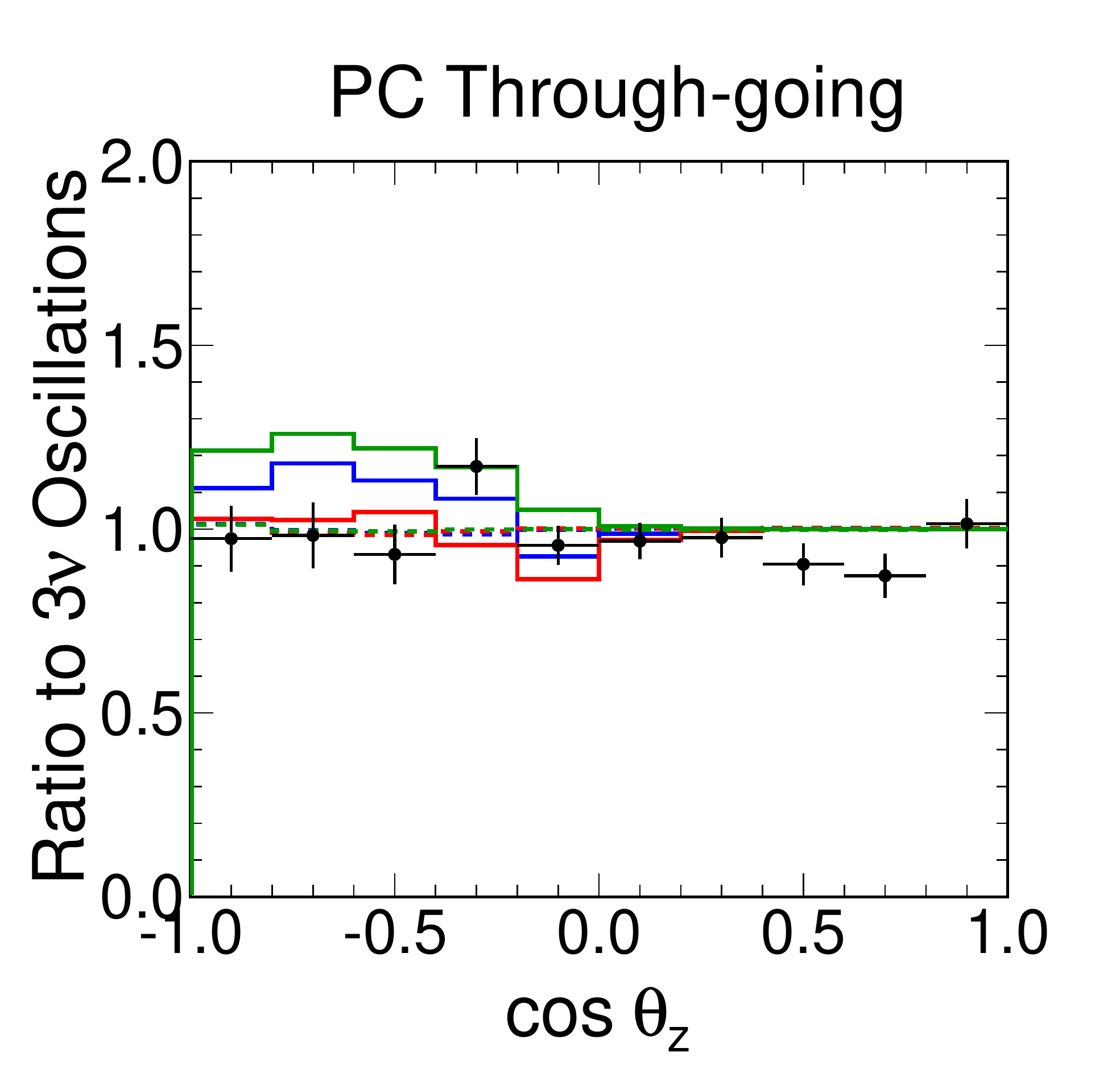}
 \includegraphics[width=\zwid,clip]{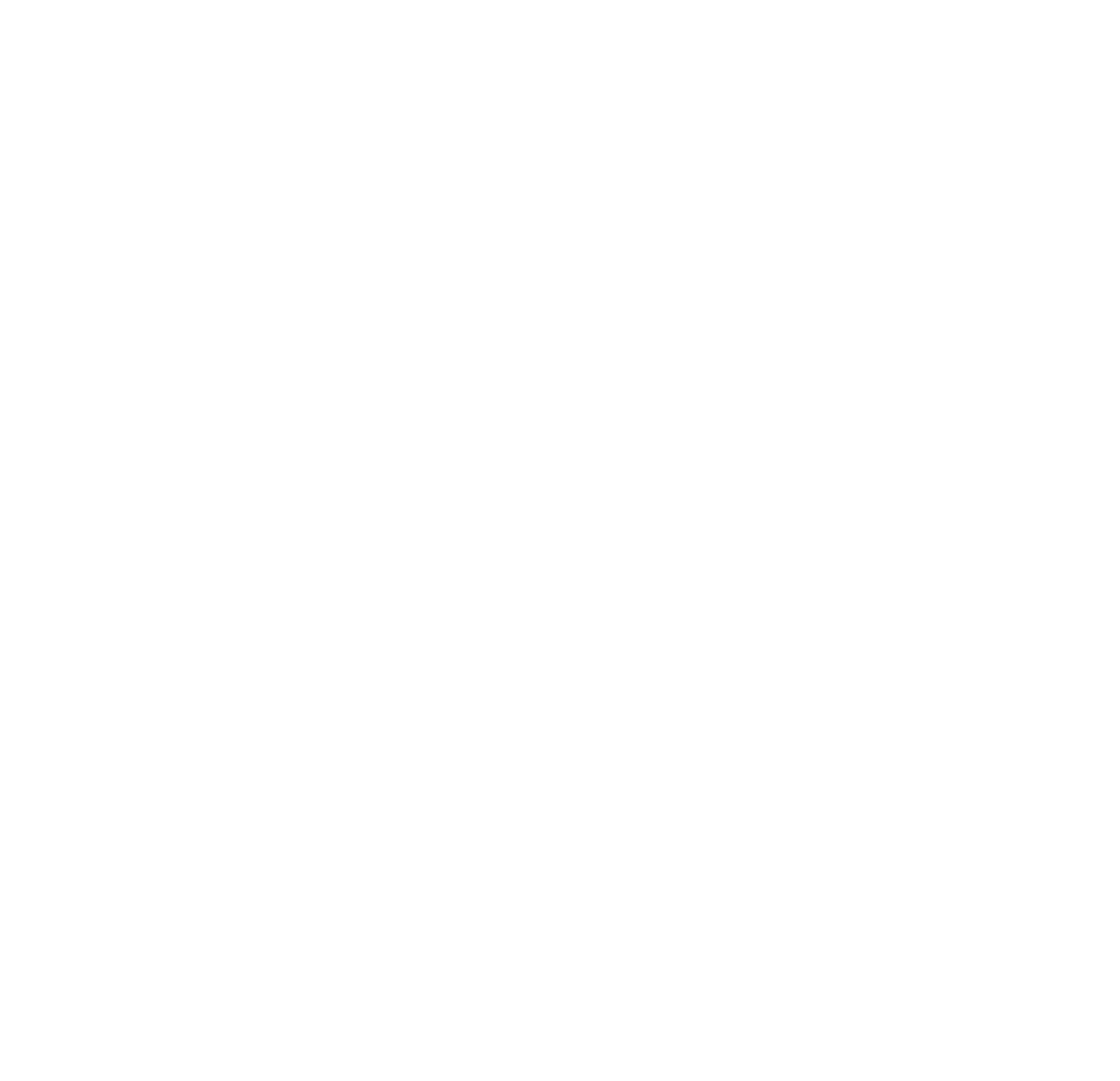} \\
 \includegraphics[width=\zwid,clip]{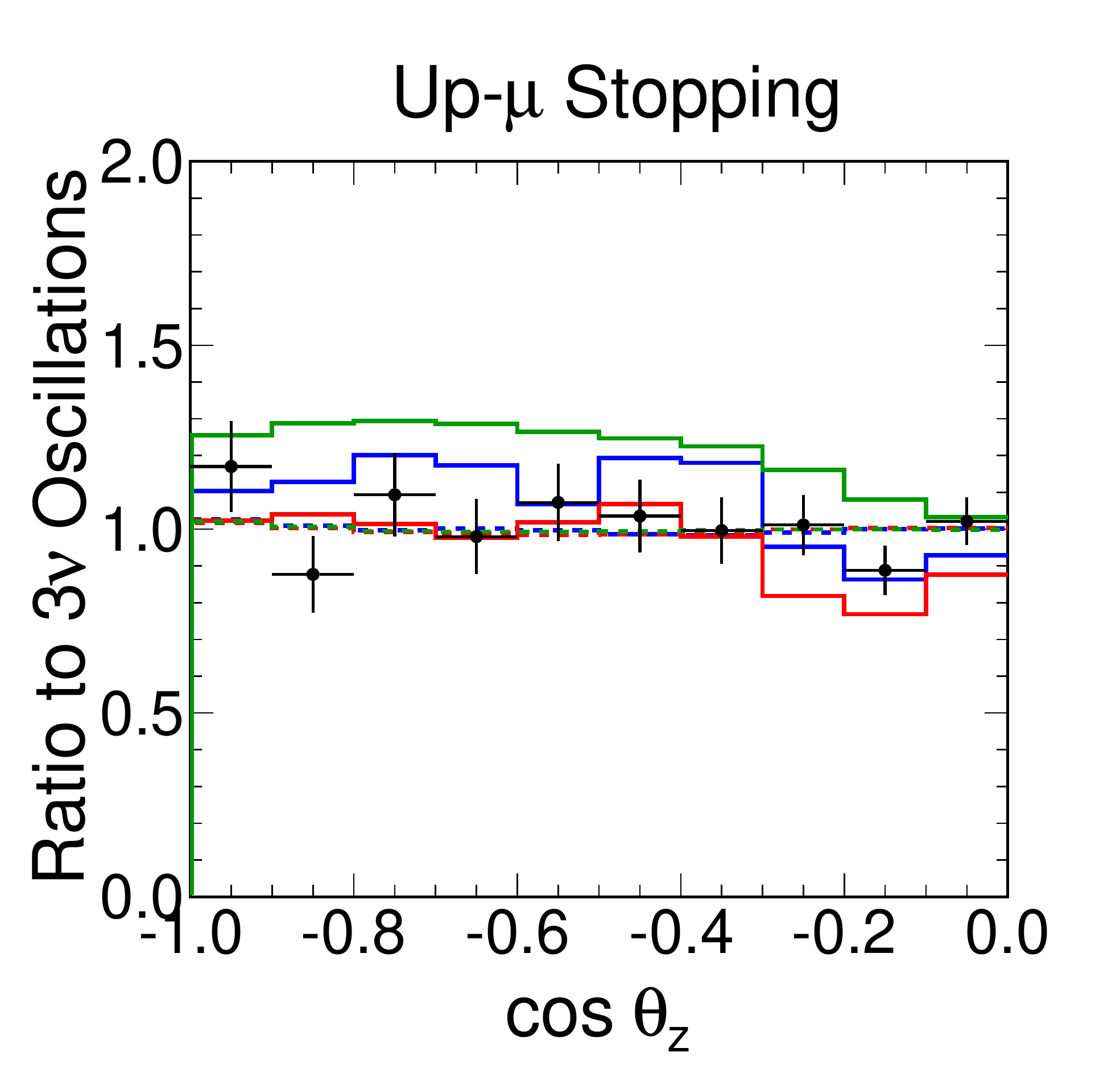}
 \includegraphics[width=\zwid,clip]{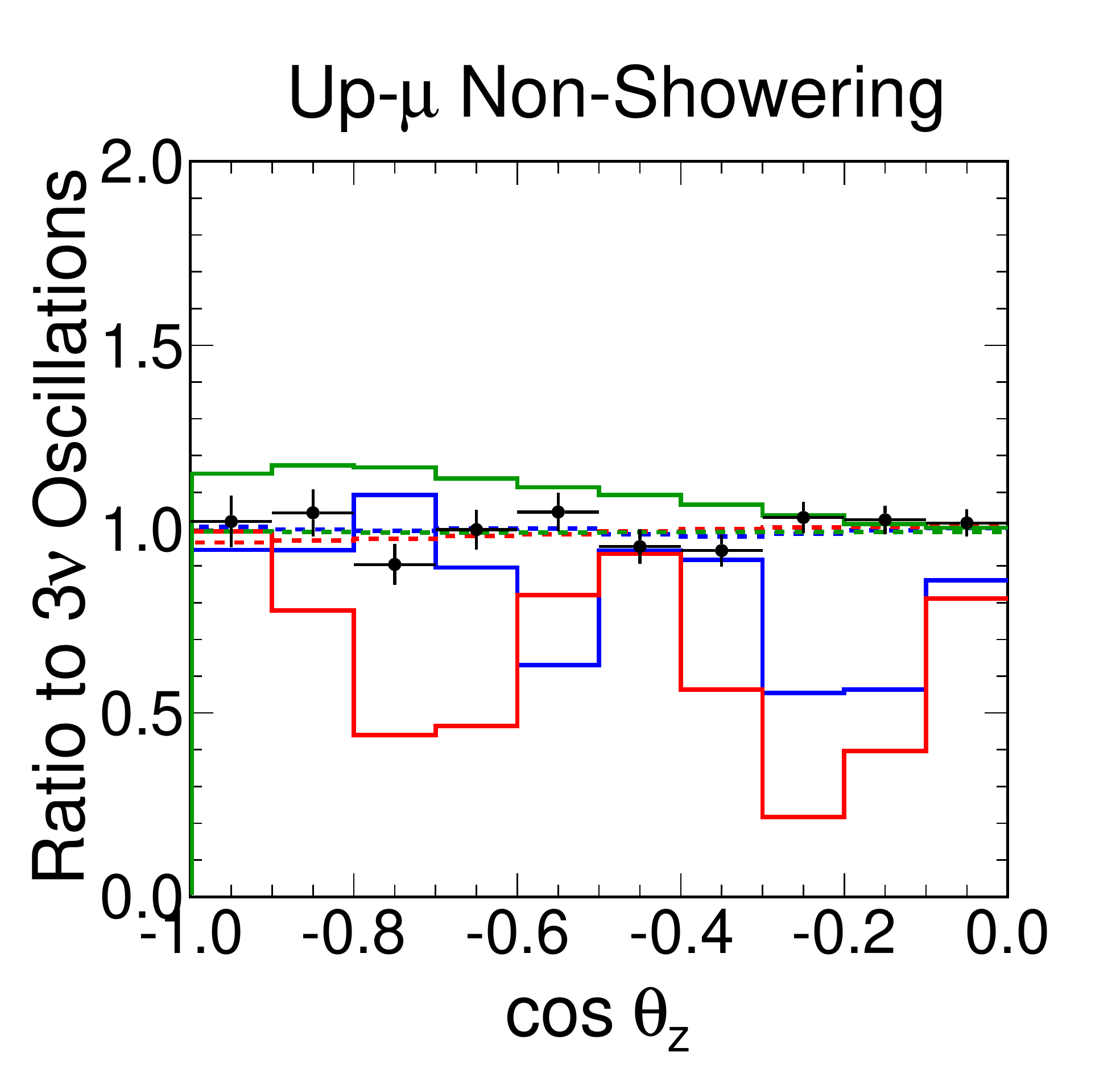}
 \includegraphics[width=\zwid,clip]{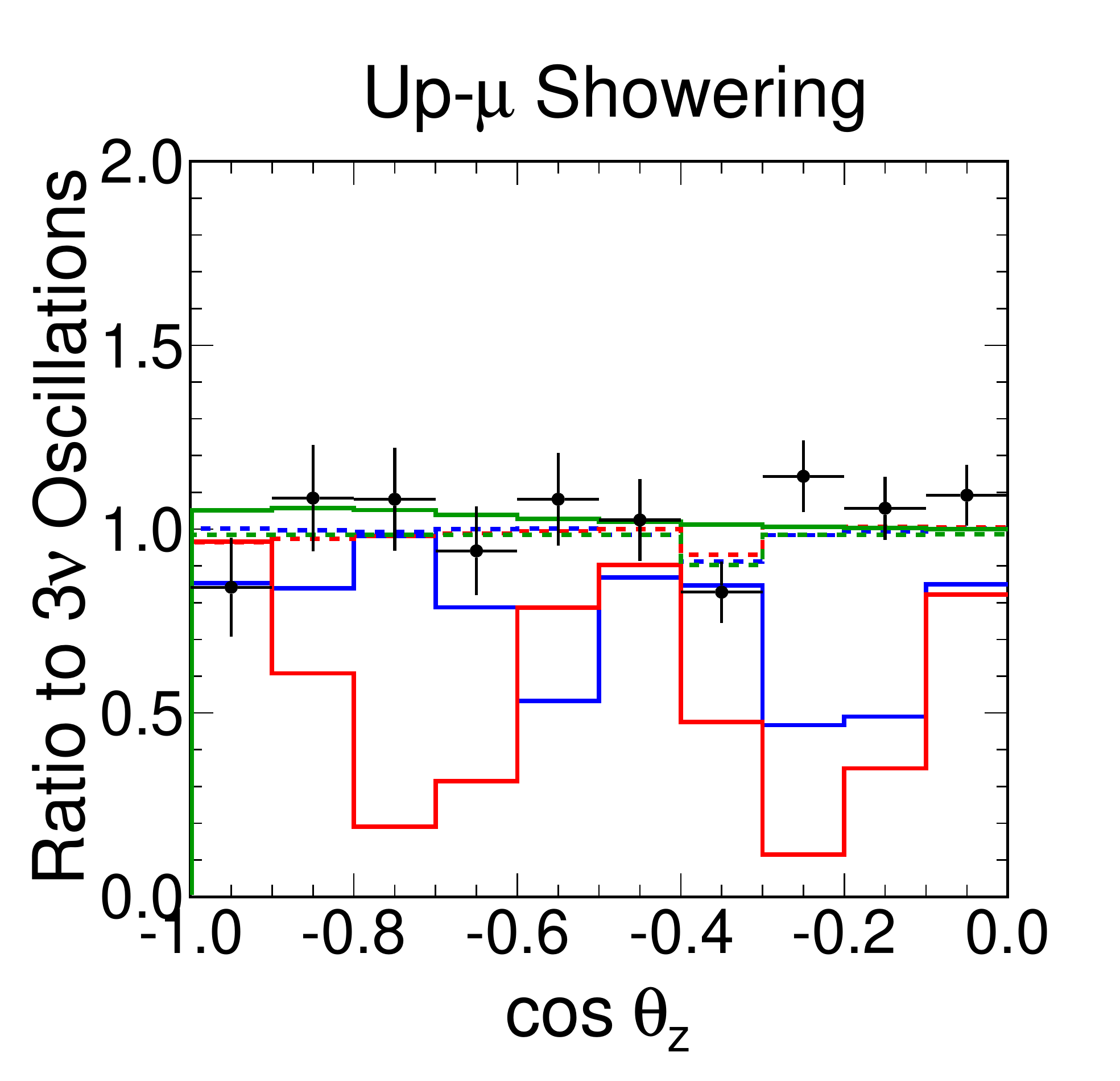}
 \caption{(color online) Ratios of the summed SK-I through SK-IV  \cz distributions relative to standard three-flavor oscillations for the $\mu$-like FC, PC, and \UP sub-samples. They are projected into \cz when binned in both momentum and angle and the Sub-GeV 2 decay-e sample is binned only in momentum. The black points represent the data with statistical errors. The dashed lines represent the best fits from the three sectors for the \at parameters and the solid lines represent examples of large Lorentz violation ($\at = \val{10^{-22}}{GeV}$, equivalent to \fig{oscillograma}). Significant deviations from unity would indicate Lorentz violation.
 }
 \label{fig:zenitha_mu}
 \end{center}
\end{figure*}

\begin{figure*}
 \begin{center}
 \includegraphics[width=\zwid,clip]{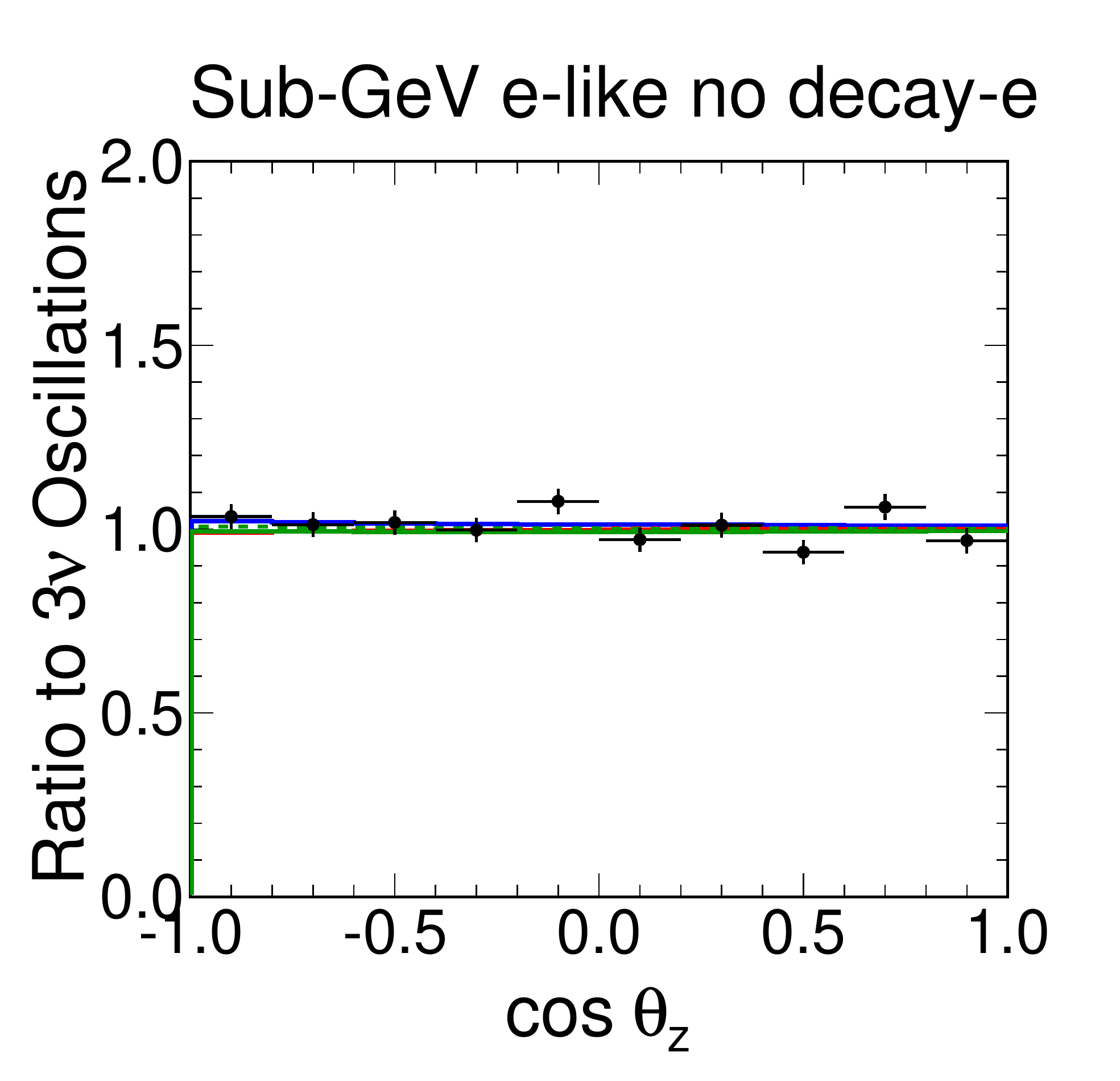}
 \includegraphics[width=\zwid,clip]{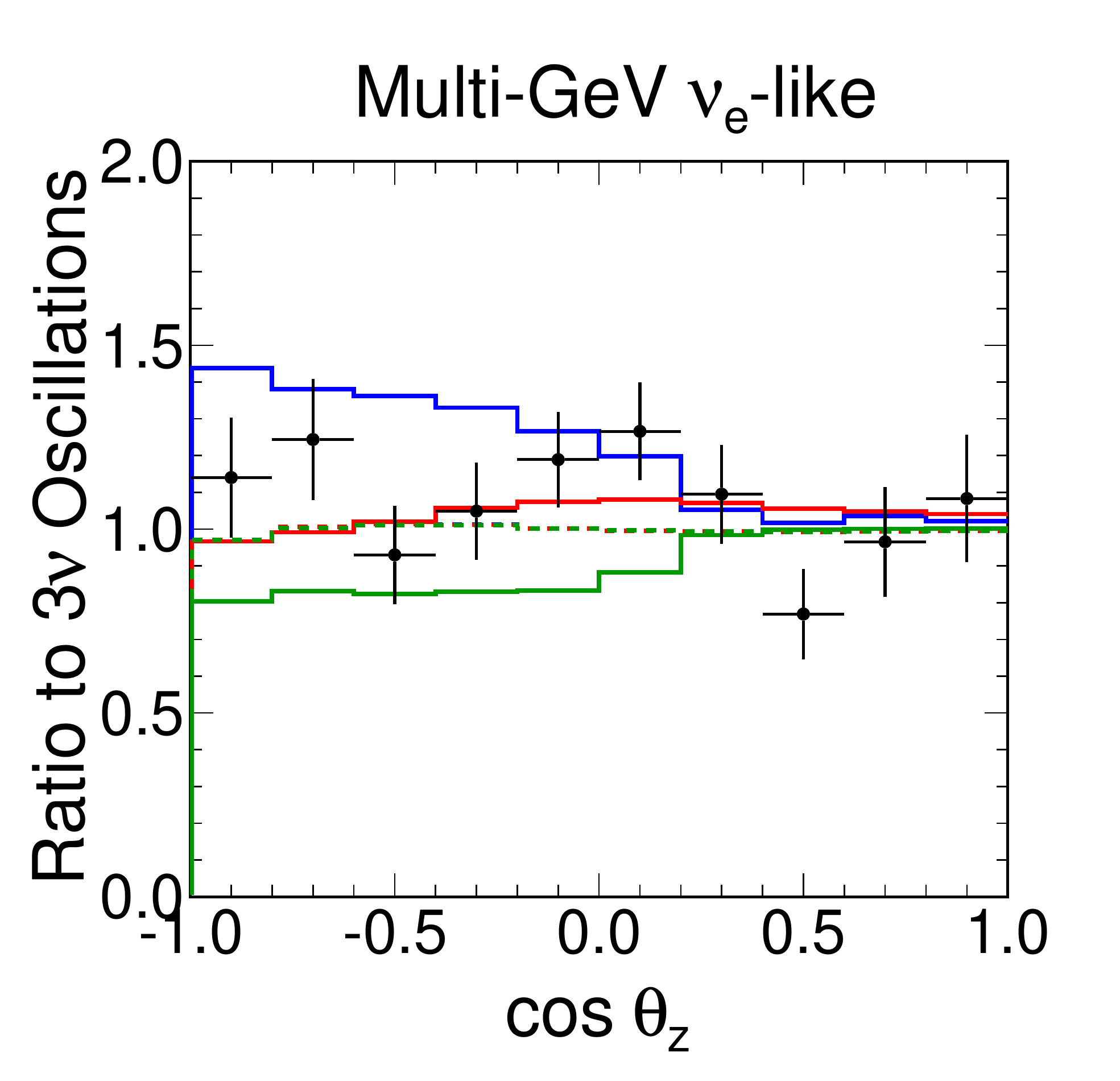}
 \includegraphics[width=\zwid,clip]{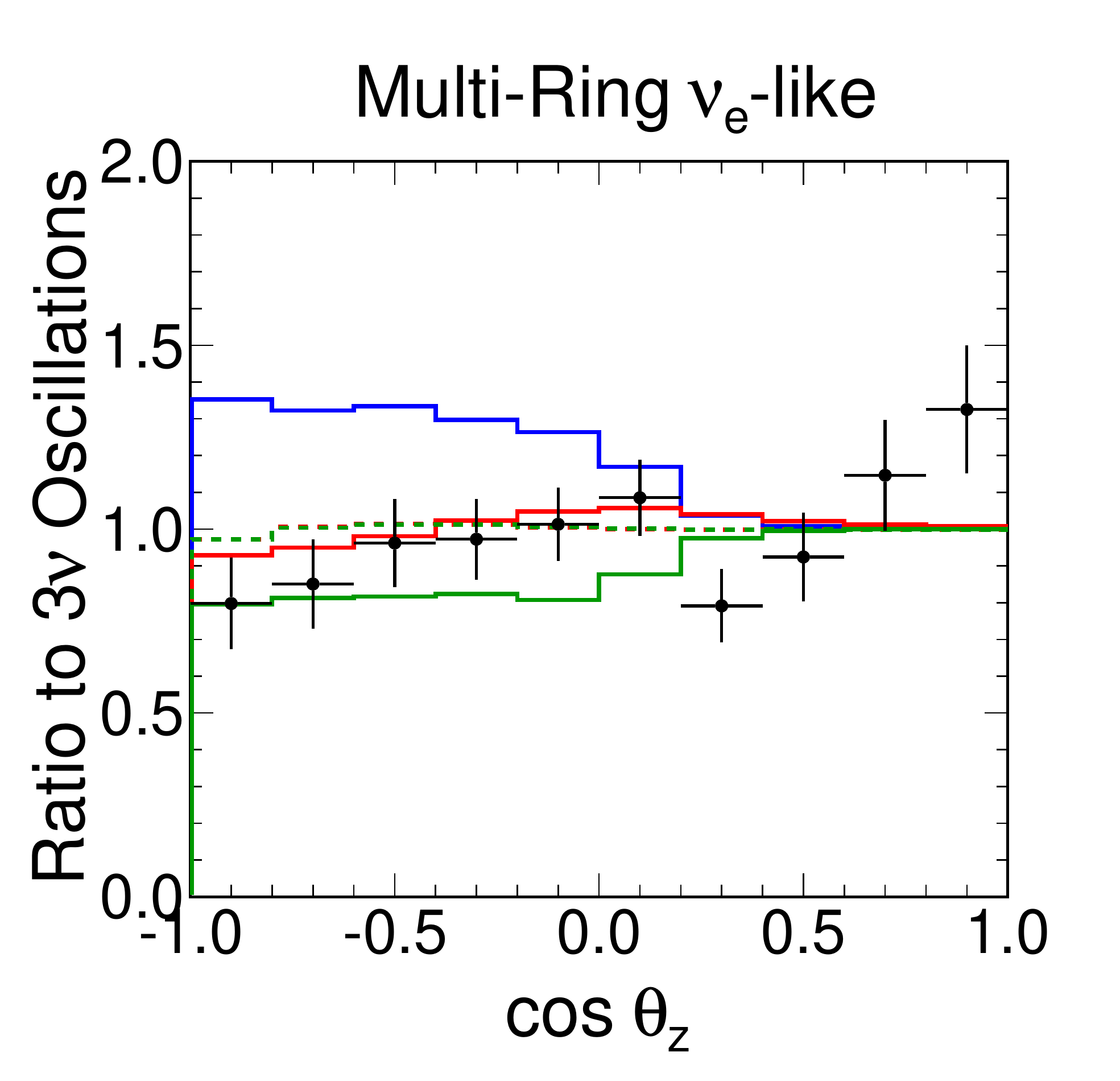} \\
 \includegraphics[width=\zwid,clip]{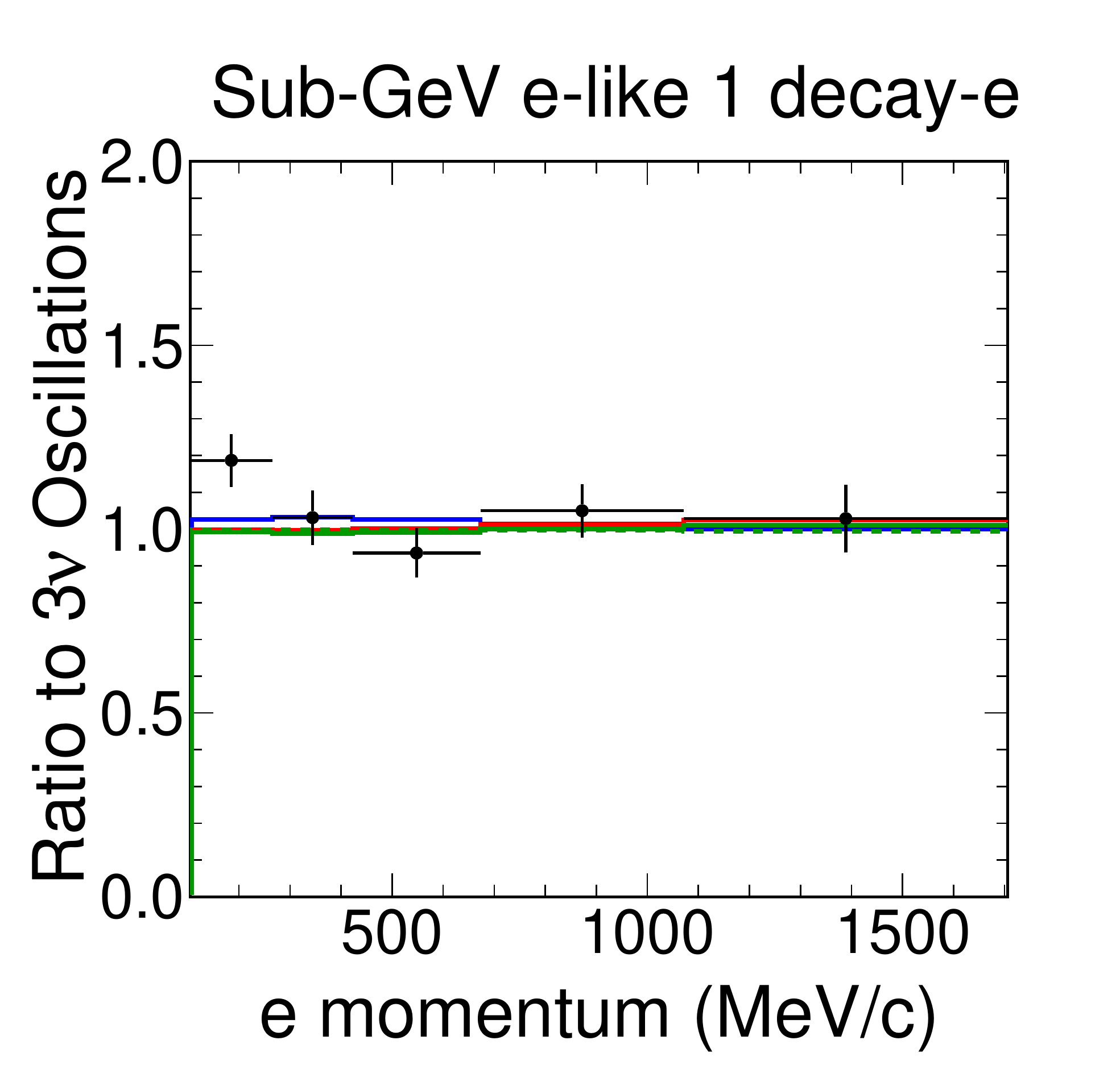} 
 \includegraphics[width=\zwid,clip]{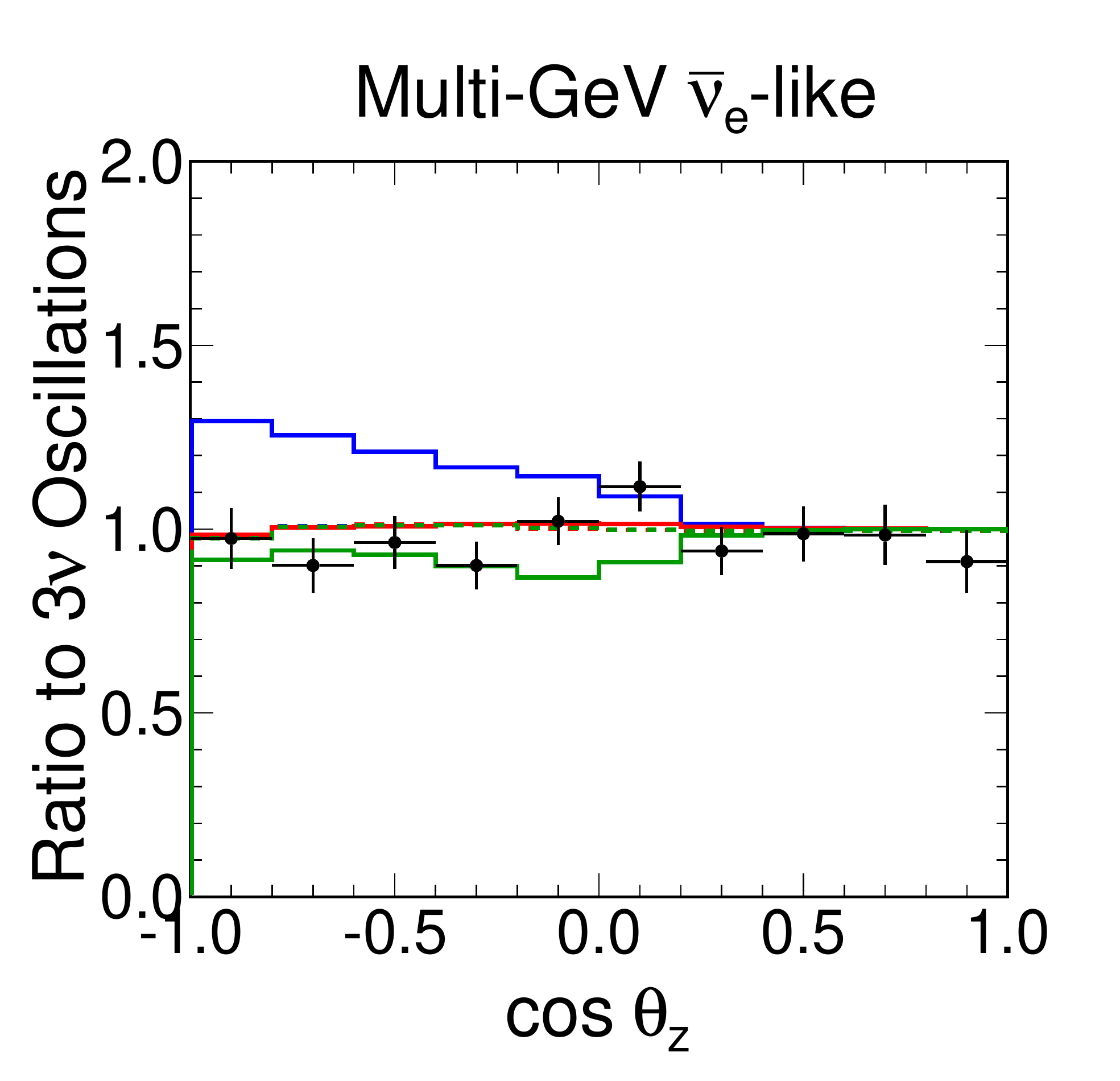}
 \includegraphics[width=\zwid,clip]{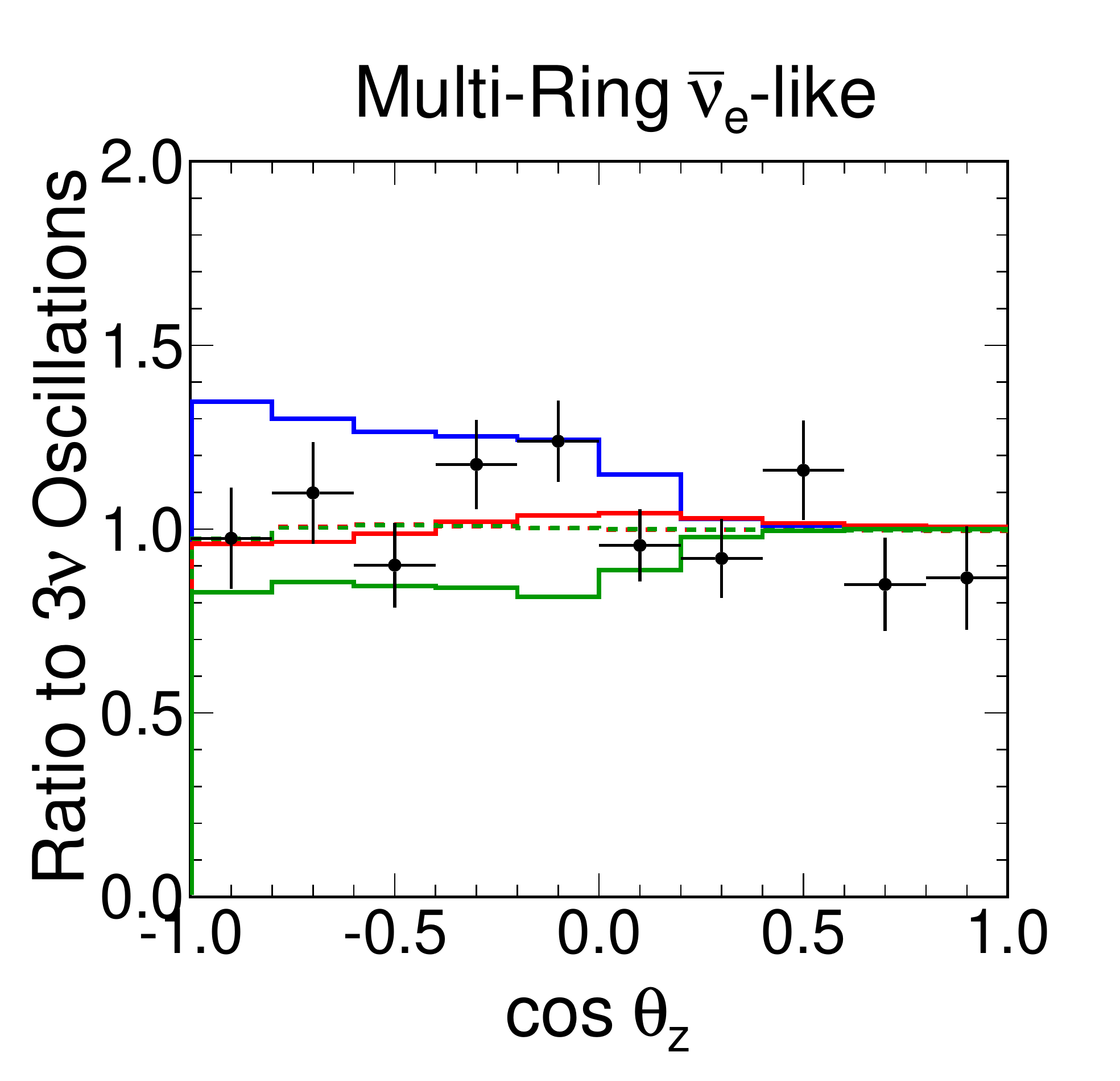}\\
 \includegraphics[width=\zwid,clip]{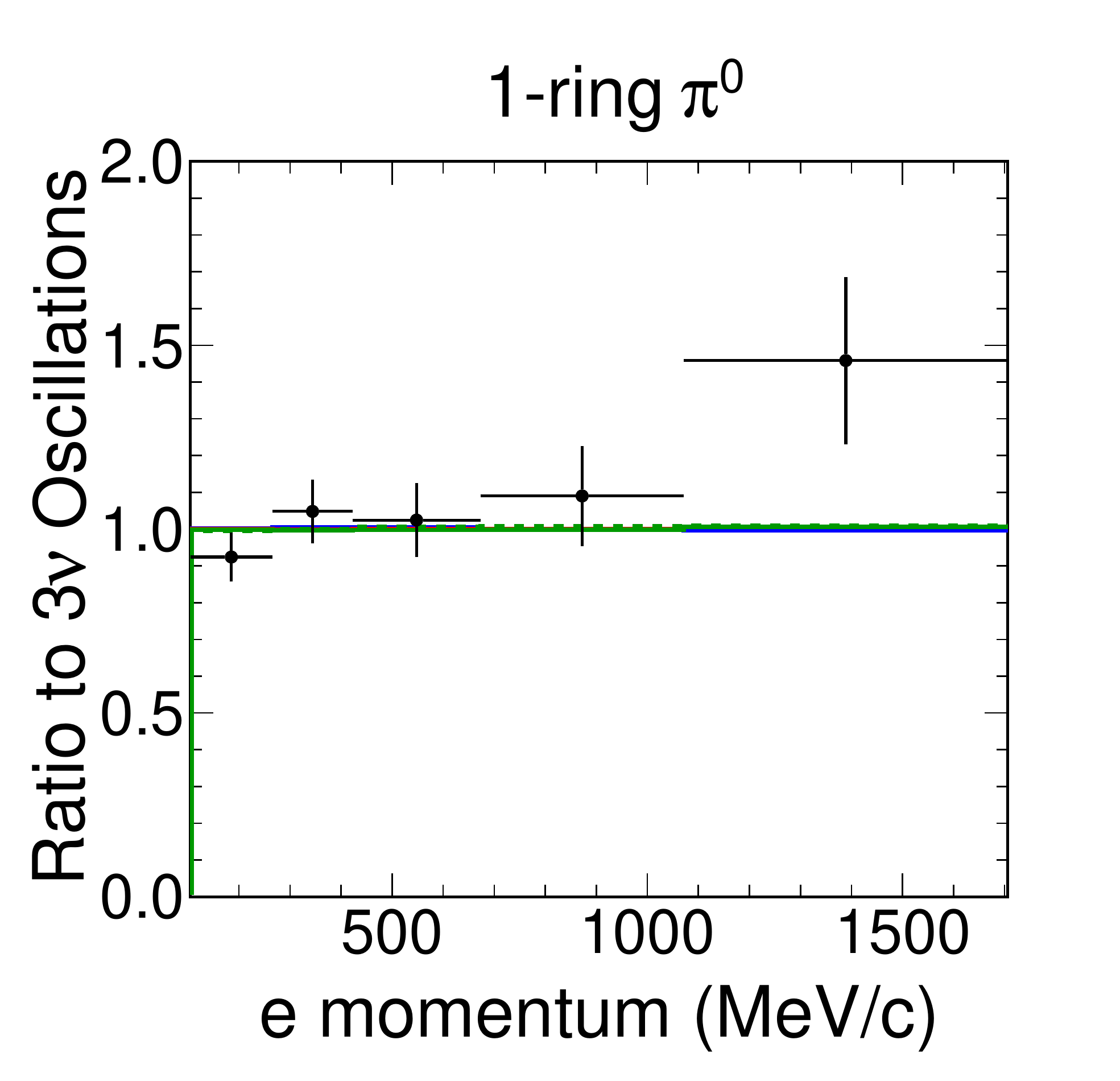}
 \includegraphics[width=\zwid,clip]{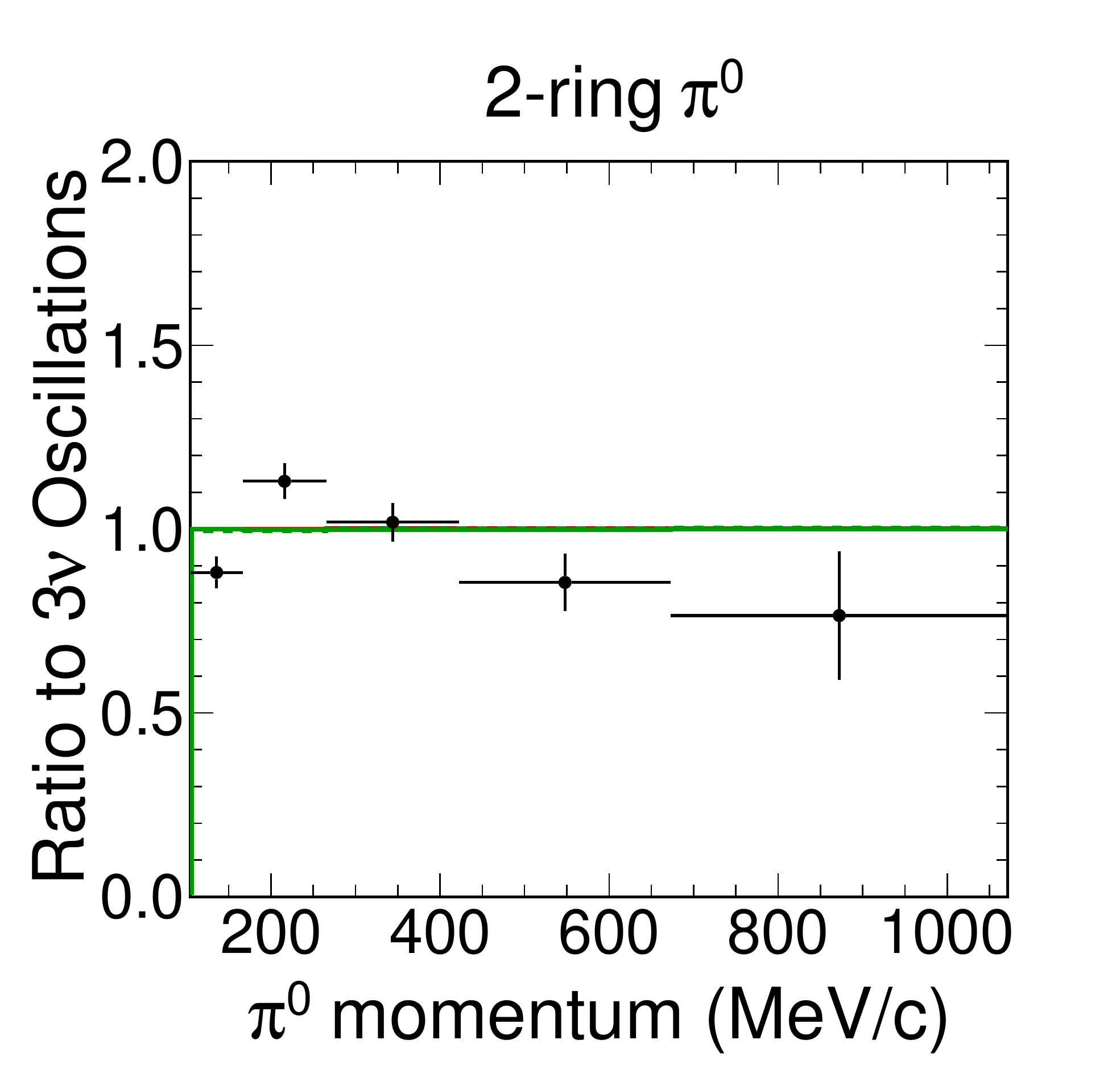}
 \includegraphics[width=\zwid,clip]{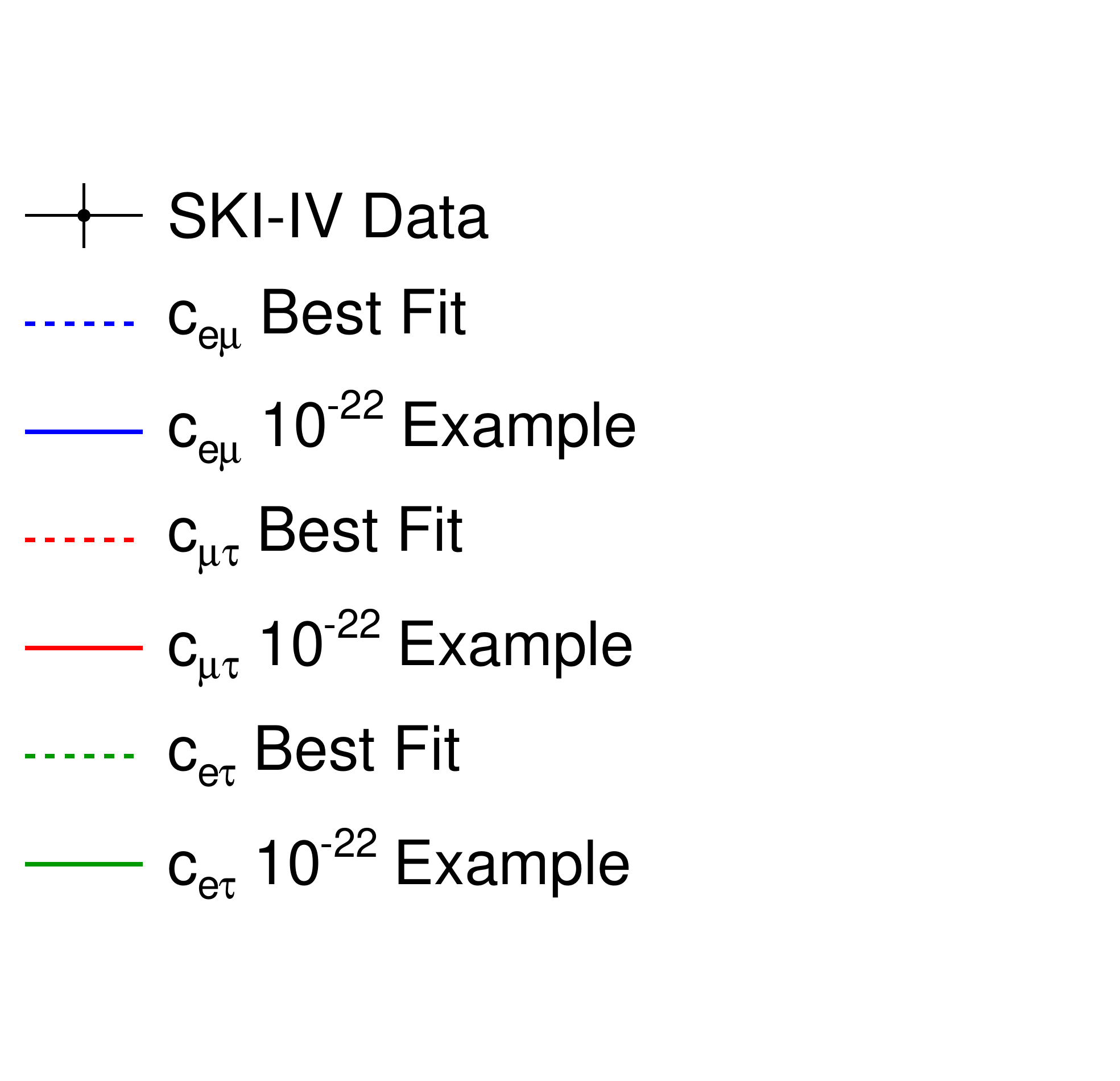}
 \caption{ (color online) Ratios of the summed SK-I through SK-IV \cz or momentum distributions relative to standard three-flavor oscillations of the $e$- and NC\pizero-like FC sub-samples. They are projected into \cz when binned in momentum and angle. The black points represent the data with statistical errors. The dashed lines represent the best fits from the three sectors for the \ctt parameters and the solid lines represent examples of large Lorentz violation ($\ctt = \sci{7.5}{-23}$, equivalent to \fig{oscillogramc}). Significant deviations from unity would indicate Lorentz violation.
 }
 \label{fig:zenithc_e}
 \end{center}
\end{figure*}

\begin{figure*}
 \begin{center}
 \includegraphics[width=\zwid,clip]{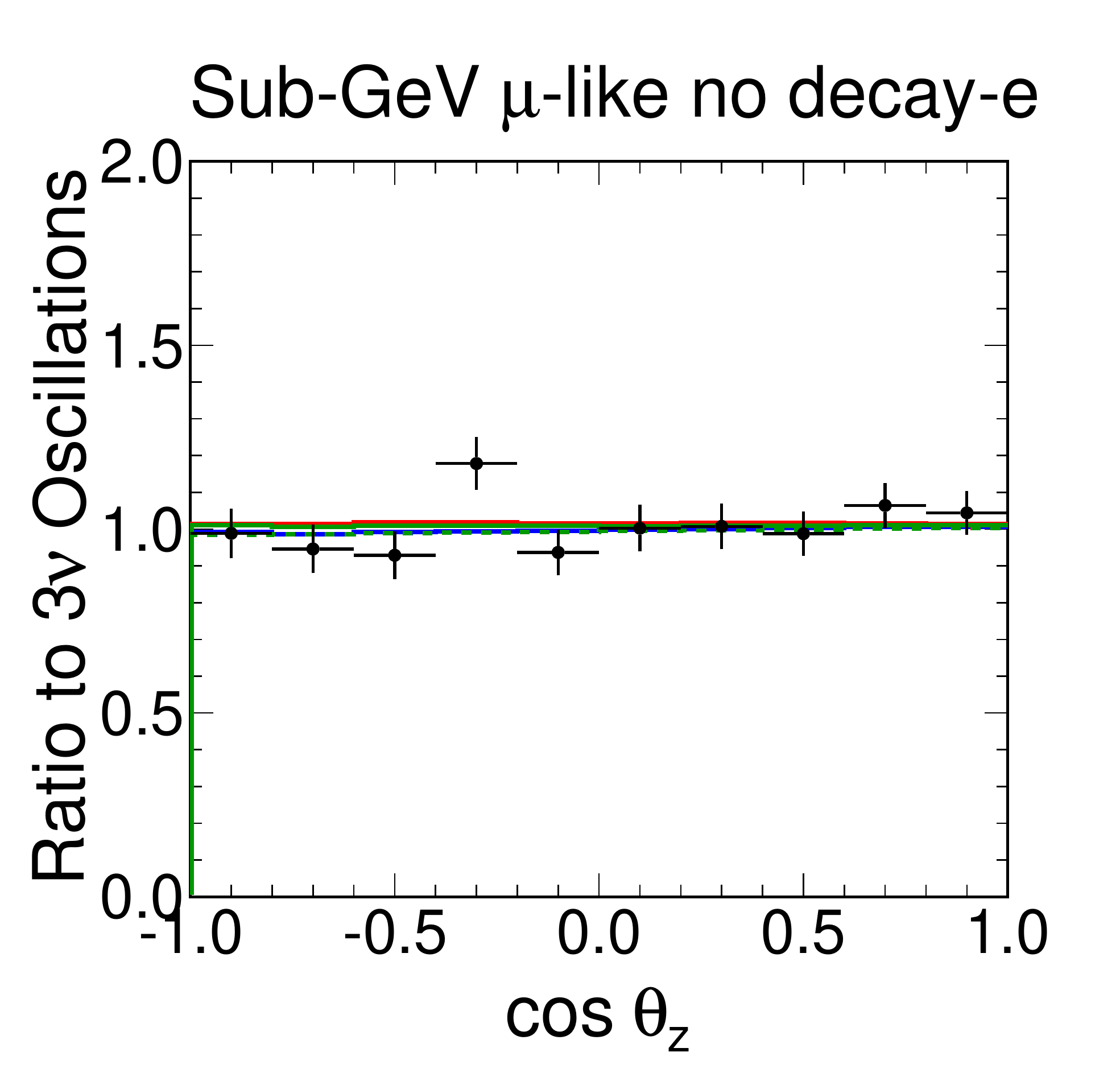}
 \includegraphics[width=\zwid,clip]{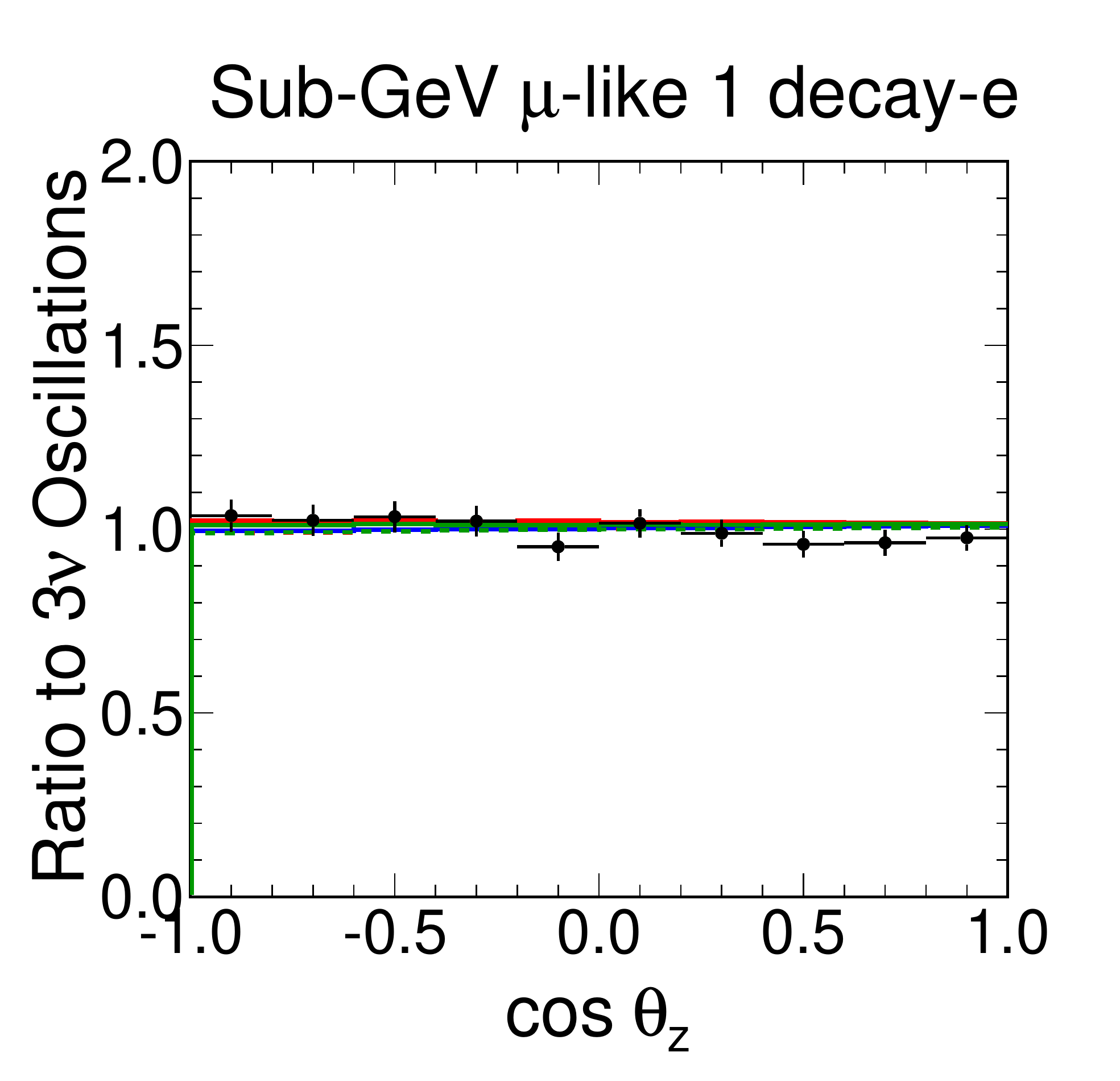}
 \includegraphics[width=\zwid,clip]{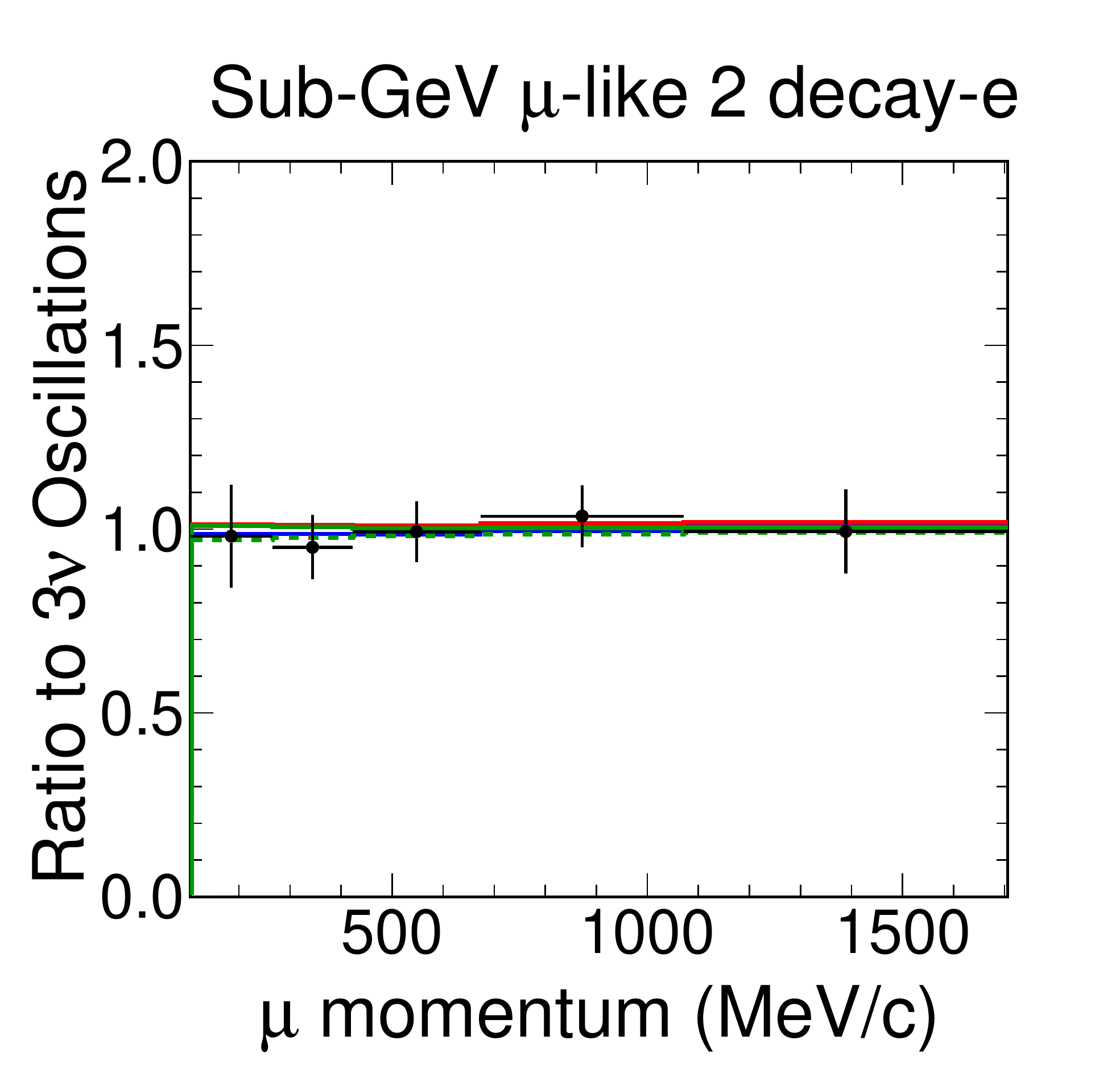} \\
 \includegraphics[width=\zwid,clip]{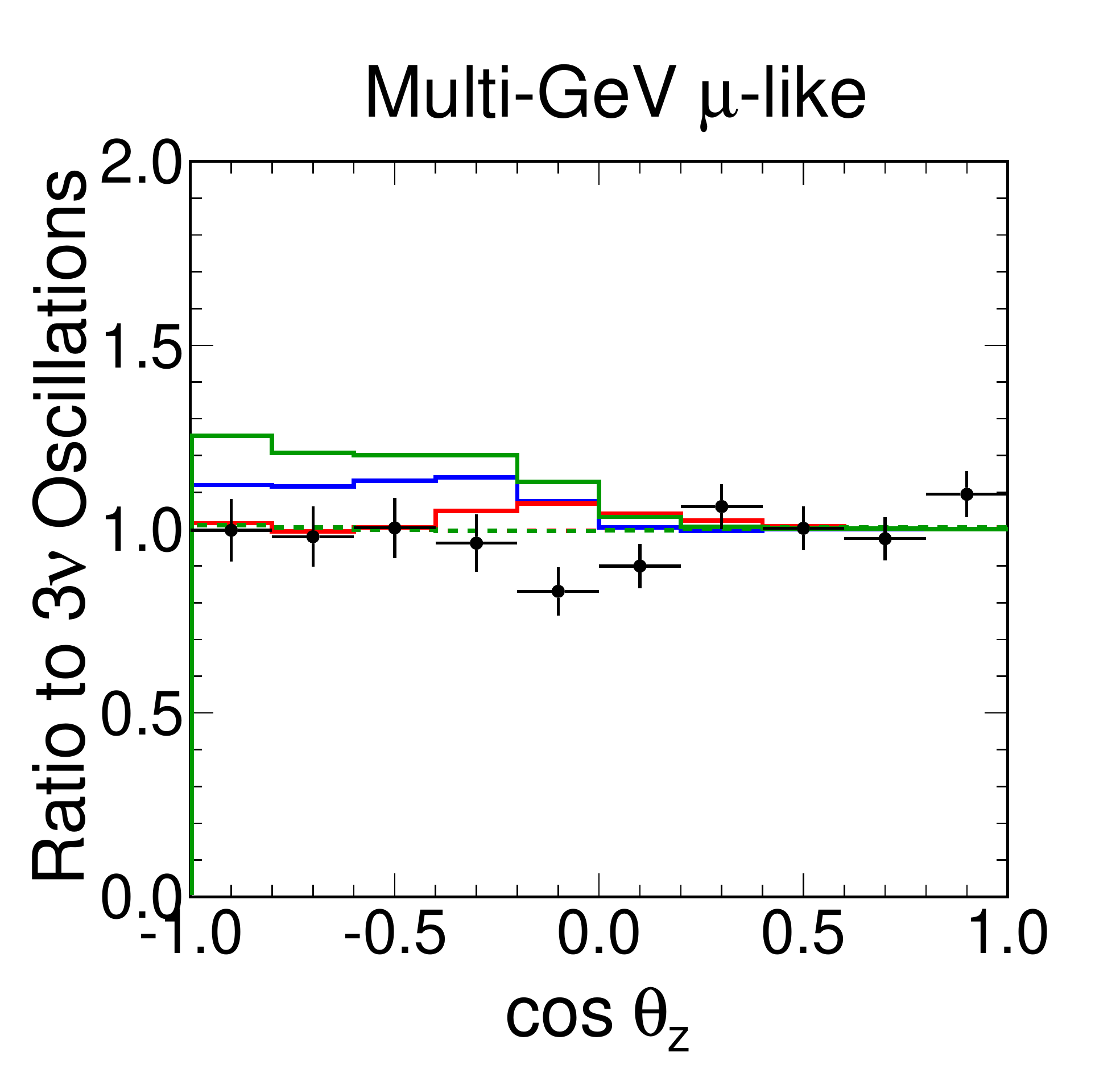}
 \includegraphics[width=\zwid,clip]{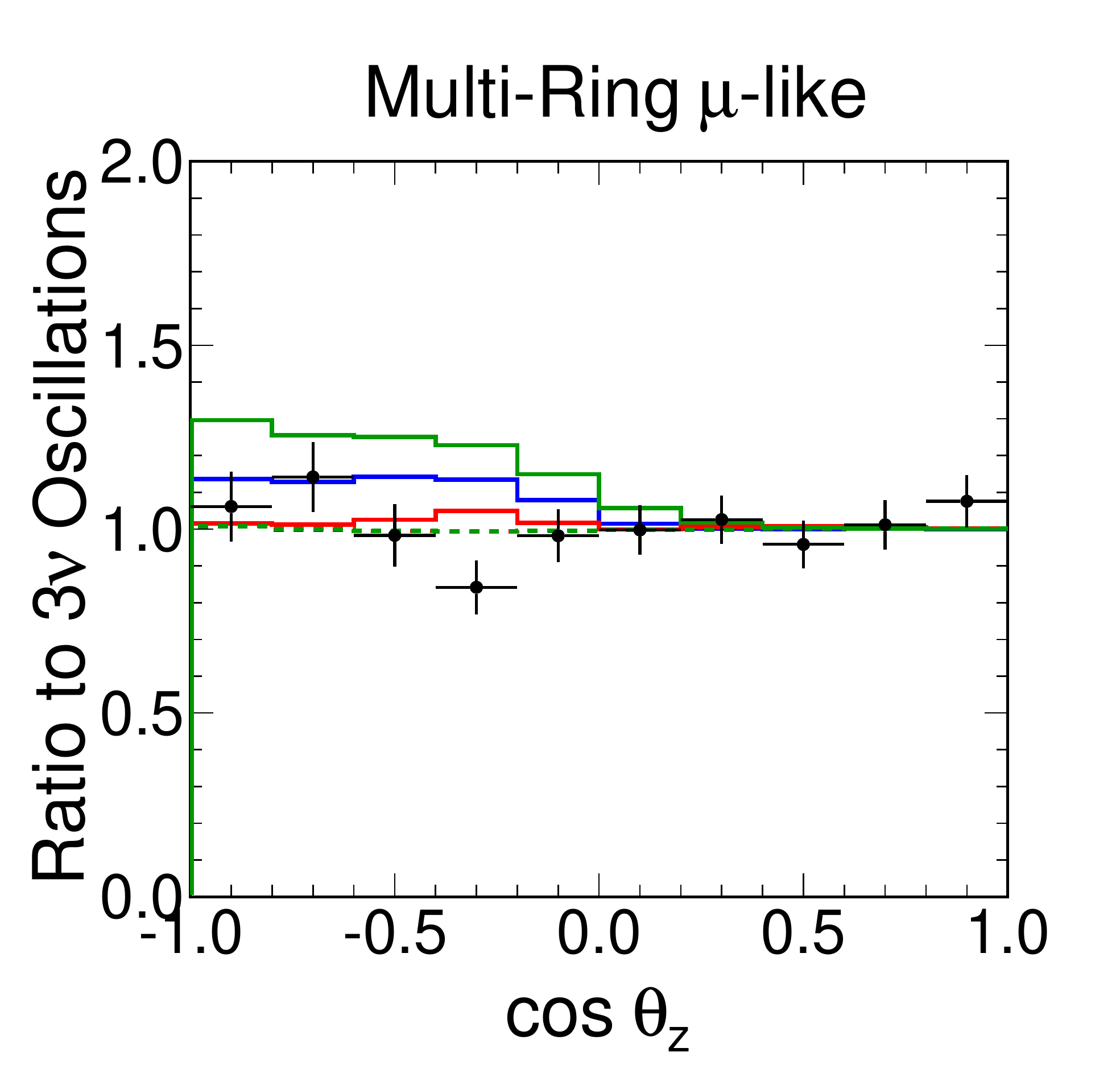}
 \includegraphics[width=\zwid,clip]{lv_ratio_c_legend.pdf} \\
 \includegraphics[width=\zwid,clip]{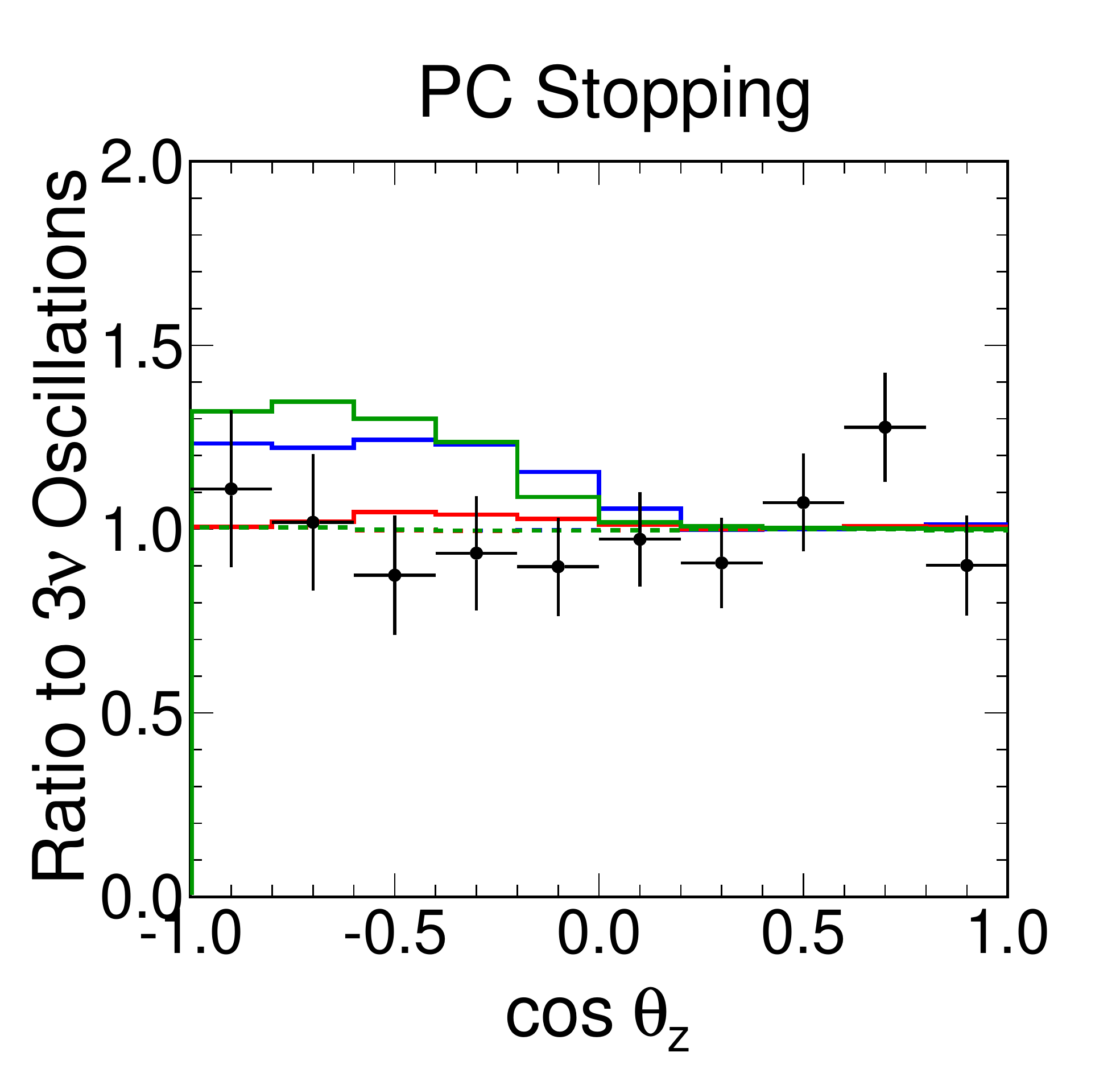}
 \includegraphics[width=\zwid,clip]{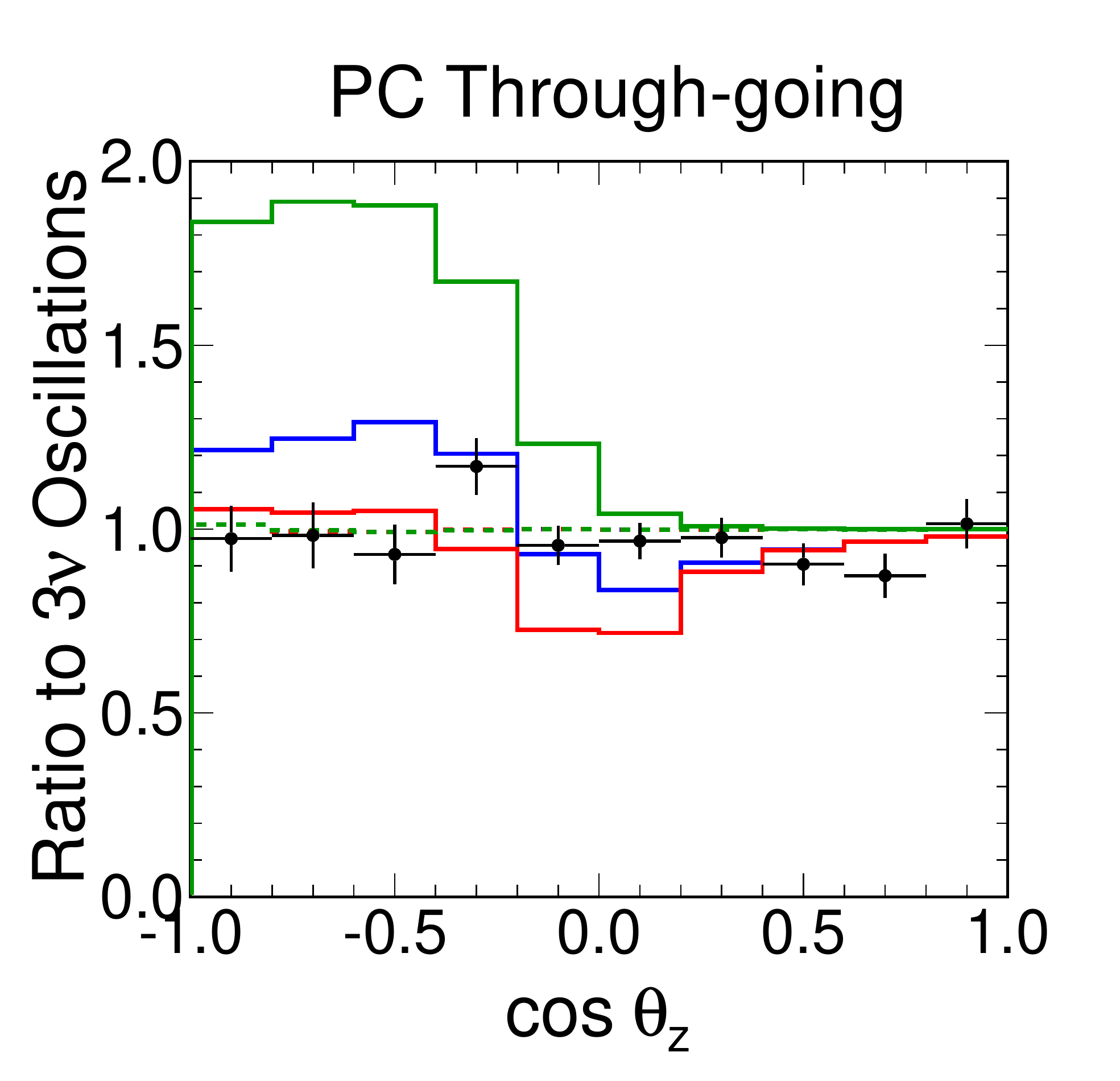}
 \includegraphics[width=\zwid,clip]{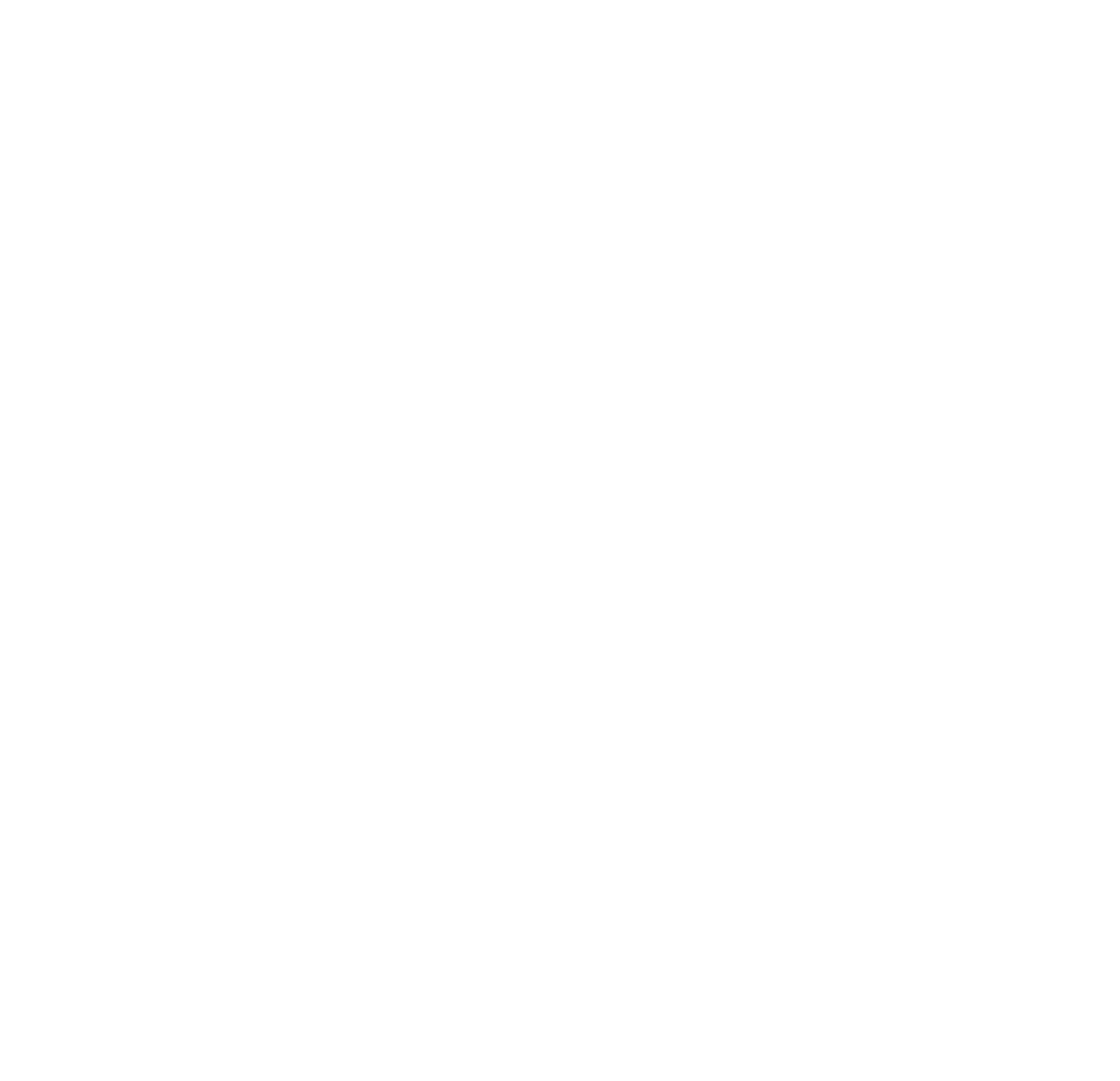} \\
 \includegraphics[width=\zwid,clip]{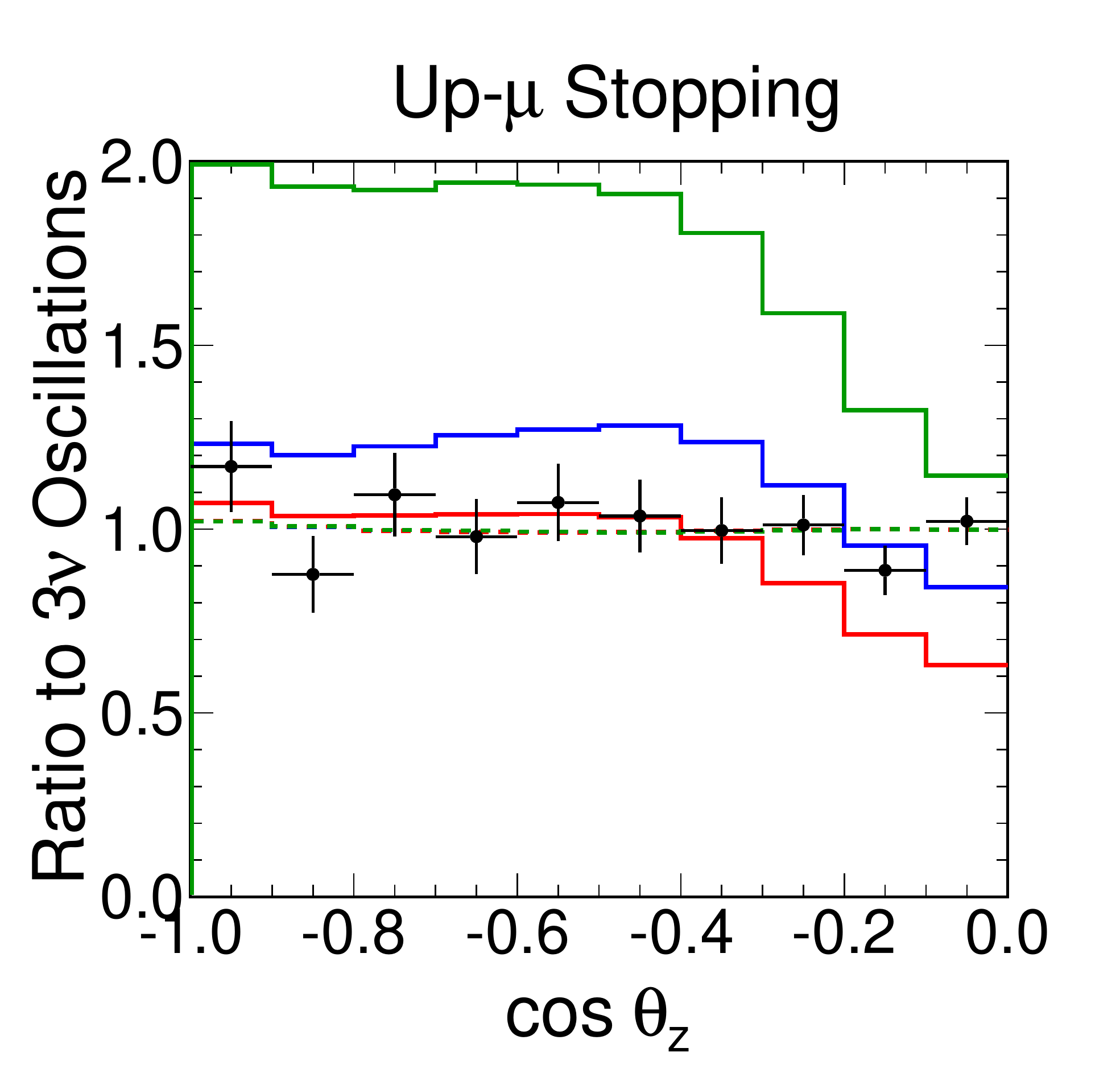}
 \includegraphics[width=\zwid,clip]{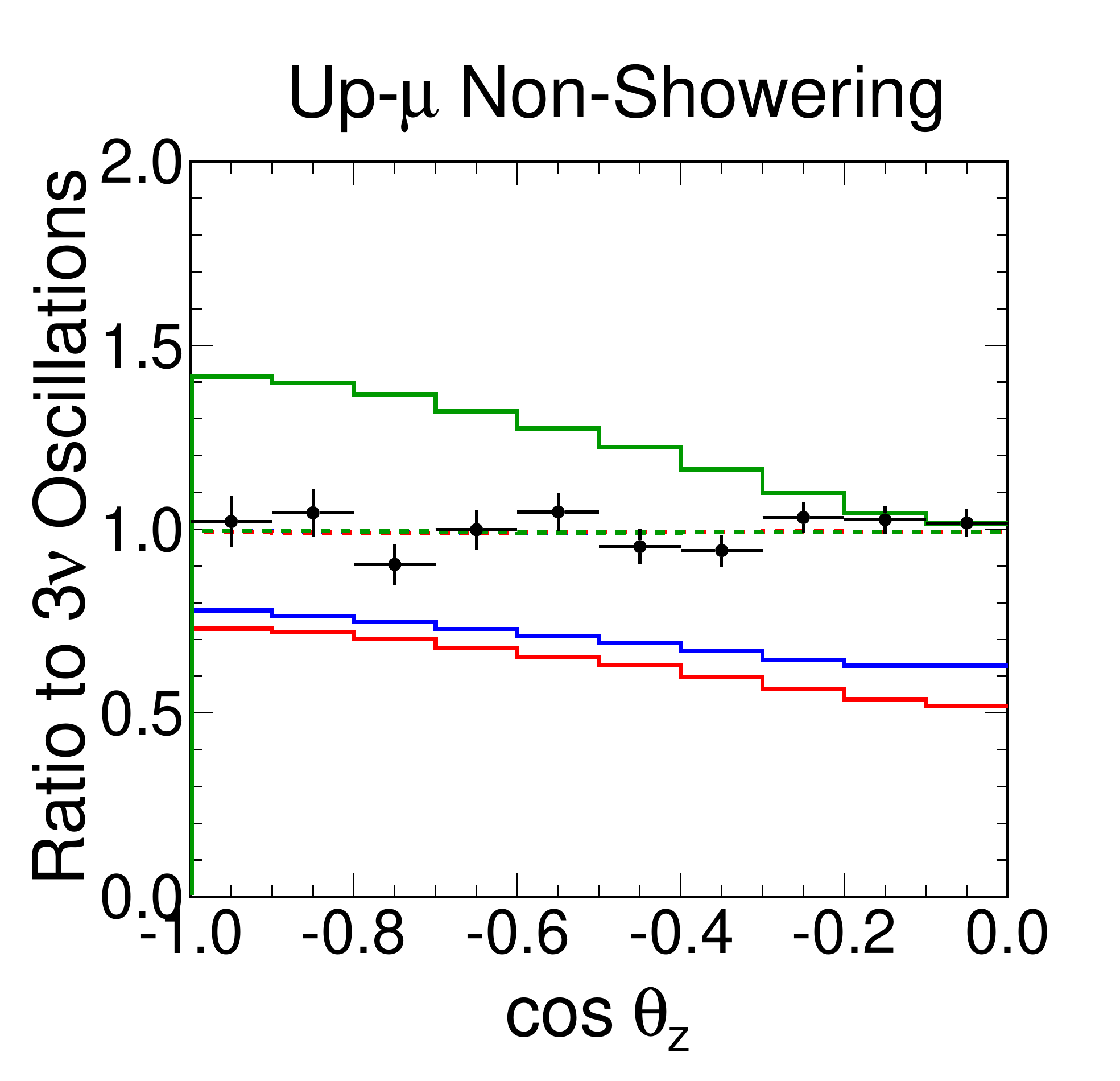}
 \includegraphics[width=\zwid,clip]{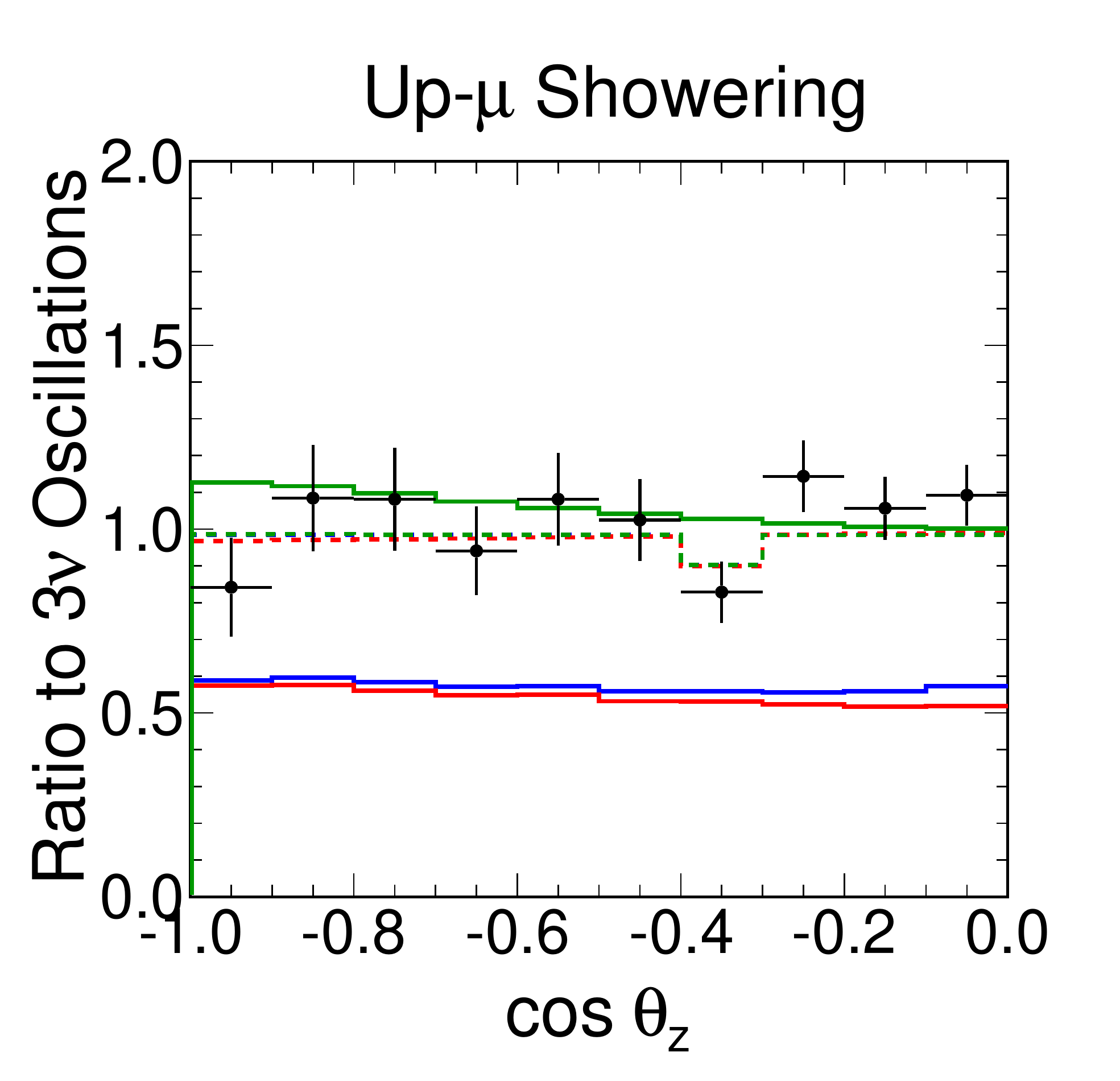}
 \caption{(color online) Ratios of the summed SK-I through SK-IV \cz distributions relative to standard three-flavor oscillations for the $\mu$-like FC, PC, and \UP sub-samples. They are projected into \cz when binned in both and the Sub-GeV 2 decay-e sample is binned only in momentum. The black points represent the data with statistical errors. The dashed lines represent the best fits from the three sectors for the \ctt parameters and the solid lines represent examples of large Lorentz violation ($\ctt = \sci{7.5}{-23}$, equivalent to \fig{oscillogramc}). Significant deviations from unity would indicate Lorentz violation.
 }
 \label{fig:zenithc_mu}
 \end{center}
\end{figure*}

\end{widetext}

\clearpage

\bibliography{lorentz}

\begin{thebibliography}{67}%
\makeatletter
\providecommand \@ifxundefined [1]{%
 \@ifx{#1\undefined}
}%
\providecommand \@ifnum [1]{%
 \ifnum #1\expandafter \@firstoftwo
 \else \expandafter \@secondoftwo
 \fi
}%
\providecommand \@ifx [1]{%
 \ifx #1\expandafter \@firstoftwo
 \else \expandafter \@secondoftwo
 \fi
}%
\providecommand \natexlab [1]{#1}%
\providecommand \enquote  [1]{``#1''}%
\providecommand \bibnamefont  [1]{#1}%
\providecommand \bibfnamefont [1]{#1}%
\providecommand \citenamefont [1]{#1}%
\providecommand \href@noop [0]{\@secondoftwo}%
\providecommand \href [0]{\begingroup \@sanitize@url \@href}%
\providecommand \@href[1]{\@@startlink{#1}\@@href}%
\providecommand \@@href[1]{\endgroup#1\@@endlink}%
\providecommand \@sanitize@url [0]{\catcode `\\12\catcode `\$12\catcode
  `\&12\catcode `\#12\catcode `\^12\catcode `\_12\catcode `\%12\relax}%
\providecommand \@@startlink[1]{}%
\providecommand \@@endlink[0]{}%
\providecommand \url  [0]{\begingroup\@sanitize@url \@url }%
\providecommand \@url [1]{\endgroup\@href {#1}{\urlprefix }}%
\providecommand \urlprefix  [0]{URL }%
\providecommand \Eprint [0]{\href }%
\providecommand \doibase [0]{http://dx.doi.org/}%
\providecommand \selectlanguage [0]{\@gobble}%
\providecommand \bibinfo  [0]{\@secondoftwo}%
\providecommand \bibfield  [0]{\@secondoftwo}%
\providecommand \translation [1]{[#1]}%
\providecommand \BibitemOpen [0]{}%
\providecommand \bibitemStop [0]{}%
\providecommand \bibitemNoStop [0]{.\EOS\space}%
\providecommand \EOS [0]{\spacefactor3000\relax}%
\providecommand \BibitemShut  [1]{\csname bibitem#1\endcsname}%
\let\auto@bib@innerbib\@empty
\bibitem [{\citenamefont {Kostelecky}\ and\ \citenamefont
  {Samuel}(1989)}]{Kostelecky:1988zi}%
  \BibitemOpen
  \bibfield  {author} {\bibinfo {author} {\bibfnamefont {V.~Alan}\ \bibnamefont
  {Kostelecky}}\ and\ \bibinfo {author} {\bibfnamefont {Stuart}\ \bibnamefont
  {Samuel}},\ }\bibfield  {title} {\enquote {\bibinfo {title} {{Spontaneous
  Breaking of Lorentz Symmetry in String Theory}},}\ }\href {\doibase
  10.1103/PhysRevD.39.683} {\bibfield  {journal} {\bibinfo  {journal}
  {Phys.Rev.}\ }\textbf {\bibinfo {volume} {D39}},\ \bibinfo {pages} {683}
  (\bibinfo {year} {1989})}\BibitemShut {NoStop}%
\bibitem [{\citenamefont {Hawking}(1976)}]{Hawking:1976ra}%
  \BibitemOpen
  \bibfield  {author} {\bibinfo {author} {\bibfnamefont {S.~W.}\ \bibnamefont
  {Hawking}},\ }\bibfield  {title} {\enquote {\bibinfo {title} {{Breakdown of
  Predictability in Gravitational Collapse}},}\ }\href {\doibase
  10.1103/PhysRevD.14.2460} {\bibfield  {journal} {\bibinfo  {journal}
  {Phys.Rev.}\ }\textbf {\bibinfo {volume} {D14}},\ \bibinfo {pages}
  {2460--2473} (\bibinfo {year} {1976})}\BibitemShut {NoStop}%
\bibitem [{\citenamefont {Brustein}\ \emph {et~al.}(2002)\citenamefont
  {Brustein}, \citenamefont {Eichler},\ and\ \citenamefont
  {Foffa}}]{Brustein:2001ik}%
  \BibitemOpen
  \bibfield  {author} {\bibinfo {author} {\bibfnamefont {Ram}\ \bibnamefont
  {Brustein}}, \bibinfo {author} {\bibfnamefont {David}\ \bibnamefont
  {Eichler}}, \ and\ \bibinfo {author} {\bibfnamefont {Stefano}\ \bibnamefont
  {Foffa}},\ }\bibfield  {title} {\enquote {\bibinfo {title} {{Probing the
  Planck scale with neutrino oscillations}},}\ }\href {\doibase
  10.1103/PhysRevD.65.105006} {\bibfield  {journal} {\bibinfo  {journal}
  {Phys.Rev.}\ }\textbf {\bibinfo {volume} {D65}},\ \bibinfo {pages} {105006}
  (\bibinfo {year} {2002})},\ \Eprint {http://arxiv.org/abs/hep-ph/0106309}
  {arXiv:hep-ph/0106309 [hep-ph]} \BibitemShut {NoStop}%
\bibitem [{\citenamefont {Colladay}\ and\ \citenamefont
  {Kostelecky}(1997)}]{Colladay:1996iz}%
  \BibitemOpen
  \bibfield  {author} {\bibinfo {author} {\bibfnamefont {Don}\ \bibnamefont
  {Colladay}}\ and\ \bibinfo {author} {\bibfnamefont {V.~Alan}\ \bibnamefont
  {Kostelecky}},\ }\bibfield  {title} {\enquote {\bibinfo {title} {{CPT
  violation and the standard model}},}\ }\href {\doibase
  10.1103/PhysRevD.55.6760} {\bibfield  {journal} {\bibinfo  {journal}
  {Phys.Rev.}\ }\textbf {\bibinfo {volume} {D55}},\ \bibinfo {pages}
  {6760--6774} (\bibinfo {year} {1997})},\ \Eprint
  {http://arxiv.org/abs/hep-ph/9703464} {arXiv:hep-ph/9703464 [hep-ph]}
  \BibitemShut {NoStop}%
\bibitem [{\citenamefont {Colladay}\ and\ \citenamefont
  {Kostelecky}(1998)}]{Colladay:1998fq}%
  \BibitemOpen
  \bibfield  {author} {\bibinfo {author} {\bibfnamefont {Don}\ \bibnamefont
  {Colladay}}\ and\ \bibinfo {author} {\bibfnamefont {V.~Alan}\ \bibnamefont
  {Kostelecky}},\ }\bibfield  {title} {\enquote {\bibinfo {title} {{Lorentz
  violating extension of the standard model}},}\ }\href {\doibase
  10.1103/PhysRevD.58.116002} {\bibfield  {journal} {\bibinfo  {journal}
  {Phys.Rev.}\ }\textbf {\bibinfo {volume} {D58}},\ \bibinfo {pages} {116002}
  (\bibinfo {year} {1998})},\ \Eprint {http://arxiv.org/abs/hep-ph/9809521}
  {arXiv:hep-ph/9809521 [hep-ph]} \BibitemShut {NoStop}%
\bibitem [{\citenamefont {Kostelecky}(2004)}]{Kostelecky:2003fs}%
  \BibitemOpen
  \bibfield  {author} {\bibinfo {author} {\bibfnamefont {V.~Alan}\ \bibnamefont
  {Kostelecky}},\ }\bibfield  {title} {\enquote {\bibinfo {title} {{Gravity,
  Lorentz violation, and the standard model}},}\ }\href {\doibase
  10.1103/PhysRevD.69.105009} {\bibfield  {journal} {\bibinfo  {journal}
  {Phys.Rev.}\ }\textbf {\bibinfo {volume} {D69}},\ \bibinfo {pages} {105009}
  (\bibinfo {year} {2004})},\ \Eprint {http://arxiv.org/abs/hep-th/0312310}
  {arXiv:hep-th/0312310 [hep-th]} \BibitemShut {NoStop}%
\bibitem [{\citenamefont {Amelino-Camelia}\ \emph {et~al.}(2005)\citenamefont
  {Amelino-Camelia}, \citenamefont {Lammerzahl}, \citenamefont {Macias},\ and\
  \citenamefont {Muller}}]{AmelinoCamelia:2005qa}%
  \BibitemOpen
  \bibfield  {author} {\bibinfo {author} {\bibfnamefont {G.}~\bibnamefont
  {Amelino-Camelia}}, \bibinfo {author} {\bibfnamefont {C.}~\bibnamefont
  {Lammerzahl}}, \bibinfo {author} {\bibfnamefont {A.}~\bibnamefont {Macias}},
  \ and\ \bibinfo {author} {\bibfnamefont {H.}~\bibnamefont {Muller}},\
  }\bibfield  {title} {\enquote {\bibinfo {title} {{The Search for quantum
  gravity signals}},}\ }\href {\doibase 10.1063/1.1900507} {\bibfield
  {journal} {\bibinfo  {journal} {AIP Conf.Proc.}\ }\textbf {\bibinfo {volume}
  {758}},\ \bibinfo {pages} {30--80} (\bibinfo {year} {2005})},\ \Eprint
  {http://arxiv.org/abs/gr-qc/0501053} {arXiv:gr-qc/0501053 [gr-qc]}
  \BibitemShut {NoStop}%
\bibitem [{\citenamefont {Bluhm}(2006)}]{Bluhm:2005uj}%
  \BibitemOpen
  \bibfield  {author} {\bibinfo {author} {\bibfnamefont {Robert}\ \bibnamefont
  {Bluhm}},\ }\bibfield  {title} {\enquote {\bibinfo {title} {{Overview of the
  SME: Implications and phenomenology of Lorentz violation}},}\ }\href
  {\doibase 10.1007/3-540-34523-X_8} {\bibfield  {journal} {\bibinfo  {journal}
  {Lect.Notes Phys.}\ }\textbf {\bibinfo {volume} {702}},\ \bibinfo {pages}
  {191--226} (\bibinfo {year} {2006})},\ \Eprint
  {http://arxiv.org/abs/hep-ph/0506054} {arXiv:hep-ph/0506054 [hep-ph]}
  \BibitemShut {NoStop}%
\bibitem [{\citenamefont {Greenberg}(2002)}]{Greenberg:2002uu}%
  \BibitemOpen
  \bibfield  {author} {\bibinfo {author} {\bibfnamefont {O.~W.}\ \bibnamefont
  {Greenberg}},\ }\bibfield  {title} {\enquote {\bibinfo {title} {{CPT
  violation implies violation of Lorentz invariance}},}\ }\href {\doibase
  10.1103/PhysRevLett.89.231602} {\bibfield  {journal} {\bibinfo  {journal}
  {Phys.Rev.Lett.}\ }\textbf {\bibinfo {volume} {89}},\ \bibinfo {pages}
  {231602} (\bibinfo {year} {2002})},\ \Eprint
  {http://arxiv.org/abs/hep-ph/0201258} {arXiv:hep-ph/0201258 [hep-ph]}
  \BibitemShut {NoStop}%
\bibitem [{Note1()}]{Note1}%
  \BibitemOpen
  \bibinfo {note} {While LV can exist without $CPT$ violation, $CPT$ violation
  requires LV.}\BibitemShut {Stop}%
\bibitem [{\citenamefont {Kostelecky}\ and\ \citenamefont
  {Mewes}(2004{\natexlab{a}})}]{Kostelecky:2004hg}%
  \BibitemOpen
  \bibfield  {author} {\bibinfo {author} {\bibfnamefont {V.~Alan}\ \bibnamefont
  {Kostelecky}}\ and\ \bibinfo {author} {\bibfnamefont {Matthew}\ \bibnamefont
  {Mewes}},\ }\bibfield  {title} {\enquote {\bibinfo {title} {{Lorentz
  violation and short-baseline neutrino experiments}},}\ }\href {\doibase
  10.1103/PhysRevD.70.076002} {\bibfield  {journal} {\bibinfo  {journal}
  {Phys.Rev.}\ }\textbf {\bibinfo {volume} {D70}},\ \bibinfo {pages} {076002}
  (\bibinfo {year} {2004}{\natexlab{a}})},\ \Eprint
  {http://arxiv.org/abs/hep-ph/0406255} {arXiv:hep-ph/0406255 [hep-ph]}
  \BibitemShut {NoStop}%
\bibitem [{\citenamefont {Kostelecky}\ and\ \citenamefont
  {Russell}(2011)}]{Kostelecky:2008ts}%
  \BibitemOpen
  \bibfield  {author} {\bibinfo {author} {\bibfnamefont {V.~Alan}\ \bibnamefont
  {Kostelecky}}\ and\ \bibinfo {author} {\bibfnamefont {Neil}\ \bibnamefont
  {Russell}},\ }\bibfield  {title} {\enquote {\bibinfo {title} {{Data Tables
  for Lorentz and CPT Violation}},}\ }\href {\doibase 10.1103/RevModPhys.83.11}
  {\bibfield  {journal} {\bibinfo  {journal} {Rev.Mod.Phys.}\ }\textbf
  {\bibinfo {volume} {83}},\ \bibinfo {pages} {11--31} (\bibinfo {year}
  {2011})},\ \bibinfo {note} {{The numbers compared here come from a newer
  version than the published version}},\ \Eprint
  {http://arxiv.org/abs/0801.0287v8} {arXiv:0801.0287v8 [hep-ph]} \BibitemShut
  {NoStop}%
\bibitem [{\citenamefont {Mattingly}(2005)}]{Mattingly:2005re}%
  \BibitemOpen
  \bibfield  {author} {\bibinfo {author} {\bibfnamefont {David}\ \bibnamefont
  {Mattingly}},\ }\bibfield  {title} {\enquote {\bibinfo {title} {{Modern tests
  of Lorentz invariance}},}\ }\href@noop {} {\bibfield  {journal} {\bibinfo
  {journal} {Living Rev.Rel.}\ }\textbf {\bibinfo {volume} {8}},\ \bibinfo
  {pages} {5} (\bibinfo {year} {2005})},\ \Eprint
  {http://arxiv.org/abs/gr-qc/0502097} {arXiv:gr-qc/0502097 [gr-qc]}
  \BibitemShut {NoStop}%
\bibitem [{\citenamefont {Kostelecky}(1999)}]{Kostelecky:1999dx}%
  \BibitemOpen
  \bibinfo {editor} {\bibfnamefont {V.Alan}\ \bibnamefont {Kostelecky}},\ ed.,\
  \href@noop {} {\emph {\bibinfo {title} {{CPT and Lorentz symmetry.
  Proceedings: CPT'98, Bloomington, USA, Nov 6-8, 1998}}}}\ (\bibinfo {year}
  {1999})\BibitemShut {NoStop}%
\bibitem [{\citenamefont {Kostelecky}(2002)}]{Kostelecky:2002zz}%
  \BibitemOpen
  \bibinfo {editor} {\bibfnamefont {V.Alan}\ \bibnamefont {Kostelecky}},\ ed.,\
  \href@noop {} {\emph {\bibinfo {title} {{CPT and Lorentz symmetry.
  Proceedings: 2nd Meeting, Bloomington, USA, Aug 15-18, 2001}}}}\ (\bibinfo
  {year} {2002})\BibitemShut {NoStop}%
\bibitem [{\citenamefont {Kostelecky}(2005)}]{Kostelecky:2005mj}%
  \BibitemOpen
  \bibinfo {editor} {\bibfnamefont {V.Alan}\ \bibnamefont {Kostelecky}},\ ed.,\
  \href@noop {} {\emph {\bibinfo {title} {{CPT and Lorentz symmetry.
  Proceedings: 3rd Meeting, Bloomington, USA, Aug 4-7, 2004}}}}\ (\bibinfo
  {year} {2005})\BibitemShut {NoStop}%
\bibitem [{\citenamefont {Kostelecky}(2008)}]{Kostelecky:2008zz}%
  \BibitemOpen
  \bibinfo {editor} {\bibfnamefont {V.Alan}\ \bibnamefont {Kostelecky}},\ ed.,\
  \href@noop {} {\emph {\bibinfo {title} {{CPT and Lorentz symmetry.
  Proceedings: 4th Meeting, Bloomington, USA, Aug 8-11, 2007}}}}\ (\bibinfo
  {year} {2008})\BibitemShut {NoStop}%
\bibitem [{\citenamefont {Kostelecky}\ and\ \citenamefont
  {Mewes}(2012)}]{Kostelecky:2011gq}%
  \BibitemOpen
  \bibfield  {author} {\bibinfo {author} {\bibfnamefont {V.~A.}\ \bibnamefont
  {Kostelecky}}\ and\ \bibinfo {author} {\bibfnamefont {M.}~\bibnamefont
  {Mewes}},\ }\bibfield  {title} {\enquote {\bibinfo {title} {{Neutrinos with
  Lorentz-violating operators of arbitrary dimension}},}\ }\href {\doibase
  10.1103/PhysRevD.85.096005} {\bibfield  {journal} {\bibinfo  {journal}
  {Phys.Rev.}\ }\textbf {\bibinfo {volume} {D85}},\ \bibinfo {pages} {096005}
  (\bibinfo {year} {2012})},\ \Eprint {http://arxiv.org/abs/1112.6395}
  {arXiv:1112.6395 [hep-ph]} \BibitemShut {NoStop}%
\bibitem [{\citenamefont {Kostelecky}\ and\ \citenamefont
  {Mewes}(2004{\natexlab{b}})}]{Kostelecky:2003xn}%
  \BibitemOpen
  \bibfield  {author} {\bibinfo {author} {\bibfnamefont {V.~Alan}\ \bibnamefont
  {Kostelecky}}\ and\ \bibinfo {author} {\bibfnamefont {Matthew}\ \bibnamefont
  {Mewes}},\ }\bibfield  {title} {\enquote {\bibinfo {title} {{Lorentz and CPT
  violation in the neutrino sector}},}\ }\href {\doibase
  10.1103/PhysRevD.70.031902} {\bibfield  {journal} {\bibinfo  {journal}
  {Phys.Rev.}\ }\textbf {\bibinfo {volume} {D70}},\ \bibinfo {pages} {031902}
  (\bibinfo {year} {2004}{\natexlab{b}})},\ \Eprint
  {http://arxiv.org/abs/hep-ph/0308300} {arXiv:hep-ph/0308300 [hep-ph]}
  \BibitemShut {NoStop}%
\bibitem [{\citenamefont {Kostelecky}\ and\ \citenamefont
  {Mewes}(2004{\natexlab{c}})}]{Kostelecky:2003cr}%
  \BibitemOpen
  \bibfield  {author} {\bibinfo {author} {\bibfnamefont {V.~Alan}\ \bibnamefont
  {Kostelecky}}\ and\ \bibinfo {author} {\bibfnamefont {Matthew}\ \bibnamefont
  {Mewes}},\ }\bibfield  {title} {\enquote {\bibinfo {title} {{Lorentz and CPT
  violation in neutrinos}},}\ }\href {\doibase 10.1103/PhysRevD.69.016005}
  {\bibfield  {journal} {\bibinfo  {journal} {Phys.Rev.}\ }\textbf {\bibinfo
  {volume} {D69}},\ \bibinfo {pages} {016005} (\bibinfo {year}
  {2004}{\natexlab{c}})},\ \Eprint {http://arxiv.org/abs/hep-ph/0309025}
  {arXiv:hep-ph/0309025 [hep-ph]} \BibitemShut {NoStop}%
\bibitem [{\citenamefont {Auerbach}\ \emph {et~al.}(2005)\citenamefont
  {Auerbach} \emph {et~al.}}]{Auerbach:2005tq}%
  \BibitemOpen
  \bibfield  {author} {\bibinfo {author} {\bibfnamefont {L.~B.}\ \bibnamefont
  {Auerbach}} \emph {et~al.} (\bibinfo {collaboration} {LSND Collaboration}),\
  }\bibfield  {title} {\enquote {\bibinfo {title} {{Tests of Lorentz violation
  in anti-nu(mu) \& anti-nu(e) oscillations}},}\ }\href {\doibase
  10.1103/PhysRevD.72.076004} {\bibfield  {journal} {\bibinfo  {journal}
  {Phys.Rev.}\ }\textbf {\bibinfo {volume} {D72}},\ \bibinfo {pages} {076004}
  (\bibinfo {year} {2005})},\ \Eprint {http://arxiv.org/abs/hep-ex/0506067}
  {arXiv:hep-ex/0506067 [hep-ex]} \BibitemShut {NoStop}%
\bibitem [{\citenamefont {Aguilar-Arevalo}\ \emph {et~al.}(2013)\citenamefont
  {Aguilar-Arevalo} \emph {et~al.}}]{AguilarArevalo:2011yi}%
  \BibitemOpen
  \bibfield  {author} {\bibinfo {author} {\bibfnamefont {A.~A.}\ \bibnamefont
  {Aguilar-Arevalo}} \emph {et~al.} (\bibinfo {collaboration} {MiniBooNE
  Collaboration}),\ }\bibfield  {title} {\enquote {\bibinfo {title} {{Test of
  Lorentz and CPT violation with Short Baseline Neutrino Oscillation
  Excesses}},}\ }\href {\doibase 10.1016/j.physletb.2012.12.020} {\bibfield
  {journal} {\bibinfo  {journal} {Phys.Lett.}\ }\textbf {\bibinfo {volume}
  {B718}},\ \bibinfo {pages} {1303--1308} (\bibinfo {year} {2013})},\ \Eprint
  {http://arxiv.org/abs/1109.3480} {arXiv:1109.3480 [hep-ex]} \BibitemShut
  {NoStop}%
\bibitem [{\citenamefont {Adamson}\ \emph {et~al.}(2008)\citenamefont {Adamson}
  \emph {et~al.}}]{Adamson:2008aa}%
  \BibitemOpen
  \bibfield  {author} {\bibinfo {author} {\bibfnamefont {P.}~\bibnamefont
  {Adamson}} \emph {et~al.} (\bibinfo {collaboration} {MINOS Collaboration}),\
  }\bibfield  {title} {\enquote {\bibinfo {title} {{Testing Lorentz Invariance
  and CPT Conservation with NuMI Neutrinos in the MINOS Near Detector}},}\
  }\href {\doibase 10.1103/PhysRevLett.101.151601} {\bibfield  {journal}
  {\bibinfo  {journal} {Phys.Rev.Lett.}\ }\textbf {\bibinfo {volume} {101}},\
  \bibinfo {pages} {151601} (\bibinfo {year} {2008})},\ \Eprint
  {http://arxiv.org/abs/0806.4945} {arXiv:0806.4945 [hep-ex]} \BibitemShut
  {NoStop}%
\bibitem [{\citenamefont {Adamson}\ \emph {et~al.}(2012)\citenamefont {Adamson}
  \emph {et~al.}}]{Adamson:2012hp}%
  \BibitemOpen
  \bibfield  {author} {\bibinfo {author} {\bibfnamefont {P.}~\bibnamefont
  {Adamson}} \emph {et~al.} (\bibinfo {collaboration} {MINOS Collaboration}),\
  }\bibfield  {title} {\enquote {\bibinfo {title} {{Search for Lorentz
  invariance and CPT violation with muon antineutrinos in the MINOS Near
  Detector}},}\ }\href {\doibase 10.1103/PhysRevD.85.031101} {\bibfield
  {journal} {\bibinfo  {journal} {Phys.Rev.}\ }\textbf {\bibinfo {volume}
  {D85}},\ \bibinfo {pages} {031101} (\bibinfo {year} {2012})},\ \Eprint
  {http://arxiv.org/abs/1201.2631} {arXiv:1201.2631 [hep-ex]} \BibitemShut
  {NoStop}%
\bibitem [{\citenamefont {Adamson}\ \emph {et~al.}(2010)\citenamefont {Adamson}
  \emph {et~al.}}]{Adamson:2010rn}%
  \BibitemOpen
  \bibfield  {author} {\bibinfo {author} {\bibfnamefont {P.}~\bibnamefont
  {Adamson}} \emph {et~al.} (\bibinfo {collaboration} {MINOS Collaboration}),\
  }\bibfield  {title} {\enquote {\bibinfo {title} {{A Search for Lorentz
  Invariance and CPT Violation with the MINOS Far Detector}},}\ }\href
  {\doibase 10.1103/PhysRevLett.105.151601} {\bibfield  {journal} {\bibinfo
  {journal} {Phys.Rev.Lett.}\ }\textbf {\bibinfo {volume} {105}},\ \bibinfo
  {pages} {151601} (\bibinfo {year} {2010})},\ \Eprint
  {http://arxiv.org/abs/1007.2791} {arXiv:1007.2791 [hep-ex]} \BibitemShut
  {NoStop}%
\bibitem [{\citenamefont {Rebel}\ and\ \citenamefont
  {Mufson}(2013)}]{Rebel:2013vc}%
  \BibitemOpen
  \bibfield  {author} {\bibinfo {author} {\bibfnamefont {B.}~\bibnamefont
  {Rebel}}\ and\ \bibinfo {author} {\bibfnamefont {S.}~\bibnamefont {Mufson}},\
  }\bibfield  {title} {\enquote {\bibinfo {title} {{The Search for
  Neutrino-Antineutrino Mixing Resulting from Lorentz Invariance Violation
  using neutrino interactions in MINOS}},}\ }\href {\doibase
  10.1016/j.astropartphys.2013.07.006} {\bibfield  {journal} {\bibinfo
  {journal} {Astropart.Phys.}\ }\textbf {\bibinfo {volume} {48}},\ \bibinfo
  {pages} {78--81} (\bibinfo {year} {2013})},\ \Eprint
  {http://arxiv.org/abs/1301.4684} {arXiv:1301.4684 [hep-ex]} \BibitemShut
  {NoStop}%
\bibitem [{\citenamefont {Abe}\ \emph {et~al.}(2012{\natexlab{a}})\citenamefont
  {Abe} \emph {et~al.}}]{Abe:2012gw}%
  \BibitemOpen
  \bibfield  {author} {\bibinfo {author} {\bibfnamefont {Y.}~\bibnamefont
  {Abe}} \emph {et~al.} (\bibinfo {collaboration} {Double Chooz
  Collaboration}),\ }\bibfield  {title} {\enquote {\bibinfo {title} {{First
  Test of Lorentz Violation with a Reactor-based Antineutrino Experiment}},}\
  }\href {\doibase 10.1103/PhysRevD.86.112009} {\bibfield  {journal} {\bibinfo
  {journal} {Phys.Rev.}\ }\textbf {\bibinfo {volume} {D86}},\ \bibinfo {pages}
  {112009} (\bibinfo {year} {2012}{\natexlab{a}})},\ \Eprint
  {http://arxiv.org/abs/1209.5810} {arXiv:1209.5810 [hep-ex]} \BibitemShut
  {NoStop}%
\bibitem [{\citenamefont {Díaz}\ \emph {et~al.}(2013)\citenamefont {Díaz},
  \citenamefont {Katori}, \citenamefont {Spitz},\ and\ \citenamefont
  {Conrad}}]{Diaz:2013iba}%
  \BibitemOpen
  \bibfield  {author} {\bibinfo {author} {\bibfnamefont {J.S.}\ \bibnamefont
  {Díaz}}, \bibinfo {author} {\bibfnamefont {T.}~\bibnamefont {Katori}},
  \bibinfo {author} {\bibfnamefont {J.}~\bibnamefont {Spitz}}, \ and\ \bibinfo
  {author} {\bibfnamefont {J.M.}\ \bibnamefont {Conrad}},\ }\bibfield  {title}
  {\enquote {\bibinfo {title} {{Search for neutrino-antineutrino oscillations
  with a reactor experiment}},}\ }\href {\doibase
  10.1016/j.physletb.2013.10.058} {\bibfield  {journal} {\bibinfo  {journal}
  {Phys.Lett.}\ }\textbf {\bibinfo {volume} {B727}},\ \bibinfo {pages}
  {412--416} (\bibinfo {year} {2013})},\ \Eprint
  {http://arxiv.org/abs/1307.5789} {arXiv:1307.5789 [hep-ex]} \BibitemShut
  {NoStop}%
\bibitem [{\citenamefont {Abbasi}\ \emph {et~al.}(2010)\citenamefont {Abbasi}
  \emph {et~al.}}]{Abbasi:2010kx}%
  \BibitemOpen
  \bibfield  {author} {\bibinfo {author} {\bibfnamefont {R.}~\bibnamefont
  {Abbasi}} \emph {et~al.} (\bibinfo {collaboration} {IceCube Collaboration}),\
  }\bibfield  {title} {\enquote {\bibinfo {title} {{Search for a
  Lorentz-violating sidereal signal with atmospheric neutrinos in IceCube}},}\
  }\href {\doibase 10.1103/PhysRevD.82.112003} {\bibfield  {journal} {\bibinfo
  {journal} {Phys.Rev.}\ }\textbf {\bibinfo {volume} {D82}},\ \bibinfo {pages}
  {112003} (\bibinfo {year} {2010})},\ \Eprint {http://arxiv.org/abs/1010.4096}
  {arXiv:1010.4096 [astro-ph.HE]} \BibitemShut {NoStop}%
\bibitem [{\citenamefont {Diaz}\ \emph {et~al.}(2009)\citenamefont {Diaz},
  \citenamefont {Kostelecky},\ and\ \citenamefont {Mewes}}]{Diaz:2009qk}%
  \BibitemOpen
  \bibfield  {author} {\bibinfo {author} {\bibfnamefont {Jorge~S.}\
  \bibnamefont {Diaz}}, \bibinfo {author} {\bibfnamefont {V.~Alan}\
  \bibnamefont {Kostelecky}}, \ and\ \bibinfo {author} {\bibfnamefont
  {Matthew}\ \bibnamefont {Mewes}},\ }\bibfield  {title} {\enquote {\bibinfo
  {title} {{Perturbative Lorentz and CPT violation for neutrino and
  antineutrino oscillations}},}\ }\href {\doibase 10.1103/PhysRevD.80.076007}
  {\bibfield  {journal} {\bibinfo  {journal} {Phys.Rev.}\ }\textbf {\bibinfo
  {volume} {D80}},\ \bibinfo {pages} {076007} (\bibinfo {year} {2009})},\
  \Eprint {http://arxiv.org/abs/0908.1401} {arXiv:0908.1401 [hep-ph]}
  \BibitemShut {NoStop}%
\bibitem [{\citenamefont {Fukuda}\ \emph {et~al.}(2003)\citenamefont {Fukuda}
  \emph {et~al.}}]{Fukuda:2002uc}%
  \BibitemOpen
  \bibfield  {author} {\bibinfo {author} {\bibfnamefont {Y.}~\bibnamefont
  {Fukuda}} \emph {et~al.} (\bibinfo {collaboration} {Super-Kamiokande
  Collaboration}),\ }\bibfield  {title} {\enquote {\bibinfo {title} {{The
  Super-Kamiokande detector}},}\ }\href {\doibase
  10.1016/S0168-9002(03)00425-X} {\bibfield  {journal} {\bibinfo  {journal}
  {Nucl.Instrum.Meth.}\ }\textbf {\bibinfo {volume} {A501}},\ \bibinfo {pages}
  {418--462} (\bibinfo {year} {2003})}\BibitemShut {NoStop}%
\bibitem [{\citenamefont {Akiri}(2013)}]{Akiri:2013hca}%
  \BibitemOpen
  \bibfield  {author} {\bibinfo {author} {\bibfnamefont {Tarek}\ \bibnamefont
  {Akiri}},\ }\href@noop {} {\enquote {\bibinfo {title} {{Sensitivity of
  atmospheric neutrinos in Super-Kamiokande to Lorentz violation}},}\ }
  (\bibinfo {year} {2013}),\ \bibinfo {note} {{Presented at the Sixth Meeting
  on CPT and Lorentz Symmetry, Bloomington, Indiana, June 17-21, 2013}},\
  \Eprint {http://arxiv.org/abs/1308.2210} {arXiv:1308.2210 [hep-ph]}
  \BibitemShut {NoStop}%
\bibitem [{\citenamefont {Fukuda}\ \emph {et~al.}(1998)\citenamefont {Fukuda}
  \emph {et~al.}}]{Fukuda:1998mi}%
  \BibitemOpen
  \bibfield  {author} {\bibinfo {author} {\bibfnamefont {Y.}~\bibnamefont
  {Fukuda}} \emph {et~al.} (\bibinfo {collaboration} {Super-Kamiokande
  Collaboration}),\ }\bibfield  {title} {\enquote {\bibinfo {title} {{Evidence
  for oscillation of atmospheric neutrinos}},}\ }\href {\doibase
  10.1103/PhysRevLett.81.1562} {\bibfield  {journal} {\bibinfo  {journal}
  {Phys.Rev.Lett.}\ }\textbf {\bibinfo {volume} {81}},\ \bibinfo {pages}
  {1562--1567} (\bibinfo {year} {1998})}\BibitemShut {NoStop}%
\bibitem [{\citenamefont {Ashie}\ \emph {et~al.}(2004)\citenamefont {Ashie}
  \emph {et~al.}}]{Ashie:2004mr}%
  \BibitemOpen
  \bibfield  {author} {\bibinfo {author} {\bibfnamefont {Y.}~\bibnamefont
  {Ashie}} \emph {et~al.} (\bibinfo {collaboration} {Super-Kamiokande
  Collaboration}),\ }\bibfield  {title} {\enquote {\bibinfo {title} {{Evidence
  for an oscillatory signature in atmospheric neutrino oscillation}},}\ }\href
  {\doibase 10.1103/PhysRevLett.93.101801} {\bibfield  {journal} {\bibinfo
  {journal} {Phys.Rev.Lett.}\ }\textbf {\bibinfo {volume} {93}},\ \bibinfo
  {pages} {101801} (\bibinfo {year} {2004})},\ \Eprint
  {http://arxiv.org/abs/hep-ex/0404034} {arXiv:hep-ex/0404034 [hep-ex]}
  \BibitemShut {NoStop}%
\bibitem [{\citenamefont {Abe}\ \emph {et~al.}(2011)\citenamefont {Abe} \emph
  {et~al.}}]{Abe:2010hy}%
  \BibitemOpen
  \bibfield  {author} {\bibinfo {author} {\bibfnamefont {K.}~\bibnamefont
  {Abe}} \emph {et~al.} (\bibinfo {collaboration} {Super-Kamiokande
  Collaboration}),\ }\bibfield  {title} {\enquote {\bibinfo {title} {{Solar
  neutrino results in Super-Kamiokande-III}},}\ }\href {\doibase
  10.1103/PhysRevD.83.052010} {\bibfield  {journal} {\bibinfo  {journal}
  {Phys.Rev.}\ }\textbf {\bibinfo {volume} {D83}},\ \bibinfo {pages} {052010}
  (\bibinfo {year} {2011})}\BibitemShut {NoStop}%
\bibitem [{\citenamefont {Cleveland}\ \emph {et~al.}(1998)\citenamefont
  {Cleveland}, \citenamefont {Daily}, \citenamefont {Davis}, \citenamefont
  {Distel}, \citenamefont {Lande} \emph {et~al.}}]{Cleveland:1998nv}%
  \BibitemOpen
  \bibfield  {author} {\bibinfo {author} {\bibfnamefont {B.~T.}\ \bibnamefont
  {Cleveland}}, \bibinfo {author} {\bibfnamefont {Timothy}\ \bibnamefont
  {Daily}}, \bibinfo {author} {\bibfnamefont {Jr.}\ \bibnamefont {Davis},
  \bibfnamefont {Raymond}}, \bibinfo {author} {\bibfnamefont {James~R.}\
  \bibnamefont {Distel}}, \bibinfo {author} {\bibfnamefont {Kenneth}\
  \bibnamefont {Lande}},  \emph {et~al.},\ }\bibfield  {title} {\enquote
  {\bibinfo {title} {{Measurement of the solar electron neutrino flux with the
  Homestake chlorine detector}},}\ }\href {\doibase 10.1086/305343} {\bibfield
  {journal} {\bibinfo  {journal} {Astrophys.J.}\ }\textbf {\bibinfo {volume}
  {496}},\ \bibinfo {pages} {505--526} (\bibinfo {year} {1998})}\BibitemShut
  {NoStop}%
\bibitem [{\citenamefont {Abdurashitov}\ \emph {et~al.}(2009)\citenamefont
  {Abdurashitov} \emph {et~al.}}]{Abdurashitov:2009tn}%
  \BibitemOpen
  \bibfield  {author} {\bibinfo {author} {\bibfnamefont {J.~N.}\ \bibnamefont
  {Abdurashitov}} \emph {et~al.} (\bibinfo {collaboration} {SAGE
  Collaboration}),\ }\bibfield  {title} {\enquote {\bibinfo {title}
  {{Measurement of the solar neutrino capture rate with gallium metal. III:
  Results for the 2002--2007 data-taking period}},}\ }\href {\doibase
  10.1103/PhysRevC.80.015807} {\bibfield  {journal} {\bibinfo  {journal}
  {Phys.Rev.}\ }\textbf {\bibinfo {volume} {C80}},\ \bibinfo {pages} {015807}
  (\bibinfo {year} {2009})},\ \Eprint {http://arxiv.org/abs/0901.2200}
  {arXiv:0901.2200 [nucl-ex]} \BibitemShut {NoStop}%
\bibitem [{\citenamefont {Altmann}\ \emph {et~al.}(2005)\citenamefont {Altmann}
  \emph {et~al.}}]{Altmann:2005ix}%
  \BibitemOpen
  \bibfield  {author} {\bibinfo {author} {\bibfnamefont {M.}~\bibnamefont
  {Altmann}} \emph {et~al.} (\bibinfo {collaboration} {GNO COLLABORATION}),\
  }\bibfield  {title} {\enquote {\bibinfo {title} {{Complete results for five
  years of GNO solar neutrino observations}},}\ }\href {\doibase
  10.1016/j.physletb.2005.04.068} {\bibfield  {journal} {\bibinfo  {journal}
  {Phys.Lett.}\ }\textbf {\bibinfo {volume} {B616}},\ \bibinfo {pages}
  {174--190} (\bibinfo {year} {2005})},\ \Eprint
  {http://arxiv.org/abs/hep-ex/0504037} {arXiv:hep-ex/0504037 [hep-ex]}
  \BibitemShut {NoStop}%
\bibitem [{\citenamefont {Hampel}\ \emph {et~al.}(1999)\citenamefont {Hampel}
  \emph {et~al.}}]{Hampel:1998xg}%
  \BibitemOpen
  \bibfield  {author} {\bibinfo {author} {\bibfnamefont {W.}~\bibnamefont
  {Hampel}} \emph {et~al.} (\bibinfo {collaboration} {GALLEX Collaboration}),\
  }\bibfield  {title} {\enquote {\bibinfo {title} {{GALLEX solar neutrino
  observations: Results for GALLEX IV}},}\ }\href {\doibase
  10.1016/S0370-2693(98)01579-2} {\bibfield  {journal} {\bibinfo  {journal}
  {Phys.Lett.}\ }\textbf {\bibinfo {volume} {B447}},\ \bibinfo {pages}
  {127--133} (\bibinfo {year} {1999})}\BibitemShut {NoStop}%
\bibitem [{\citenamefont {Aharmim}\ \emph {et~al.}(2013)\citenamefont {Aharmim}
  \emph {et~al.}}]{Aharmim:2011vm}%
  \BibitemOpen
  \bibfield  {author} {\bibinfo {author} {\bibfnamefont {B.}~\bibnamefont
  {Aharmim}} \emph {et~al.} (\bibinfo {collaboration} {SNO Collaboration}),\
  }\bibfield  {title} {\enquote {\bibinfo {title} {{Combined Analysis of all
  Three Phases of Solar Neutrino Data from the Sudbury Neutrino
  Observatory}},}\ }\href {\doibase 10.1103/PhysRevC.88.025501} {\bibfield
  {journal} {\bibinfo  {journal} {Phys.Rev.}\ }\textbf {\bibinfo {volume}
  {C88}},\ \bibinfo {pages} {025501} (\bibinfo {year} {2013})},\ \Eprint
  {http://arxiv.org/abs/1109.0763} {arXiv:1109.0763 [nucl-ex]} \BibitemShut
  {NoStop}%
\bibitem [{\citenamefont {Abe}\ \emph {et~al.}(2008)\citenamefont {Abe} \emph
  {et~al.}}]{Abe:2008aa}%
  \BibitemOpen
  \bibfield  {author} {\bibinfo {author} {\bibfnamefont {S.}~\bibnamefont
  {Abe}} \emph {et~al.} (\bibinfo {collaboration} {KamLAND Collaboration}),\
  }\bibfield  {title} {\enquote {\bibinfo {title} {{Precision Measurement of
  Neutrino Oscillation Parameters with KamLAND}},}\ }\href {\doibase
  10.1103/PhysRevLett.100.221803} {\bibfield  {journal} {\bibinfo  {journal}
  {Phys.Rev.Lett.}\ }\textbf {\bibinfo {volume} {100}},\ \bibinfo {pages}
  {221803} (\bibinfo {year} {2008})}\BibitemShut {NoStop}%
\bibitem [{\citenamefont {An}\ \emph {et~al.}(2012)\citenamefont {An} \emph
  {et~al.}}]{An:2012eh}%
  \BibitemOpen
  \bibfield  {author} {\bibinfo {author} {\bibfnamefont {F.~P.}\ \bibnamefont
  {An}} \emph {et~al.} (\bibinfo {collaboration} {DAYA-BAY Collaboration}),\
  }\bibfield  {title} {\enquote {\bibinfo {title} {{Observation of
  electron-antineutrino disappearance at Daya Bay}},}\ }\href {\doibase
  10.1103/PhysRevLett.108.171803} {\bibfield  {journal} {\bibinfo  {journal}
  {Phys.Rev.Lett.}\ }\textbf {\bibinfo {volume} {108}},\ \bibinfo {pages}
  {171803} (\bibinfo {year} {2012})}\BibitemShut {NoStop}%
\bibitem [{\citenamefont {Abe}\ \emph {et~al.}(2013{\natexlab{a}})\citenamefont
  {Abe} \emph {et~al.}}]{Abe:2013sxa}%
  \BibitemOpen
  \bibfield  {author} {\bibinfo {author} {\bibfnamefont {Y.}~\bibnamefont
  {Abe}} \emph {et~al.} (\bibinfo {collaboration} {Double Chooz
  Collaboration}),\ }\bibfield  {title} {\enquote {\bibinfo {title} {{First
  Measurement of $\theta_{13}$ from Delayed Neutron Capture on Hydrogen in the
  Double Chooz Experiment}},}\ }\href {\doibase 10.1016/j.physletb.2013.04.050}
  {\bibfield  {journal} {\bibinfo  {journal} {Phys.Lett.}\ }\textbf {\bibinfo
  {volume} {B723}},\ \bibinfo {pages} {66--70} (\bibinfo {year}
  {2013}{\natexlab{a}})},\ \Eprint {http://arxiv.org/abs/1301.2948}
  {arXiv:1301.2948 [hep-ex]} \BibitemShut {NoStop}%
\bibitem [{\citenamefont {Ahn}\ \emph {et~al.}(2012)\citenamefont {Ahn} \emph
  {et~al.}}]{Ahn:2012nd}%
  \BibitemOpen
  \bibfield  {author} {\bibinfo {author} {\bibfnamefont {J.~K.}\ \bibnamefont
  {Ahn}} \emph {et~al.} (\bibinfo {collaboration} {RENO collaboration}),\
  }\bibfield  {title} {\enquote {\bibinfo {title} {{Observation of Reactor
  Electron Antineutrino Disappearance in the RENO Experiment}},}\ }\href
  {\doibase 10.1103/PhysRevLett.108.191802} {\bibfield  {journal} {\bibinfo
  {journal} {Phys.Rev.Lett.}\ }\textbf {\bibinfo {volume} {108}},\ \bibinfo
  {pages} {191802} (\bibinfo {year} {2012})},\ \Eprint
  {http://arxiv.org/abs/1204.0626} {arXiv:1204.0626 [hep-ex]} \BibitemShut
  {NoStop}%
\bibitem [{\citenamefont {Ahn}\ \emph {et~al.}(2006)\citenamefont {Ahn} \emph
  {et~al.}}]{Ahn:2006zza}%
  \BibitemOpen
  \bibfield  {author} {\bibinfo {author} {\bibfnamefont {M.~H.}\ \bibnamefont
  {Ahn}} \emph {et~al.} (\bibinfo {collaboration} {K2K Collaboration}),\
  }\bibfield  {title} {\enquote {\bibinfo {title} {{Measurement of Neutrino
  Oscillation by the K2K Experiment}},}\ }\href {\doibase
  10.1103/PhysRevD.74.072003} {\bibfield  {journal} {\bibinfo  {journal}
  {Phys.Rev.}\ }\textbf {\bibinfo {volume} {D74}},\ \bibinfo {pages} {072003}
  (\bibinfo {year} {2006})},\ \Eprint {http://arxiv.org/abs/hep-ex/0606032}
  {arXiv:hep-ex/0606032 [hep-ex]} \BibitemShut {NoStop}%
\bibitem [{\citenamefont {Adamson}\ \emph {et~al.}(2013)\citenamefont {Adamson}
  \emph {et~al.}}]{Adamson:2013whj}%
  \BibitemOpen
  \bibfield  {author} {\bibinfo {author} {\bibfnamefont {P.}~\bibnamefont
  {Adamson}} \emph {et~al.} (\bibinfo {collaboration} {MINOS Collaboration}),\
  }\bibfield  {title} {\enquote {\bibinfo {title} {{Measurement of Neutrino and
  Antineutrino Oscillations Using Beam and Atmospheric Data in MINOS}},}\
  }\href {\doibase 10.1103/PhysRevLett.110.251801} {\bibfield  {journal}
  {\bibinfo  {journal} {Phys.Rev.Lett.}\ }\textbf {\bibinfo {volume} {110}},\
  \bibinfo {pages} {251801} (\bibinfo {year} {2013})},\ \Eprint
  {http://arxiv.org/abs/1304.6335} {arXiv:1304.6335 [hep-ex]} \BibitemShut
  {NoStop}%
\bibitem [{\citenamefont {Abe}\ \emph {et~al.}(2012{\natexlab{b}})\citenamefont
  {Abe} \emph {et~al.}}]{Abe:2012gx}%
  \BibitemOpen
  \bibfield  {author} {\bibinfo {author} {\bibfnamefont {K.}~\bibnamefont
  {Abe}} \emph {et~al.} (\bibinfo {collaboration} {T2K Collaboration}),\
  }\bibfield  {title} {\enquote {\bibinfo {title} {{First Muon-Neutrino
  Disappearance Study with an Off-Axis Beam}},}\ }\href {\doibase
  10.1103/PhysRevD.85.031103} {\bibfield  {journal} {\bibinfo  {journal}
  {Phys.Rev.}\ }\textbf {\bibinfo {volume} {D85}},\ \bibinfo {pages} {031103}
  (\bibinfo {year} {2012}{\natexlab{b}})}\BibitemShut {NoStop}%
\bibitem [{\citenamefont {Abe}\ \emph {et~al.}(2013{\natexlab{b}})\citenamefont
  {Abe} \emph {et~al.}}]{Abe:2012jj}%
  \BibitemOpen
  \bibfield  {author} {\bibinfo {author} {\bibfnamefont {K.}~\bibnamefont
  {Abe}} \emph {et~al.} (\bibinfo {collaboration} {Super-Kamiokande
  Collaboration}),\ }\bibfield  {title} {\enquote {\bibinfo {title} {{Evidence
  for the Appearance of Atmospheric Tau Neutrinos in Super-Kamiokande}},}\
  }\href {\doibase 10.1103/PhysRevLett.110.181802} {\bibfield  {journal}
  {\bibinfo  {journal} {Phys.Rev.Lett.}\ }\textbf {\bibinfo {volume} {110}},\
  \bibinfo {pages} {181802} (\bibinfo {year} {2013}{\natexlab{b}})}\BibitemShut
  {NoStop}%
\bibitem [{\citenamefont {Abe}\ \emph {et~al.}(2013{\natexlab{c}})\citenamefont
  {Abe} \emph {et~al.}}]{Abe:2013xua}%
  \BibitemOpen
  \bibfield  {author} {\bibinfo {author} {\bibfnamefont {K.}~\bibnamefont
  {Abe}} \emph {et~al.} (\bibinfo {collaboration} {T2K Collaboration}),\
  }\bibfield  {title} {\enquote {\bibinfo {title} {{Evidence of Electron
  Neutrino Appearance in a Muon Neutrino Beam}},}\ }\href {\doibase
  10.1103/PhysRevD.88.032002} {\bibfield  {journal} {\bibinfo  {journal}
  {Phys.Rev.}\ }\textbf {\bibinfo {volume} {D88}},\ \bibinfo {pages} {032002}
  (\bibinfo {year} {2013}{\natexlab{c}})},\ \Eprint
  {http://arxiv.org/abs/1304.0841} {arXiv:1304.0841 [hep-ex]} \BibitemShut
  {NoStop}%
\bibitem [{\citenamefont {Agafonova}\ \emph {et~al.}(2013)\citenamefont
  {Agafonova} \emph {et~al.}}]{Agafonova:2013dtp}%
  \BibitemOpen
  \bibfield  {author} {\bibinfo {author} {\bibfnamefont {N.}~\bibnamefont
  {Agafonova}} \emph {et~al.} (\bibinfo {collaboration} {OPERA
  Collaboration}),\ }\bibfield  {title} {\enquote {\bibinfo {title} {{New
  results on $\nu_\mu \to \nu_\tau$ appearance with the OPERA experiment in the
  CNGS beam}},}\ }\href {\doibase 10.1007/JHEP11(2013)036} {\bibfield
  {journal} {\bibinfo  {journal} {JHEP}\ }\textbf {\bibinfo {volume} {1311}},\
  \bibinfo {pages} {036} (\bibinfo {year} {2013})},\ \Eprint
  {http://arxiv.org/abs/1308.2553} {arXiv:1308.2553 [hep-ex]} \BibitemShut
  {NoStop}%
\bibitem [{\citenamefont {Wolfenstein}(1978)}]{Wolfenstein:1977ue}%
  \BibitemOpen
  \bibfield  {author} {\bibinfo {author} {\bibfnamefont {L.}~\bibnamefont
  {Wolfenstein}},\ }\bibfield  {title} {\enquote {\bibinfo {title} {{Neutrino
  Oscillations in Matter}},}\ }\href {\doibase 10.1103/PhysRevD.17.2369}
  {\bibfield  {journal} {\bibinfo  {journal} {Phys.Rev.}\ }\textbf {\bibinfo
  {volume} {D17}},\ \bibinfo {pages} {2369--2374} (\bibinfo {year}
  {1978})}\BibitemShut {NoStop}%
\bibitem [{\citenamefont {Mikheyev}\ and\ \citenamefont
  {Smirnov}(1989)}]{Mikheyev:1989dy}%
  \BibitemOpen
  \bibfield  {author} {\bibinfo {author} {\bibfnamefont {S.~P.}\ \bibnamefont
  {Mikheyev}}\ and\ \bibinfo {author} {\bibfnamefont {A.~Yu.}\ \bibnamefont
  {Smirnov}},\ }\bibfield  {title} {\enquote {\bibinfo {title} {{Resonant
  neutrino oscillations in matter}},}\ }\href {\doibase
  10.1016/0146-6410(89)90008-2} {\bibfield  {journal} {\bibinfo  {journal}
  {Prog.Part.Nucl.Phys.}\ }\textbf {\bibinfo {volume} {23}},\ \bibinfo {pages}
  {41--136} (\bibinfo {year} {1989})}\BibitemShut {NoStop}%
\bibitem [{\citenamefont {Desai}\ \emph {et~al.}(2008)\citenamefont {Desai}
  \emph {et~al.}}]{Desai:2007ra}%
  \BibitemOpen
  \bibfield  {author} {\bibinfo {author} {\bibfnamefont {S.}~\bibnamefont
  {Desai}} \emph {et~al.} (\bibinfo {collaboration} {Super-Kamiokande
  Collaboration}),\ }\bibfield  {title} {\enquote {\bibinfo {title} {{Study of
  TeV neutrinos with upward showering muons in Super-Kamiokande}},}\ }\href
  {\doibase 10.1016/j.astropartphys.2007.11.005} {\bibfield  {journal}
  {\bibinfo  {journal} {Astropart.Phys.}\ }\textbf {\bibinfo {volume} {29}},\
  \bibinfo {pages} {42--54} (\bibinfo {year} {2008})},\ \Eprint
  {http://arxiv.org/abs/0711.0053} {arXiv:0711.0053 [hep-ex]} \BibitemShut
  {NoStop}%
\bibitem [{\citenamefont {Ashie}\ \emph {et~al.}(2005)\citenamefont {Ashie}
  \emph {et~al.}}]{Ashie:2005ik}%
  \BibitemOpen
  \bibfield  {author} {\bibinfo {author} {\bibfnamefont {Y.}~\bibnamefont
  {Ashie}} \emph {et~al.} (\bibinfo {collaboration} {Super-Kamiokande
  Collaboration}),\ }\bibfield  {title} {\enquote {\bibinfo {title} {{A
  Measurement of atmospheric neutrino oscillation parameters by
  SUPER-KAMIOKANDE I}},}\ }\href {\doibase 10.1103/PhysRevD.71.112005}
  {\bibfield  {journal} {\bibinfo  {journal} {Phys.Rev.}\ }\textbf {\bibinfo
  {volume} {D71}},\ \bibinfo {pages} {112005} (\bibinfo {year}
  {2005})}\BibitemShut {NoStop}%
\bibitem [{\citenamefont {Wendell}\ \emph {et~al.}(2010)\citenamefont {Wendell}
  \emph {et~al.}}]{Wendell:2010md}%
  \BibitemOpen
  \bibfield  {author} {\bibinfo {author} {\bibfnamefont {R.}~\bibnamefont
  {Wendell}} \emph {et~al.} (\bibinfo {collaboration} {Super-Kamiokande
  Collaboration}),\ }\bibfield  {title} {\enquote {\bibinfo {title}
  {{Atmospheric neutrino oscillation analysis with sub-leading effects in
  Super-Kamiokande I, II, and III}},}\ }\href {\doibase
  10.1103/PhysRevD.81.092004} {\bibfield  {journal} {\bibinfo  {journal}
  {Phys.Rev.}\ }\textbf {\bibinfo {volume} {D81}},\ \bibinfo {pages} {092004}
  (\bibinfo {year} {2010})}\BibitemShut {NoStop}%
\bibitem [{\citenamefont {Abe}\ \emph {et~al.}(2014{\natexlab{a}})\citenamefont
  {Abe} \emph {et~al.}}]{sterilepaper}%
  \BibitemOpen
  \bibfield  {author} {\bibinfo {author} {\bibfnamefont {K.}~\bibnamefont
  {Abe}} \emph {et~al.} (\bibinfo {collaboration} {Super-Kamiokande
  Collaboration}),\ }\href@noop {} {\enquote {\bibinfo {title} {{Limits on
  Sterile Neutrino Mixing using Atmospheric Neutrinos in Super-Kamiokande}},}\
  } (\bibinfo {year} {2014}{\natexlab{a}}),\ \Eprint
  {http://arxiv.org/abs/1410.2008} {arXiv:1410.2008 [hep-ph]} \BibitemShut
  {NoStop}%
\bibitem [{\citenamefont {Abe}\ \emph {et~al.}(2013{\natexlab{d}})\citenamefont
  {Abe}, \citenamefont {Hayato}, \citenamefont {Iida}, \citenamefont {Iyogi},
  \citenamefont {Kameda} \emph {et~al.}}]{Abe:2013gga}%
  \BibitemOpen
  \bibfield  {author} {\bibinfo {author} {\bibfnamefont {K.}~\bibnamefont
  {Abe}}, \bibinfo {author} {\bibfnamefont {Y.}~\bibnamefont {Hayato}},
  \bibinfo {author} {\bibfnamefont {T.}~\bibnamefont {Iida}}, \bibinfo {author}
  {\bibfnamefont {K.}~\bibnamefont {Iyogi}}, \bibinfo {author} {\bibfnamefont
  {J.}~\bibnamefont {Kameda}},  \emph {et~al.},\ }\bibfield  {title} {\enquote
  {\bibinfo {title} {{Calibration of the Super-Kamiokande Detector}},}\
  }\href@noop {} {\  (\bibinfo {year} {2013}{\natexlab{d}})},\ \Eprint
  {http://arxiv.org/abs/hep-ex/1307.0162} {arXiv:hep-ex/1307.0162
  [physics.ins-det]} \BibitemShut {NoStop}%
\bibitem [{\citenamefont {Maki}\ \emph {et~al.}(1962)\citenamefont {Maki},
  \citenamefont {Nakagawa},\ and\ \citenamefont {Sakata}}]{Maki:1962mu}%
  \BibitemOpen
  \bibfield  {author} {\bibinfo {author} {\bibfnamefont {Ziro}\ \bibnamefont
  {Maki}}, \bibinfo {author} {\bibfnamefont {Masami}\ \bibnamefont {Nakagawa}},
  \ and\ \bibinfo {author} {\bibfnamefont {Shoichi}\ \bibnamefont {Sakata}},\
  }\bibfield  {title} {\enquote {\bibinfo {title} {{Remarks on the unified
  model of elementary particles}},}\ }\href {\doibase 10.1143/PTP.28.870}
  {\bibfield  {journal} {\bibinfo  {journal} {Prog.Theor.Phys.}\ }\textbf
  {\bibinfo {volume} {28}},\ \bibinfo {pages} {870--880} (\bibinfo {year}
  {1962})}\BibitemShut {NoStop}%
\bibitem [{\citenamefont {Dziewonski}\ and\ \citenamefont
  {Anderson}(1981)}]{Dziewonski:1981xy}%
  \BibitemOpen
  \bibfield  {author} {\bibinfo {author} {\bibfnamefont {A.~M.}\ \bibnamefont
  {Dziewonski}}\ and\ \bibinfo {author} {\bibfnamefont {D.~L.}\ \bibnamefont
  {Anderson}},\ }\bibfield  {title} {\enquote {\bibinfo {title} {{Preliminary
  reference earth model}},}\ }\href {\doibase 10.1016/0031-9201(81)90046-7}
  {\bibfield  {journal} {\bibinfo  {journal} {Phys.Earth Planet.Interiors}\
  }\textbf {\bibinfo {volume} {25}},\ \bibinfo {pages} {297--356} (\bibinfo
  {year} {1981})}\BibitemShut {NoStop}%
\bibitem [{\citenamefont {Wang}(2007)}]{Wang:2007zzl}%
  \BibitemOpen
  \bibfield  {author} {\bibinfo {author} {\bibfnamefont {Wei}\ \bibnamefont
  {Wang}},\ }\emph {\bibinfo {title} {{Studies of non-standard effects in
  atmospheric neutrino oscillations of Super-Kamiokande}}},\ \href@noop {}
  {Ph.D. thesis},\ \bibinfo  {school} {{Boston University}} (\bibinfo {year}
  {2007})\BibitemShut {NoStop}%
\bibitem [{\citenamefont {Barger}\ \emph {et~al.}(1980)\citenamefont {Barger},
  \citenamefont {Whisnant}, \citenamefont {Pakvasa},\ and\ \citenamefont
  {Phillips}}]{Barger:1980tf}%
  \BibitemOpen
  \bibfield  {author} {\bibinfo {author} {\bibfnamefont {Vernon~D.}\
  \bibnamefont {Barger}}, \bibinfo {author} {\bibfnamefont {K.}~\bibnamefont
  {Whisnant}}, \bibinfo {author} {\bibfnamefont {S.}~\bibnamefont {Pakvasa}}, \
  and\ \bibinfo {author} {\bibfnamefont {R.~J.~N.}\ \bibnamefont {Phillips}},\
  }\bibfield  {title} {\enquote {\bibinfo {title} {{Matter Effects on
  Three-Neutrino Oscillations}},}\ }\href {\doibase 10.1103/PhysRevD.22.2718}
  {\bibfield  {journal} {\bibinfo  {journal} {Phys.Rev.}\ }\textbf {\bibinfo
  {volume} {D22}},\ \bibinfo {pages} {2718} (\bibinfo {year}
  {1980})}\BibitemShut {NoStop}%
\bibitem [{\citenamefont {Katori}(2012)}]{Katori:2012pe}%
  \BibitemOpen
  \bibfield  {author} {\bibinfo {author} {\bibfnamefont {Teppei}\ \bibnamefont
  {Katori}} (\bibinfo {collaboration} {MiniBooNE Collaboration}),\ }\bibfield
  {title} {\enquote {\bibinfo {title} {{Tests of Lorentz and CPT violation with
  MiniBooNE neutrino oscillation excesses}},}\ }\href {\doibase
  10.1142/S0217732312300248} {\bibfield  {journal} {\bibinfo  {journal}
  {Mod.Phys.Lett.}\ }\textbf {\bibinfo {volume} {A27}},\ \bibinfo {pages}
  {1230024} (\bibinfo {year} {2012})},\ \Eprint
  {http://arxiv.org/abs/1206.6915} {arXiv:1206.6915 [hep-ex]} \BibitemShut
  {NoStop}%
\bibitem [{\citenamefont {Katori}\ and\ \citenamefont
  {Spitz}(2014)}]{Katori:2013jca}%
  \BibitemOpen
  \bibfield  {author} {\bibinfo {author} {\bibfnamefont {Teppei}\ \bibnamefont
  {Katori}}\ and\ \bibinfo {author} {\bibfnamefont {Joshua}\ \bibnamefont
  {Spitz}},\ }\bibfield  {title} {\enquote {\bibinfo {title} {{Testing Lorentz
  Symmetry with the Double Chooz Experiment}},}\ }in\ \href@noop {} {\emph
  {\bibinfo {booktitle} {{CPT and Lorentz Symmetry VI}}}}\ (\bibinfo
  {publisher} {{World Scientific}},\ \bibinfo {address} {{Singapore}},\
  \bibinfo {year} {2014})\ \Eprint {http://arxiv.org/abs/1307.5805}
  {arXiv:1307.5805 [hep-ph]} \BibitemShut {NoStop}%
\bibitem [{\citenamefont {Fogli}\ \emph {et~al.}(2002)\citenamefont {Fogli},
  \citenamefont {Lisi}, \citenamefont {Marrone}, \citenamefont {Montanino},\
  and\ \citenamefont {Palazzo}}]{Fogli:2002pt}%
  \BibitemOpen
  \bibfield  {author} {\bibinfo {author} {\bibfnamefont {G.~L.}\ \bibnamefont
  {Fogli}}, \bibinfo {author} {\bibfnamefont {E.}~\bibnamefont {Lisi}},
  \bibinfo {author} {\bibfnamefont {A.}~\bibnamefont {Marrone}}, \bibinfo
  {author} {\bibfnamefont {D.}~\bibnamefont {Montanino}}, \ and\ \bibinfo
  {author} {\bibfnamefont {A.}~\bibnamefont {Palazzo}},\ }\bibfield  {title}
  {\enquote {\bibinfo {title} {{Getting the most from the statistical analysis
  of solar neutrino oscillations}},}\ }\href {\doibase
  10.1103/PhysRevD.66.053010} {\bibfield  {journal} {\bibinfo  {journal}
  {Phys.Rev.}\ }\textbf {\bibinfo {volume} {D66}},\ \bibinfo {pages} {053010}
  (\bibinfo {year} {2002})},\ \Eprint {http://arxiv.org/abs/hep-ph/0206162}
  {arXiv:hep-ph/0206162 [hep-ph]} \BibitemShut {NoStop}%
\bibitem [{\citenamefont {Abe}\ \emph {et~al.}(2014{\natexlab{b}})\citenamefont
  {Abe} \emph {et~al.}}]{Abe:2014ugx}%
  \BibitemOpen
  \bibfield  {author} {\bibinfo {author} {\bibfnamefont {K.}~\bibnamefont
  {Abe}} \emph {et~al.} (\bibinfo {collaboration} {T2K Collaboration}),\
  }\bibfield  {title} {\enquote {\bibinfo {title} {{Precise Measurement of the
  Neutrino Mixing Parameter $\theta_{23}$ from Muon Neutrino Disappearance in
  an Off-axis Beam}},}\ }\href {\doibase 10.1103/PhysRevLett.112.181801}
  {\bibfield  {journal} {\bibinfo  {journal} {Phys.Rev.Lett.}\ }\textbf
  {\bibinfo {volume} {112}},\ \bibinfo {pages} {181801} (\bibinfo {year}
  {2014}{\natexlab{b}})},\ \Eprint {http://arxiv.org/abs/1403.1532}
  {arXiv:1403.1532 [hep-ex]} \BibitemShut {NoStop}%
\bibitem [{\citenamefont {Beringer}\ \emph {et~al.}(2012)\citenamefont
  {Beringer} \emph {et~al.}}]{PDG}%
  \BibitemOpen
  \bibfield  {author} {\bibinfo {author} {\bibfnamefont {J.}~\bibnamefont
  {Beringer}} \emph {et~al.} (\bibinfo {collaboration} {Particle Data Group}),\
  }\bibfield  {title} {\enquote {\bibinfo {title} {{Review of Particle Physics
  (RPP)}},}\ }\href {\doibase 10.1103/PhysRevD.86.010001} {\bibfield  {journal}
  {\bibinfo  {journal} {Phys.Rev.}\ }\textbf {\bibinfo {volume} {D86}},\
  \bibinfo {pages} {010001} (\bibinfo {year} {2012})}\BibitemShut {NoStop}%
\bibitem [{\citenamefont {Diaz}(2014)}]{JSDiaz}%
  \BibitemOpen
  \bibfield  {author} {\bibinfo {author} {\bibfnamefont {Jorge~S.}\
  \bibnamefont {Diaz}},\ }\href@noop {} {} (\bibinfo {year} {2014}),\ \bibinfo
  {note} {in preparation, {IUHET} 585}\BibitemShut {NoStop}%
\end{thebibliography}%


\end{document}